\documentclass[fleqn,usenatbib]{mnras}
\usepackage{newtxtext,newtxmath,hyperref}
\usepackage[T1]{fontenc}
\usepackage{graphicx}	
\usepackage{amsmath}	
\usepackage{enumitem}
\usepackage[english]{babel}
\usepackage{gensymb}
\usepackage{hyperref}
\usepackage{comment}
\usepackage{xcolor}
\usepackage{soul}
\usepackage{algorithm2e}

\newcommand{\secref}[1]{\hyperref[#1]{section~\ref*{#1}}}
\newcommand{\appref}[1]{\hyperref[#1]{appendix~\ref*{#1}}}

\title[LEDA antenna modelling]{Antenna beam characterisation for the global 21cm experiment LEDA and its impact on signal model parameter reconstruction}

\author[M. Spinelli et al.]{M.~Spinelli$^{1,2,3}$\thanks{E-mail: mspinelli@phys.ethz.ch}, G.~Kyriakou$^{4,5}$\thanks{E-mail: georgios.kyriakou@inaf.it}, G.~Bernardi$^{6,7,8}$, P.~Bolli$^{4}$, L.J.~Greenhill$^{9}$, A.~Fialkov$^{10,11}$ \and H.~Garsden$^{9,12}$\\
$^{1}$Institute for Particle Physics and Astrophysics,
ETH Z{\"u}rich, Wolfgang Pauli Strasse 27, 8093 Z{\"u}rich, Switzerland\\
$^{2}$Department of Physics and Astronomy, University of the Western Cape, Robert Sobukhwe Road, Bellville, 7535, South Africa\\
$^{3}$INAF-Osservatorio Astronomico di Trieste, Via G.B. Tiepolo 11, 34143 Trieste, Italy\\
$^{4}$INAF-Osservatorio Astronomico di Arcetri, Largo Enrico Fermi 5, 50125, Firenze, Italy\\
$^{5}$Department of Physics and Astronomy, University of Florence, P.za di San Marco, 4, 50121, Firenze, Italy \\
$^{6}$INAF-Istituto di Radioastronomia, via Gobetti 101, 40129, Bologna, Italy\\
$^{7}$Department of Physics \& Electronics, Artillery Road, Rhodes University, Makhanda, South Africa\\
$^{8}$South African Radio Astronomy Observatory, FIR street, Observatory, Cape Town, South Africa\\
$^{9}$Center for Astrophysics | Harvard \& Smithsonian, 60 Garden Street, Cambridge MA 02138 USA\\
 $^{10}$Kavli Institute for Cosmology, Madingley Road, Cambridge CB3 0HA, UK \\
 $^{11}$Institute of Astronomy, University of Cambridge, Madingley Road, Cambridge CB3 0HA, UK\\
 $^{12}$Astronomy Unit, School of Physical and Chemical Sciences, Queen Mary University of London,
Mile End Road, London, E1 4NS, UK\\}
\date{Accepted XXX. Received YYY; in original form ZZZ}
\date{\today}

\begin{document}
\label{firstpage}
\pagerange{\pageref{firstpage}--\pageref{lastpage}}
\maketitle

\begin{abstract} 
Cosmic Dawn, the onset of star formation in the early universe, can in principle be studied via the 21cm transition of neutral hydrogen, for which a sky-averaged absorption signal, redshifted to MHz frequencies, is predicted to be {\it O}(10-100)\,mK. Detection requires separation of the 21cm signal from bright chromatic foreground emission due to Galactic structure, and the characterisation of how it couples to instrumental response.
In this work, we present characterisation of antenna gain patterns for the Large-aperture Experiment to detect the Dark Ages (LEDA) via simulations, assessing the effects of the antenna ground-plane geometries used, and measured soil properties.  We then investigate the impact of beam pattern uncertainties on the reconstruction of a Gaussian absorption feature.
Assuming the pattern is known and correcting for the chromaticity of the instrument, the foregrounds can be modelled with a log-polynomial, and the 21cm signal identified with high accuracy. However, uncertainties on the soil properties lead to \textperthousand\ changes in the chromaticity that can bias the signal recovery. The bias can be up to a factor of two in amplitude and up to few \% in the frequency location.
These effects do not appear to be mitigated by larger ground planes, conversely gain patterns with larger ground planes exhibit more complex frequency structure, significantly compromising the parameter reconstruction.
Our results, consistent with findings from other antenna design studies, emphasise the importance of chromatic response and suggest caution in assuming log-polynomial foreground models in global signal experiments. 
\end{abstract}



\section{Introduction}\label{sec:intro}
The 21cm transition of neutral hydrogen (HI) is predicted to trace cold diffuse gas during Cosmic Dawn, the epoch during which the first generation of stars formed, $\sim 100$ million years after the Big Bang.  Prior to this, the spin temperature of the transition was likely in equilibrium with the Cosmic Microwave Background, well above the gas kinetic temperature.  The rise of Ly-$\alpha$ background radiation from pockets of star formation coupled the spin temperature to the gas kinetic temperature via the Wouthuysen-Field effect \citep[WF,][]{Wouthuysen1952, Field1958}.  The growing population of stellar remnants created an X-ray background that drove the gas kinetic and the spin temperatures higher \citep[e.g.][]{Furlanetto2006,Pritchard2010}. Averaged over the sky, the relative proportion of Ly-$\alpha$ coupling and X-ray heating varied with redshift and could create a broad absorption trough in the spectrum of the Cosmic Microwave Background.

Detection of the predicted trough would provide unique information about the formation of the first luminous structures in the Universe \citep[e.g.,][]{Barkana2005,Furlanetto2006,Fialkov2013,Mirocha2014,Mesinger2016,Mirocha2019, Reis2020,Magg2021, Gessey2022, Reis2022}, and of the thermal history of the intergalactic medium \citep{Pritchard2007,Mesinger2013,Fialkov2014,Reis2021}. 

The signal redshifted to radio frequencies $\lesssim 100$~MHz and of order $-100$~mK in amplitude would in principle be detectable using meter-scale antennas and an integration time on the order of 100 hours, for sufficiently accurate radiometric calibration, and well understood celestial foregrounds (1000-10.000~K) and antenna gain patterns as a function of frequency.

The scientific importance of a measurement of the 21cm global signal absorption feature has motivated the effort to build several experiments in different locations and with different technical approaches.

The Experiment to Detect the Global Epoch-of-Reionization Signature (EDGES) relies upon a broad-band horizontal ‘blade’ dipole design and is deployed at the Murchison Radio-astronomy Observatory (MRO) in Western Australia. It reported the detection of a broad $-520$~mK absorption profile centered at $\sim78$~MHz \citep{Bowman2018} supported by validation tests also described by \citet{Mahesh2021}. 
The profile, more than a factor two deeper than predicted from theory based on standard physics \citep[e.g. ][]{Cohen2017,Reis2021}, has triggered several studies aimed at an understanding of its origin.
From a theoretical point of view, such a result could imply that either the temperature of the radio background is higher than the  CMB \citep[e.g.][]{Bowman2018, Feng2018, Reis2020} or the neutral gas at redshift $\sim 17$ was colder than expected, possibly due to the interaction with dark matter  \citep[e.g.][]{Barkana2018, Munoz2018, Fialkov2019, Liu2019}. Other studies suggested the presence of un-modelled systematics \citep{Hills2018,Singh2019,Spinelli2019,Bevins2020}, weaknesses in the analysis pipeline due to inaccurate modelling of the foregrounds \citep[e.g.][]{Singh2019}, along with flaws in the statistical interpretation of the data \citep[e.g.][]{Sims2020}. 

The third generation of the Shaped Antenna measurement of the background RAdio Spectrum (SARAS) features a spectral radiometer based on a monocone antenna and receiver floated on a lake in Southern India \citep{Nambissan2021}, has recently presented an analysis of their data \citep{Singh2021}, rejecting the EDGES absorption  profile at 95.3$\%$ confidence level.

Besides EDGES and SARAS, there are several 21cm global signal experiments underway, such as the Probing Radio Intensity at high-Z from Marion (PRIZM) experiment \citep{Philip2019}, the Broadband Instrument for Global HydrOgen ReioNisation Signal \citep[BIGHORNS;][]{Sokolowski2015}, the Radio Experiment for the Analysis of Cosmic Hydrogen (REACH)
\citep{Cumner2021} and the  Mapper of the IGM Spin Temperature (MIST)\footnote{\hyperlink{http://www.physics.mcgill.ca/mist/}{http://www.physics.mcgill.ca/mist/}}.

\medskip
This work focuses on the Large-aperture Experiment to detect the Dark Age \citep[LEDA;][]{Price2018} equipped and operated between two and five dual polarisation antennas within the Owens Valley Radio Observatory Long Wavelength Array station (OVRO-LWA) for radiometry, using custom RF and digital signal processing to enable the requisite timing, calibration, and stability (2013-2020).

Using early radiometric data, \citet{Bernardi2016} set a coarse upper limit for the amplitude and the width for the anticipated absorption trough.  Using later data (Dec. 2018 - May 2019),
\citet{Spinelli2021} constrained the spectral index, $\beta$, of Galactic radio emission in the northern sky (60-87~MHz), obtaining values compatible with expectation from simulations and other measurements.
We note that study by \citet{Garsden2021} was distinct, using contemporaneous interferometric data, generated by the LEDA correlator that was part of OVRO-LWA, to characterize the effect of systematic errors in calibration on dynamic range for 2D cylindrical spatial power spectra.  (See also \citet{Eastwood2018} regarding an alternate approach, also using LEDA interferometric data.)

The critical challenge in analysing radiometric data is 
accurate
subtraction of the bright foregrounds, normally attempted by modelling the spectrally smooth 
emission with an N-term log-polynomial and performing Bayesian
inference for both foreground and background signals \citep[e.g.,][]{Bernardi2015,Bernardi2016,Bowman2018,Singh2021}.

Chromaticity is a key factor  in antenna design \citep[e.g.,][]{Mozdzen2016,Nambissan2021,Cumner2021}
and chromatic beam effects  can limit the effectiveness of foreground subtraction \citep[e.g.][]{Vedantham2014}. 
Accurate antenna beam simulations,  including realistic ground plane and soil descriptions, are thus fundamental to understand
radiometric data. 
\citet{Mahesh2021} explored antenna beam modelling as a source of uncertainty in global signal measurement, checking the 
stability of the signal reported by EDGES
with respect to choice of numerical electromagnetic solver code. \citet{Bradley2019} instead investigated how a possible systematic artefact within
the antenna ground plane may produce broad absorption
features in the 
spectra. 

In preparation for future data analysis, the REACH project 
 has developed a software pipeline that can incorporate more efficiently the effect of beam chromaticity coupled with a non-trivial scaling in frequency of the foreground \citep[][]{Anstey2020}. This 
strategy, although computationally costly, leads to robust 21cm global signal extraction in
simulations and can thus 
enable better
informed optimisation of antenna design \citep{Cumner2021,Anstey2022}. Other 
techniques that incorporate beam effects in the foreground model using machine learning methods have been
proposed by \citet[e.g.,][]{Tauscher2020,Tauscher2021}.

\medskip In this paper, we make use of the
common log-polynomial model to parametrise LEDA mock simulated spectra and we analyse in detail its limitations when  
a refined soil modelling with realistic values of its dielectric properties and conductivity is considered.  We explore different ground planes, focusing on the ones used in the field
during data acquisition and discuss the
impact on the Bayesian reconstruction of signal and foreground parameters.

This paper is organised as follows. We describe {\it in situ} 
measurements of soil complex-permittivity and improved electromagnetic simulations in \autoref{sec:beam}, including 
the details of the multi-layer modelling of the ground.
We describe the construction of simulated spectra in \secref{sec:Tobs} and the computation of the
chromaticity correction factor in \secref{sec:chrom}. In \secref{sec:model}, we discuss the accuracy of the smooth foreground approximation in the presence of realistic beam modelling.  We analyse in \secref{sec:recon} the impact of our realistic beam on the reconstruction of the parameters describing a neutral hydrogen absorption feature.
A discussion of the results and our conclusions are presented in \secref{sec:conclusions}.

\section{Modelling LEDA antennas}\label{sec:beam}

In this section we present in detail improved electromagnetic models for antennas used for radiometry by LEDA. 
Driven by science requirements, the frequency range of interest is between 50 and 87 MHz (although some of the simulation results are shown in a larger frequency range). 
In \secref{sec:ground}, we present the antenna geometry and the ground planes we have considered. Note that the models reflect as-built designs. In \secref{sec:soilmes}, we describe the {\it in situ} soil permittivty data gathering. In \secref{sec:oldbeam}, we review the analytic beam model used in previous analyses of LEDA data, while the focus of \secref{sec:FEKO} is incorporation of the soil permittivty data and alternate ground-plane geometries.

\subsection{Antenna and ground plane description}\label{sec:ground}

\begin{figure}
        \includegraphics[width=\columnwidth]{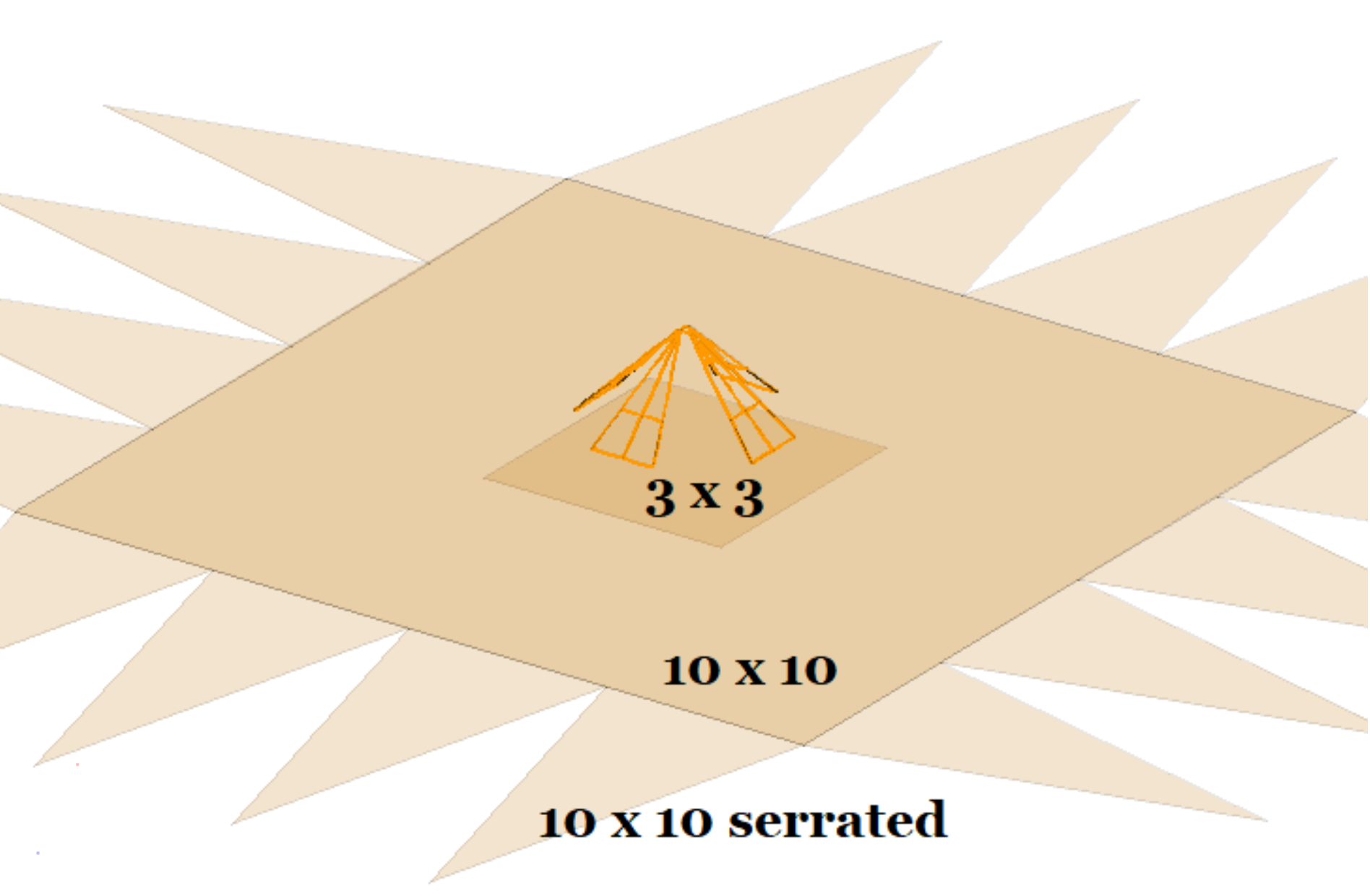}
        \caption{LEDA antenna 3D view of geometry with the ground planes to be used in the analysis. Dimensions are shown in meters.}
\label{fig:example_ground}
\end{figure}

We outline here the basic geometrical properties of the antennas used by LEDA for radiometry, which excluding the ground plane are similar to those used in general for construction of Long Wavelength Array stations \citep[][and references therein]{Taylor2012} and OVRO-LWA specifically.

Each antenna comprises two pairs of triangular dipole arms 1.50~m long, angled downward by $45^\circ$ (\autoref{fig:example_ground}). 
We focus our analysis on the east-west orientation.

The default OVRO-LWA 
antenna ground plane \citep{Paravastu2007, Schmitt2009} is a 3\,m $\times$ 3\,m galvanized welded steel mesh with 2.51~mm wire diameter (12.5 gauge) and $10.2$~cm spacing. 
In this analysis, we modelled the beam for one antenna, numbered 252 \citep{Price2018}, with three ground plane configurations that corresponded to a series of test modifications made in the field \citep{Spinelli2021}. The 3\,m $\times$ 3\,m ground plane was replaced first by a 10\,m $\times$ 10\,m patch of mesh comprising the same material but with a 3.06~mm wire diameter (11 gauge), and later by the same but with serrations as represented by four 5~m long, 1.25~m wide isosceles triangles positioned on each side (referred to as the serrated ground plane). 
The arrangement is shown schematically 
in \autoref{fig:example_ground}.
As in \citet{Bowman2018}, peripheral serrations are sometimes added to ground planes to reduce coherence among currents proximate to the edge discontinuity.

\subsection{Characterization of soil complex permittivity}\label{sec:soilmes}

Measurements of complex permittivity, using coaxial impedance dielectric reflectometry \citep{Seyfried2004}, were made at three depths (4, 14, and 21$\pm0.5$ inches, corresponding to 10.16 cm, 35.56 cm and 53.34 cm) in a test pit dug at a midpoint $\sim 170$~m from antennas 252 and 254 and backfilled.  Three 50~MHz sensors were inserted into the undisturbed strata revealed along one wall of the pit.  Sensor firmware provided temperature-corrected estimates of permittivity using the US Department of Agriculture calibration for soil comprising sand, silt, and clay with conductivity $<1.5$ S\,m$^{-1}$  \citep{Seyfried2005}.  

Data were collected episodically from May 2019 to Jan. 2020, at epochs spaced in time so as to track complex permittivity during the precipitation-free summer and fall seasons, as it approached baseline values. Scatter in conductivity measurements was 0.001 S\,m$^{-1}$ and $\leq 0.02$ in the real part of permittivity.

The soil composition observed at the test pit above 53~cm depth was similar to that removed from a 1.5~m deep vertical auger hole drilled $\sim 10$\,m SE of antenna 252: sand and a mix of sand and pebbles ($<5$\,mm), depending on depth.  These observations are consistent with the broader geological context of this part of the Owens Valley. \citet{Tallyn2002} classify the soil deposits on level terrain near the steep front of the Inyo range to the east as deep, well-drained Mazourka-Cajon-Hessica formations of sandy soil.  With finer positional resolution, \citet{Danskin1998} describe well-sorted and unconsolidated lacustrine deposits of sand, gravelly sand, and silty sand to depths on the order of 100~m in the vicinity of Black Mountain, which overlooks the site from the ENE. This layering and favourable drainage across a broad area, and the absence of bedrock, buried boulders, and volcanic debris suggest that even without the benefit of ground penetrating radar analyses, it is reasonable to anticipate that the antenna sites are likely to be good ones from a geological perspective.

\subsection{Analytic beam model}\label{sec:oldbeam}
We summarise in this section the analytical beam model of the LWA dipole used in previous works. For a more extensive description see \citet{Dowell2011,Taylor2012,Ellingson2013}. Previous LEDA studies \citep{Bernardi2015, Spinelli2019} used this beam modelling.
This model allows the reconstruction of the antenna beam pattern in every azimuth direction, using two principle antenna planes ($E$ and $H$, effectively taking into account the antenna symmetries) and reads:
\begin{equation}
A(\theta,\phi,\nu)=\sqrt{[p_E(\theta,\nu) \cos{\phi}]^2 + [p_H(\theta,\nu) \sin{\phi}]^2}.
\notag
\end{equation}
The pattern in the $E$- and $H$-plane is given by:
\begin{equation}
p_i(\nu,\theta)=\left[ 1 - \left( \frac{\theta}{\pi/2} \right)^{\alpha_i(\nu)} \right](\cos{\theta})^{\beta_i(\nu)} + \gamma_i(\nu) \left(\frac{\theta}{\pi/2} \right)(\cos\theta)^{\delta_i(\nu)}
\label{eq:analytic_beam}
\end{equation}
where $i=E, H$ and $\theta$ is the elevation angle. The behavior of the coefficients $[\alpha_i,\beta_i,\gamma_i, \delta_i]$ with respect to frequency is fitted with a polynomial of $n^{\rm th}$-order\footnote{Different works throughout the years have use different value for the polynomial order. \citet{Bernardi2016,Spinelli2019} used a 3$^{\rm rd}$ order polynomial while \citet{Spinelli2021} used a 13$^{\rm th}$ order polynomial.} to NEC4 simulations \citep{Hicks2012}. 
Note that for this simulation only the $3\times3$ ground plane case is available.


\begin{table}
\centering
\caption{Soil parameters for the one-layer and the multi-layer model, extracted from measurements of soil at the LWA site during both dry and wet conditions at depths ${\rm z}_i,\ i=1,2,3$.}\label{tab:soil_params}
\begin{tabular}{ |p{2cm}|p{1cm}|p{1cm}|p{1cm}|p{1cm}| }
 \multicolumn{5}{|c|}{Soil layer parameters (\( \sigma \) in S/m, \( \epsilon_r \) dimensionless)} \\
 \hline
  & \( \sigma_{\rm dry} \) & \( \sigma_{\rm wet} \) & \( \epsilon_{ r, {\rm dry}} \) & \( \epsilon_{ r,{\rm wet}} \) \\
 \hline 
 one layer & 0.004 & 0.01 & 4.4 & 6.5 \\
 \hline \hline
$ {\rm z}_1=10.16\ cm $ & 0.0013 & 0.005 & 3.73 & 8.09 \\
 \hline
$  {\rm z}_2=35.56\ cm $ & 0.004 & 0.0068 & 4.25 & 6.45 \\
 \hline
$ {\rm z}_3=53.34\ cm $ & 0.0187 & 0.0388 & 7.58 & 20.56 \\

\end{tabular}
\end{table}

\subsection{Improving the beam model}\label{sec:FEKO}
\begin{figure}
        \includegraphics[width=\columnwidth,height=0.2\textheight]{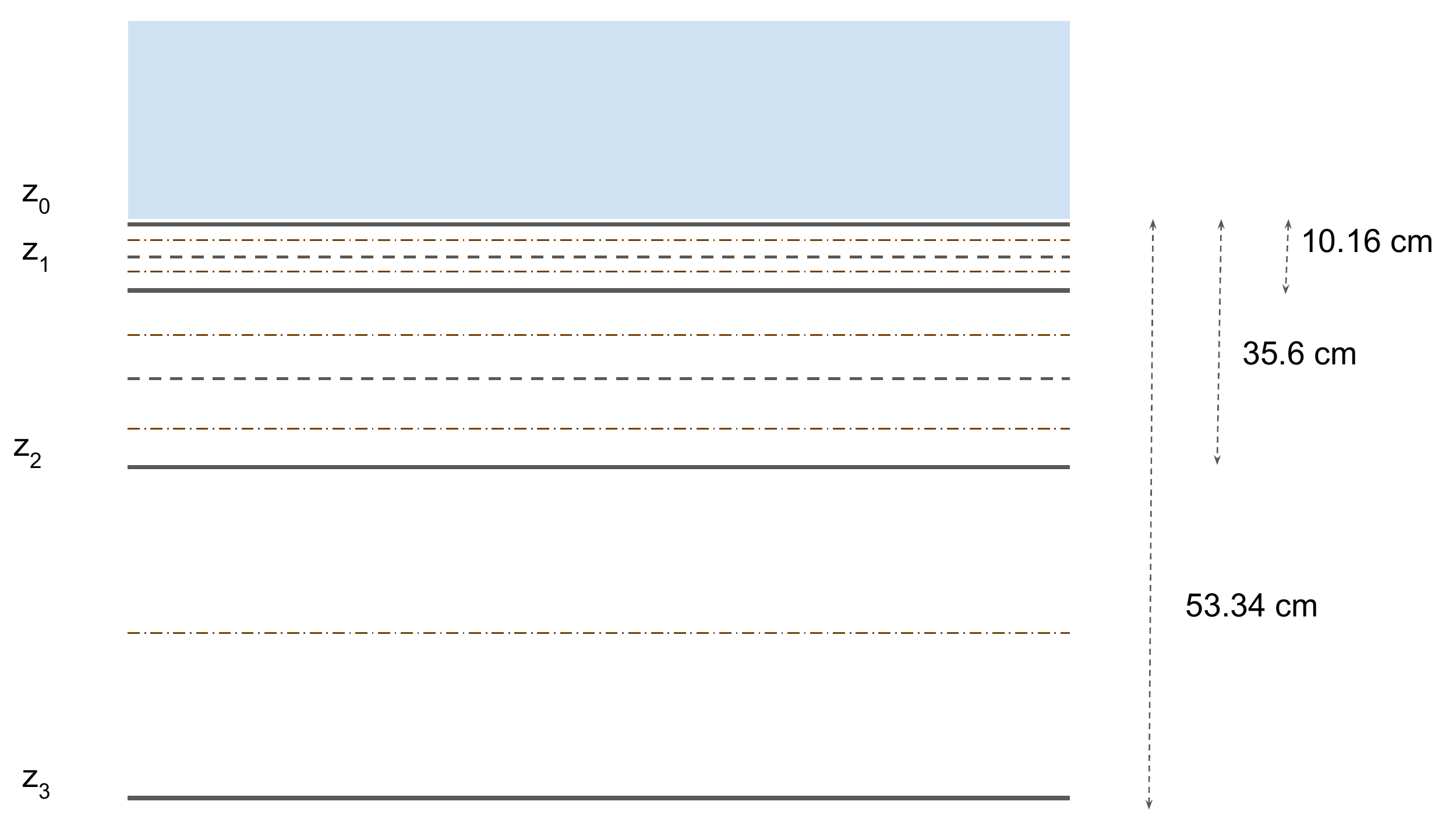}
        \caption{A 2D schematic view of
        the multi-layer splitting of soil half-infinite space $ \rm z<0 $,
        described in \autoref{sec:FEKO}.
        A number of different measurements were made at 3 specific depths $ \rm z_i$ (10.16 cm, 35.56 cm and 53.34 cm) and then ensemble averaged. A numerical EM driven multi-layer modelling is then constructed with an iterative sub-layering scheme. See text and \autoref{alg:sublayers} for details.}
\label{fig:example_layers}
\end{figure}

In this section, we incorporate the soil properties measured in-situ, describing more realistically the environmental conditions in which the data have been collected. 
Although not previously explored for the LEDA antennas, there are examples in the literature of similar studies where the soil is analysed as an homogeneous dielectric semi-infinite volume, discriminating only between dry and wet soil conditions \citep[e.g.,][]{Sutinjo2015}, or a combination of both, as in \citet{Bradley2019}.

\paragraph*{One layer baseline model.} 
For our new simulation we use FEKO\footnote{\url{https://altairhyperworks.com/feko/}}, a commercial software widely employed in numerical EM simulations and based on the Method of Moments (MoM, e.g. \citet{Davidson2010}).

We model the soil as a single half-infinite layer of constant complex permittivity and conductivity, namely: 
\begin{equation}
    \epsilon=\epsilon_0\left(\epsilon_r + i\frac{\sigma}{2\pi\nu\epsilon_0}\right)
    \label{eqn:complex_epsilon}
\end{equation}
where $\epsilon_0$ is the electrical permittivity of vacuum, $\epsilon_r$ is the dimensionless relative permittivity, $\sigma$ is the conductivity in S/m, and $\nu$ is the frequency in Hz. This quantity had been measured on site for different soil moisture condition corresponding to both dry and wet soil (see \autoref{tab:soil_params}). We use this one-layer model as our baseline one since it is simple and relatively immune to numerical artefacts. We simulate the three different ground planes - $3\times3$, $10\times10$ and serrated - described in  \secref{sec:ground} and \autoref{fig:example_ground}.

\paragraph*{Measurement driven three-layer modelling.} 
We update the modelling just described adding the new available measurements for the soil complex permittivity. In constructing our first layered geometrical model, we adopted  values at the nominal depths of \autoref{tab:soil_params}, according to \secref{sec:soilmes}. A thickness was also chosen for the simulated layers according to the measurement depths, and we ensemble averaged the permittivities of the available measurements.

Note that the thickness of these three consecutive layers does not match the standard numerical accuracy criteria required by EM solvers. Despite this, we use these measurements to construct a three-layer soil model for the FEKO simulation (``measurements'' column of \autoref{tab:beam_type}).

\paragraph*{Numerical EM driven multi-layer modelling.}
The thicknesses of the three-layer model are imposed by the depth of the available measurements and might not be sufficient for accurately representing a varying soil permittivity gradient. In order to solve the possible numerical issues connected to the thicknesses of the three-layer model we refine our electromagnetic simulations using more layers. Although the separation between two consecutive layers is subject to different discretisation rules according to each method, requesting a separation smaller than $\lambda/10$ is a widely used criterion.
The standard FEKO method for any multi-layer substrate is a planar Green's function analytical solution embedded in the MoM formalism. A continuously varying permittivity is treated by means of properly discretising in layers, and a Transmission Line Green's Function (TLGF) approach is used by the solver. Details for this can be found in \citet{Michalski1997}.

\medskip
When considering the $\lambda/10$ rule to discretise the half-infinite space, we have to take into account the different phase velocity which implies a different wavelength in each layer. For a low-loss medium, such as soil ($\sigma\ll 2\pi\nu\epsilon_0$ when considering our available measurements), the phase velocity is $c_p=c_0/\epsilon_r$, where $c_0$ is the speed of light in vacuum. We shall split each of the previous three layers into sub-layers such that their thickness is less than or equal to $\lambda_p/10$, and then double these layers iteratively until a certain convergence criterion is satisfied. The algorithm which implements this iteration is presented in \autoref{alg:sublayers}. In each iteration, we interpolate linearly to compute the new values of complex permittivity (\texttt{linInterp}), we merge them with the values of the previous iteration (\texttt{merge}), and we run a new FEKO simulation to calculate the gain pattern (\texttt{Gain}).
More details on the choice of the sub-layering scheme are explained in \autoref{app:multi}. 

This iteration is applied both for relative permittivity and for conductivity in a square lattice fashion, and a maximum frequency $\nu_{\rm max}=100$~MHz is used which provides a strong bound. By performing the first step of discretization, we conclude that each of the previous three layers of the dry soil should be divided into one or two sub-layers (see the grey dashed lines in \autoref{fig:example_layers}). We then follow an iterative splitting by doubling these sub-layers (red dotted-dash lines in \autoref{fig:example_layers}), in order to make an ever finer discretization. Wet soil conditions require a slightly finer splitting and are not reported in the schematic figure. The number of sub-layers used in the first iteration is reported in \autoref{tab:beam_type}. 

\begin{algorithm}
\caption{Sub-layer splitting algorithm (see also \autoref{fig:example_layers}) to achieve gain at zenith convergence of 0.1 dBi. [arg] indicates a vector-valued argument, while a subscript is used to index these values ($i_1:i_2$ also indicates parsing from initial index $i_1$ to final index $i_2$). See text and appendix for details.}\label{alg:sublayers}
$[\epsilon]\gets [\epsilon_1, \epsilon_2, \epsilon_3]$\;
$[\lambda_p]\gets\frac{c_0\epsilon_0}{\nu_{\rm max}\Re\{[\epsilon]\}}$\;
$[\Delta\epsilon]\gets [\epsilon_1, \epsilon_2-\epsilon_1, \epsilon_3- \epsilon_2]$\;
$[N] \gets [1, 1, 1]$\;
$G_{0}\gets$ Gain($[\epsilon]$)\;
\For{$i=0,1,2$}{
$N_{init}\gets\left\lceil\frac{|z_{i+1}-z_i|}{[\lambda_p]_{i+1}/10}\right\rceil$\;
$[\Delta\epsilon]_{i+1}\gets\frac{[\Delta\epsilon]_{i+1}}{N_{init}}$\;
$[N]_{i+1} \gets N_{init}$\;
$[\epsilon]\gets$ merge($[\epsilon]_{1:i}$,linInterp($[\epsilon]_{i+1},[N]_{i+1},[\Delta\epsilon]_{i+1}$)\;
}
$G_1 \gets$ Gain($[\epsilon]$)\;
\While{$ |G_1-G_0| \begingroup>\endgroup 0.1 $ dBi}{
  $G_0\gets G_1$\;
  \For{$i=0,1,2$}{
    \begin{align}
    [\epsilon]\gets \text{merge}&([\epsilon]_{1:[N]_{i}-1},\hdots \nonumber \\
    \text{linInterp}&([\epsilon]_{[N]_{i}:[N]_{i+1}-1},2[N]_{i+1},[\Delta\epsilon]_{i+1}/2))\text{\;} \nonumber
    \end{align}
  }
  $[\Delta\epsilon]\gets\frac{[\Delta\epsilon]}{2}$\;
  $[N]\gets 2[N]$\;
  $G_1\gets$ Gain($[\epsilon]$)
}
\end{algorithm}

\begin{figure}
\includegraphics[width=\columnwidth, keepaspectratio]{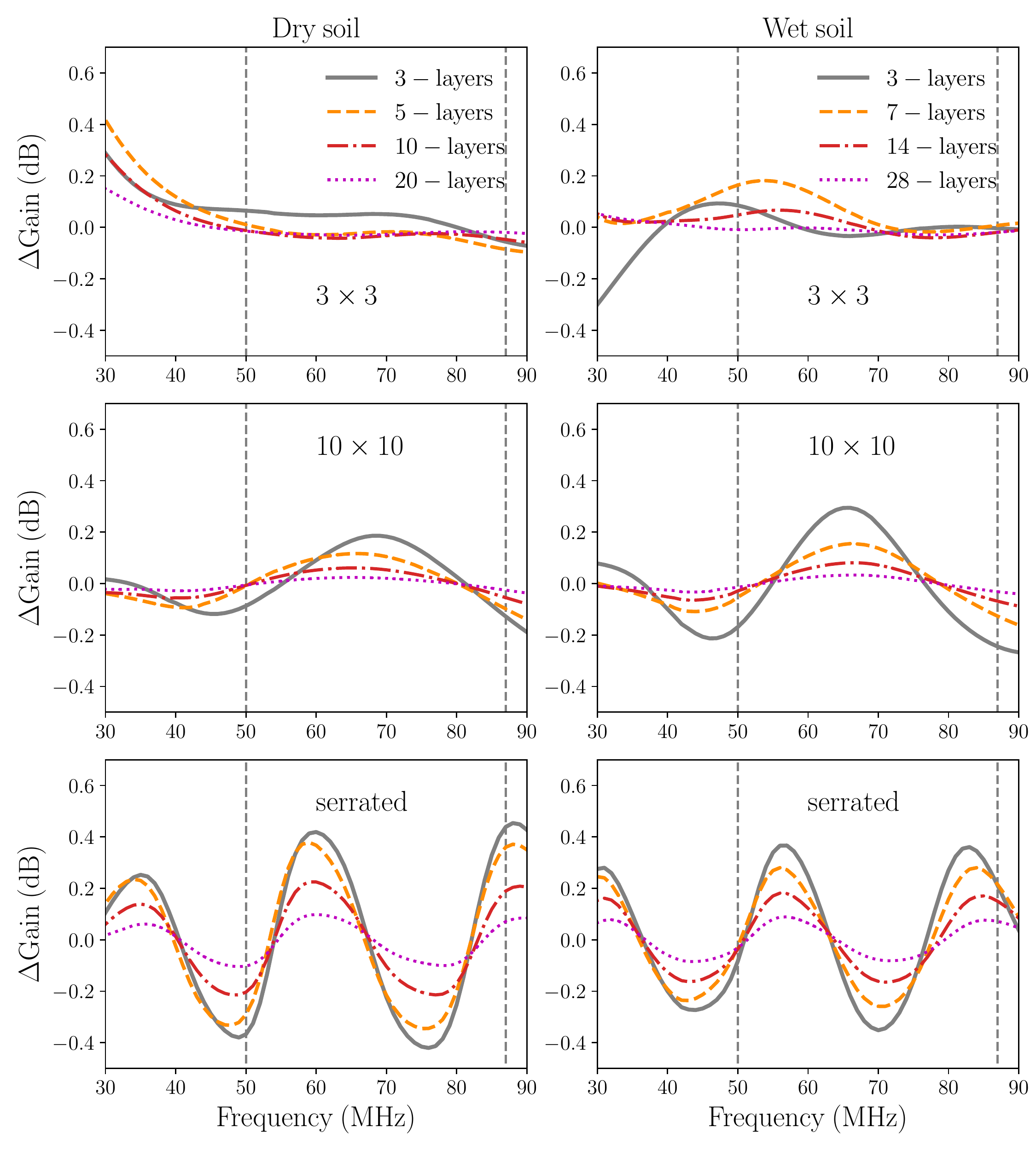}
\caption{Gain at zenith dB difference of multi-layer implementations, calculated between the converged solution with 40 and 56 layers for dry and wet soil conditions, respectively, and solutions with fewer layers. Six cases are examined: different types of soil conditions (dry, wet) and different types of ground plane $3 \times 3$,$10 \times 10$, serrated). The vertical dashed lines highlight the region where LEDA data are available.}
\label{fig:dgaindB_convergence}
\end{figure}

\begin{figure}
\includegraphics[width=\columnwidth]{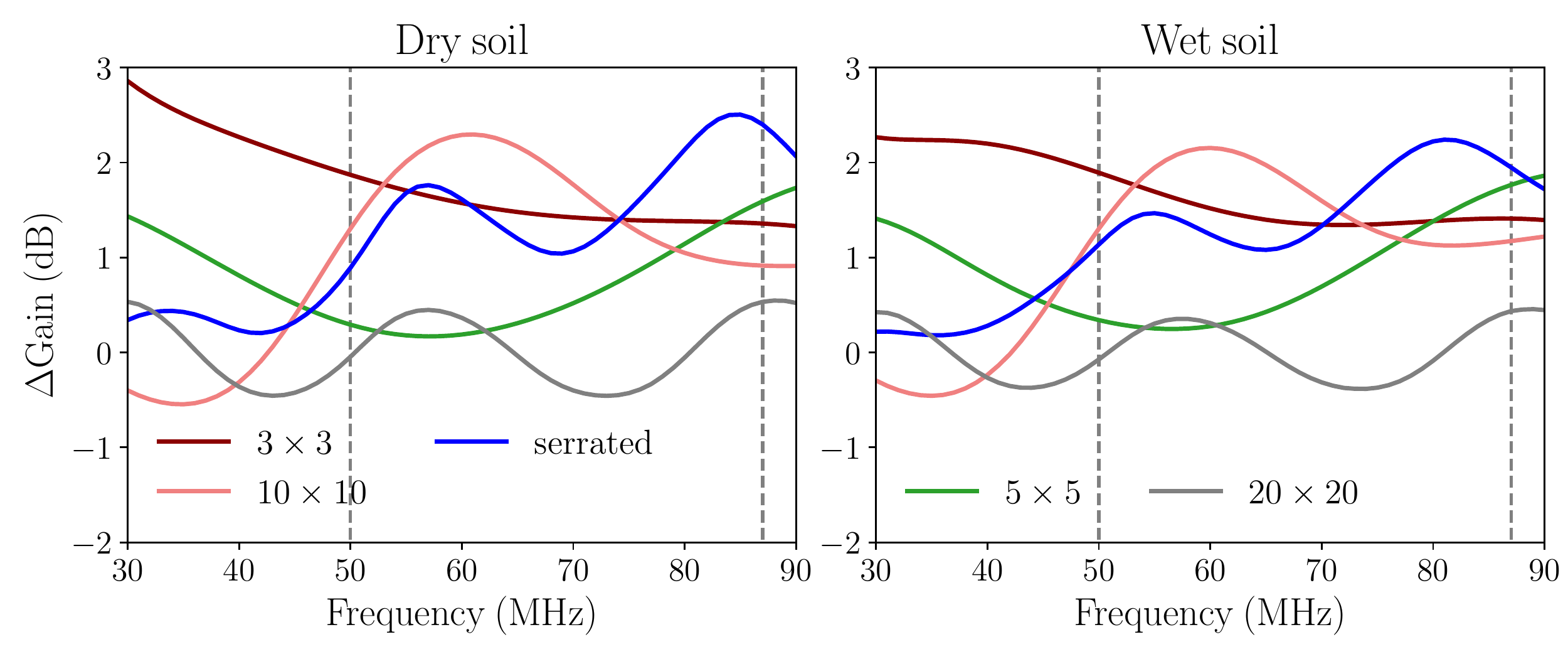}
\caption{Gain at zenith dB difference of ground plane implementations, calculated between a reference model of an infinite PEC plane and the converged model of each ground plane as reported in the legend, for dry and wet soil conditions, respectively. An intermediate case of a $5 \times 5$ ground plane and a large $20 \times 20$ one, not examined elsewhere, are also shown. The vertical dashed lines in each panel highlight the region where LEDA data are available.}
\label{fig:groundplanedB_convergence}
\end{figure}

\medskip 

The iterative algorithm stops when a convergence threshold for the value of the gain at zenith is satisfied. We choose 0.1 dB which should be roughly similar to the numerical accuracy of FEKO. The number of layers needed for convergence are 40 and 56 layers for dry and wet conditions, respectively. 
The high number of layers to reach convergence is needed only for the serrated case but we assume it for all ground planes for convenience. We refer to this implementation as the converged model.
\autoref{tab:beam_type} summarises the different multi-layer approaches listing their number of layers. \autoref{fig:dgaindB_convergence} shows the gain differences (in dB) between the converged case and the solution obtained with a smaller number of layers at all previous iterations. The analysis is repeated for all three different ground planes. We note the existence of periodic oscillations for the case of the $10 \times 10$ ground plane, with or without serrations. This fact highlights that successive implementations of sub-layers differ roughly by sinusoidal factors, which are still present when the ground plane gets bigger. This is a counter-intuitive conclusion, which might originate from sharp permittivity value transitions between consecutive layers (expressed in boundary conditions of the TLGF). The $3 \times 3$ ground plane performs better with respect to the oscillation, apart from a low-frequency drift off.

\begin{table}
\caption{Number of soil layers that were used in each of our 3 multi-layer models: $\lambda_{ p}/10$ is for the first iteration of the algorithm, and converged is the final step of doubling that allows gain convergence to 0.1dB.}\label{tab:beam_type}
\begin{center}
\begin{tabular}{|c|c|c|c|} 
 & \multicolumn{3}{c}{number of soil layers for multi-layer schemes}\\
\hline
& measurements & $\lambda_{ p}/10$ & converged\\
\hline
dry & 3 & 5 & 40\\
 \hline
wet & 3 & 7 & 56\\
\end{tabular}
\end{center}
\end{table}

\paragraph*{Finite size of the ground plane.}
Oscillation effects are also observed when comparing all finite sized ground planes with respect to an infinite Perfect Electric Conductor (PEC) plane solution. To demonstrate this, in \autoref{fig:groundplanedB_convergence} we subtract from the converged case the solution for an infinite PEC plane, for each of the considered ground planes. We also include a $5 \times 5$ and a $20 \times 20$ ground plane, as an intermediate and extreme case. The sinusoidal variations that can be seen imply that there are finite-size truncation/diffraction effects which need long electrical distances (i.e., distances as multiples of $\lambda$) to diminish. The amplitude and periodicity of these oscillations are different for each ground plane and depend on their size. 

\begin{figure}
\includegraphics[width=\columnwidth, keepaspectratio]{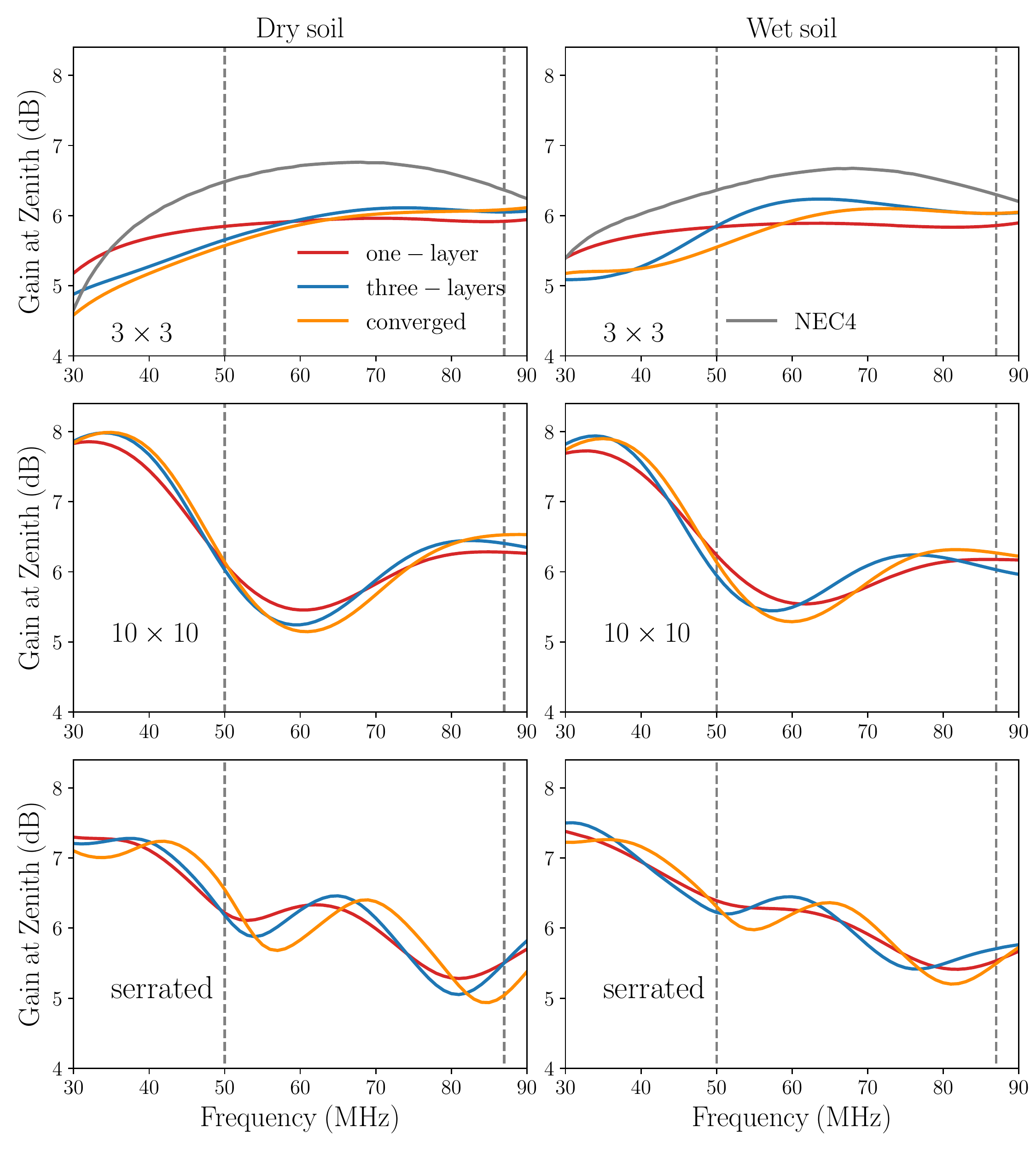} 
\caption{Gain at zenith of different multi-layer implementations, including the baseline model, a three-layer model, and the converged multi-layer model (40 and 56 layers for dry and wet soil conditions, respectively). Six cases are examined, as in \autoref{fig:groundplanedB_convergence}. For the $3\times3$ ground plane, the NEC4 simulated model is also shown. The vertical dashed lines in each panel highlight the region where LEDA data are available.}
\label{fig:gaindb_zenith}
\end{figure}

\medskip
Having examined the effect of different ground planes, soil conditions as well as soil layering options, a comparative plot of zenith gain can best illustrate their effects before any chromaticity correction is calculated. For each ground plane and soil conditions, we present in \autoref{fig:gaindb_zenith} the baseline one-layer, the three-layer, and the converged multi-layer FEKO model. As expected, the 3x3 ground plane provides in most cases a lower gain, since there are more losses related to the part of the soil not covered by the ground plane. It is, however, the best one in terms of ground plane induced ripple, which has a lower frequency due to its smaller dimension. Note that concerns over the spectral structure of an underlying ground plane have already been addressed in other experiments such as SARAS3 \citep{Nambissan2021,Singh2021} that moved the antenna on a lake to minimise the chromatic response. The MIST experiment has opted instead for not deploying a ground plane and carefully characterise the ground properties. For an antenna configuration less sensitive to the ground properties than LEDA (which is facing downwards and is more coupled to any underlying structure), a larger ground plane is also expected to alleviate the problem. The EDGES team, for example, has used a $30\times 30$ ground plane for their results \citep{Bowman2018}.

We include in \autoref{fig:gaindb_zenith} for comparison the $3 \times 3$ ground plane solution obtained with the NEC4 pattern used in previous analysis. The NEC4 patterns have been scaled down using the radiation efficiency as calculated with FEKO data (see the next paragraph) since the NEC4 model was overestimating the gain by omitting the inclusion of soil losses in its gain calculation \citep{Weiner1991}. Despite this correction, the FEKO and NEC4 models differ by a factor of up to 1 dB across the frequency range of interest, a result that showcases the differences between numerical solvers. 

\begin{figure}
\includegraphics[width=\columnwidth, keepaspectratio]{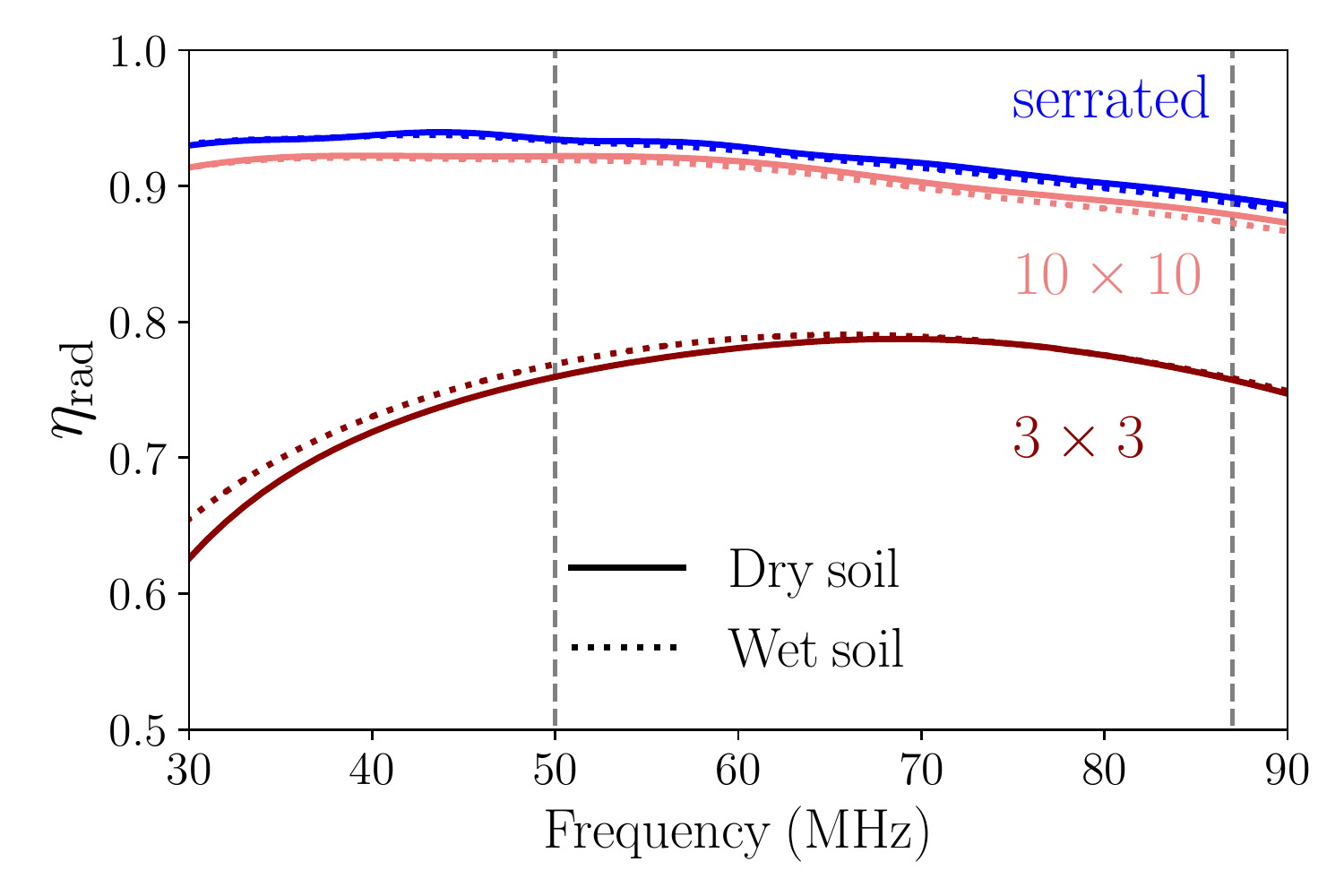}
\caption{Radiation efficiency over frequency calculated for the three different cases of ground plane, using the FEKO solution outputs. Dry (solid lines) and wet (dotted lines) conditions for the baseline one-layer model are presented. The vertical dashed lines highlight the region where LEDA data are available.}
\label{fig:rad_eff}
\end{figure}

\paragraph*{Radiation efficiency.} 
We present here the radiation efficiency $\eta_{\rm rad}$ as a function of frequency, which takes into account losses over all of the 3D antenna pattern. The radiation efficiency is calculated using the integral of the upper hemisphere far field patterns and the input power at the antenna port, given by FEKO.
In \autoref{fig:rad_eff}, $\eta_{\rm rad}$ is shown for the three examined ground planes. As expected, the $3\times 3$ ground plane presents more losses, with a clearer frequency dependence as well, since the efficiency is poorer at lower frequencies. A $10\times 10$ ground plane with or without serrations is more appropriate to keep losses smaller than $10\%$ in most of the frequency band, and quite more stable at the extremes, as they have a larger extent and allow for more power radiated below the horizon ($>90^\circ$) to be reflected back, contributing to the gain. The radiation efficiency is not constant and this affects the gain integral which is not $4\pi$ and varies with frequency. The beam integral is expected to vary significantly in the case of $3 \times 3$ ground planes, such that any normalisation of the beam pattern should be made separately for each frequency (see also \secref{sec:chrom}).

\begin{figure}
\includegraphics[width=\columnwidth]{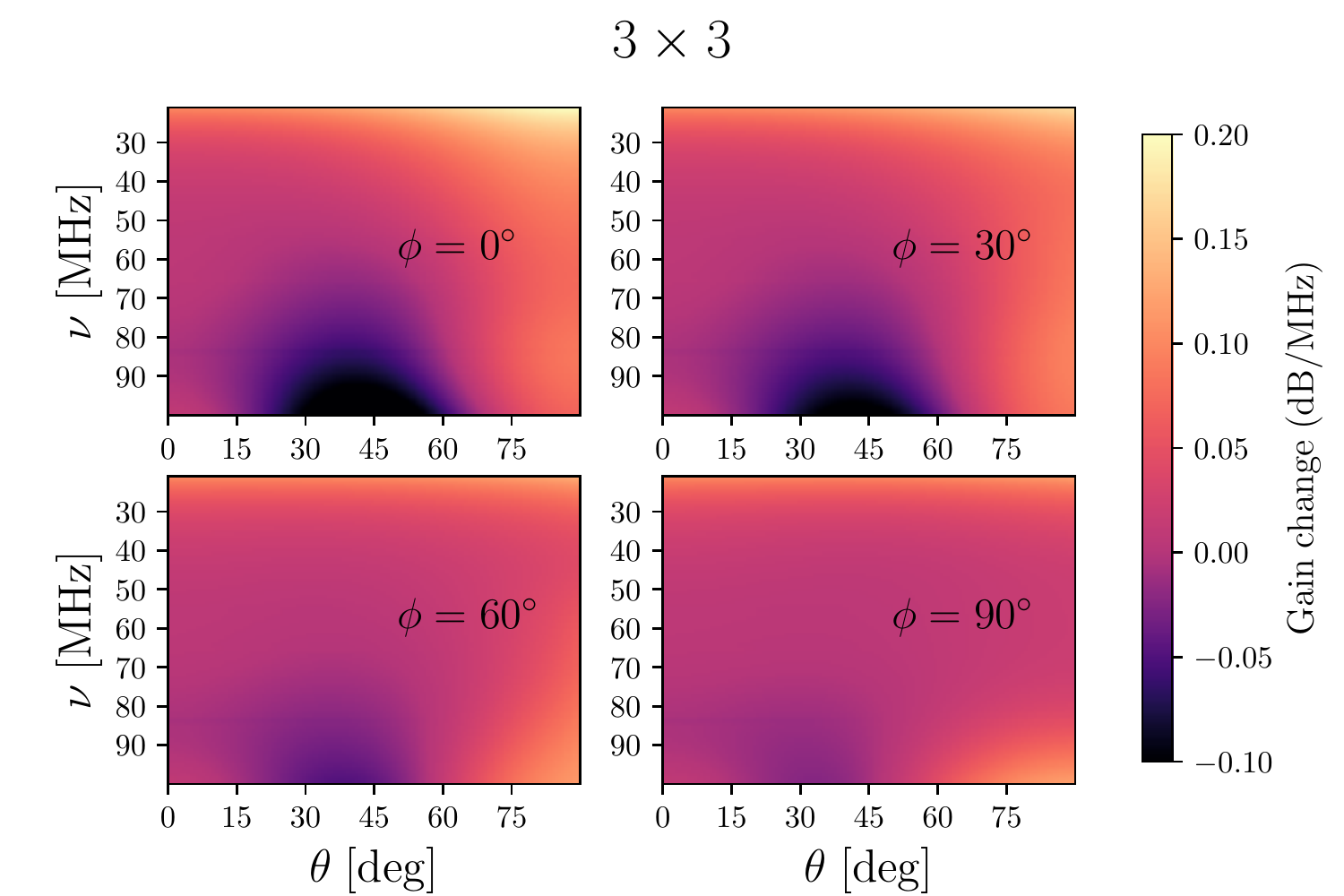}
\includegraphics[width=\columnwidth]{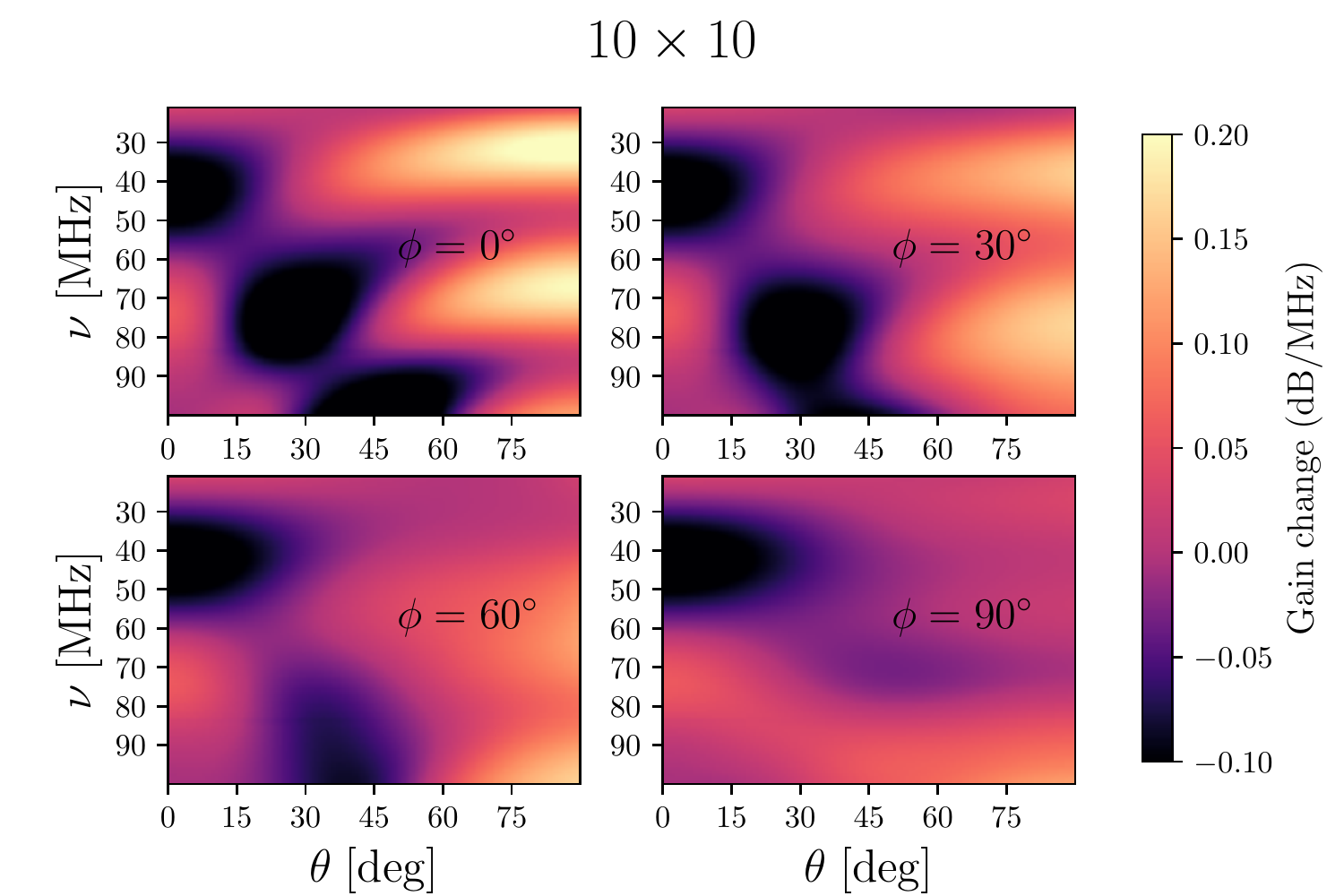}
\includegraphics[width=\columnwidth]{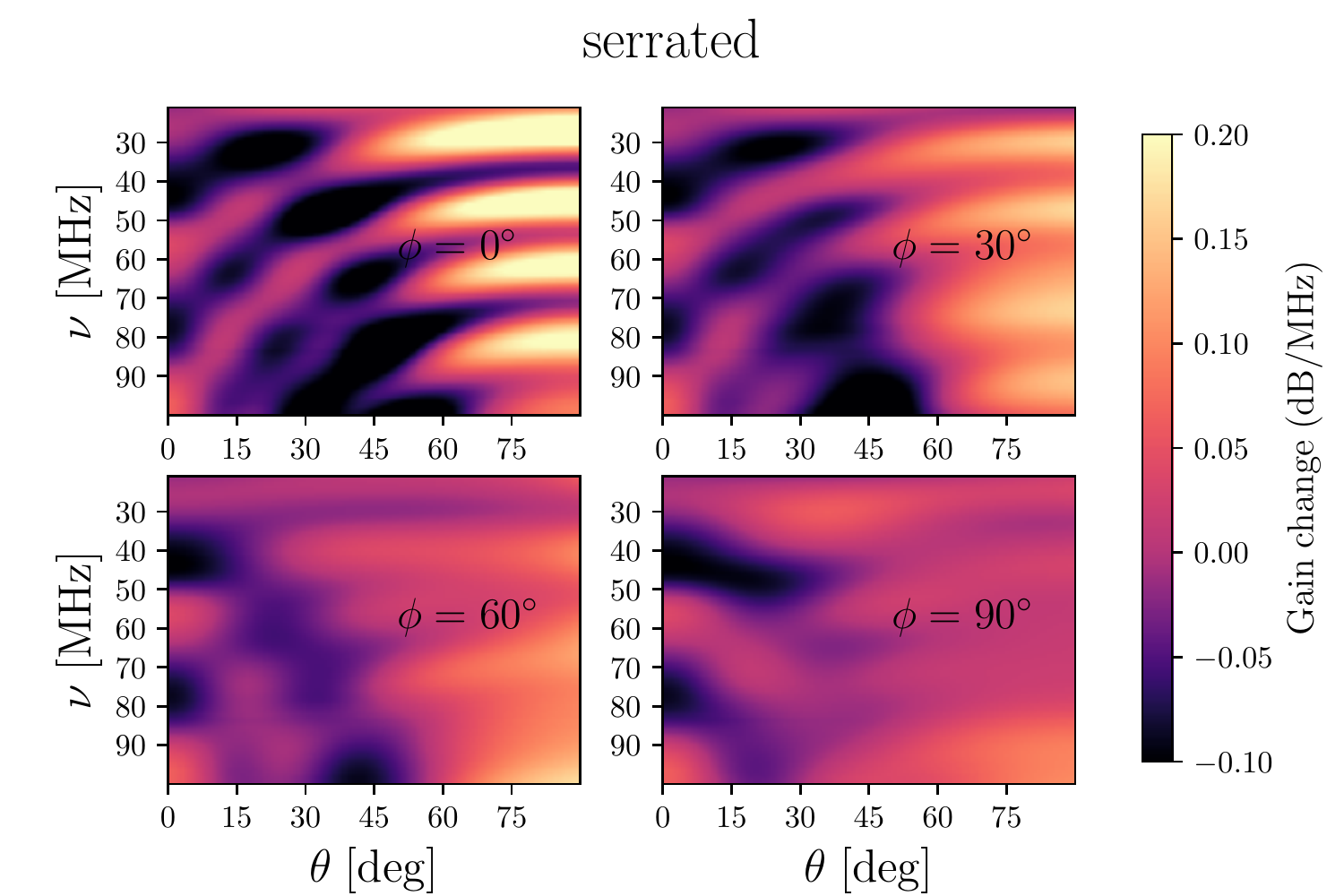}
\caption{Beam gain change per unit frequency as a function of the angle $\theta$ and for four selected $\phi$ values (0, 30,60 and 90 degrees). We show here the one-layer model for dry soil condition.}
\label{fig:beamdir}
\end{figure}

\paragraph*{Beam gain variation.} As a final assessment of the radiation pattern spectral robustness, we present a number of 2D colour-maps of the beam dB change in gain with respect to frequency in \autoref{fig:beamdir}, for 4 azimuth angles $\phi=0^{\circ}$,$30^{\circ}$, $60^{\circ}$,$90^{\circ}$, a criterion similar to that evaluated by \citet{Mahesh2021}. This kind of plot is useful not only to confirm the sinusoidal-factor spectral periodicity due to the ground plane structure, but any other spatial effects which predominantly appear across zenith angle $\theta$. It can be seen from the figure that these spatial variations are significant even for $\theta < 40^\circ$ , which compares with the Half-Power Beam Width of the antenna. Another interesting phenomenon is that the ``phase'' of this sinusoidal-factor variation is different for each elevation, which means that any analytic approach such as that of \autoref{eq:analytic_beam} with $\theta\times\cos(\theta)$ polynomial terms is not adequate to describe the complexity of the beam pattern.

The amplitude in azimuth angle diminishes when we cross from the $E$-plane ($ \phi=0^{\circ}$) to the $H$-plane ($ \phi=90^{\circ}$). The greatest variation both in terms of amplitude and number of complete cycles in the frequency range of interest is found for the serrated ground plane, which reaches as high as 0.2 dB/MHz in many $\nu -\theta$ sample points.

\section{Simulated observations}\label{sec:sim}

In this section we describe the construction of LEDA mock measured spectra, the correction factor for the beam-induced chromaticity of the simulated spectra and discuss qualitatively its impact on the accuracy of the smooth foreground model. As we already mentioned, we make use of a baseline, single-layer model, since the previous analysis has shown that the three-layer and multilayer models suffer from some uncertainties: lack of more measurements for the former, and interpolations assumptions for the latter.

\subsection{Modelling the sky observed temperature}\label{sec:Tobs}

In order to simulate the spectra measured by LEDA we compute the beam-averaged sky brightness temperature as seen by a single antenna $T(\hat{\boldsymbol{n}}_0,\nu,t)$ at the time $t$ and direction $\hat{\boldsymbol{n}}_0$ \citep[e.g.,][]{Bernardi2015}:
\begin{equation}\label{eq:meas}
T_{\rm obs}(\hat{\boldsymbol{r}}_0,\nu,t) = \frac{\int_{\Omega} B(\hat{\boldsymbol{n}}',\nu) \, T_{\rm sky}( \hat{\boldsymbol{n}}',\nu,t) \, d\hat{\boldsymbol{n}}'}{\int_{\Omega} B(\hat{\boldsymbol{n}}',\nu) \, d\hat{\boldsymbol{n}}'} + T_{\rm N}(\nu) + T_{21}(\nu)
\end{equation}
where 
$B$ is the antenna gain pattern and $T_{\rm N}$ the instrumental noise. 
$T_{\rm sky}$ is the sky brightness temperature, which changes with time as the sky drifts over the dipole.
To model this latter we simply consider the Haslam $408$~MHz full-sky map $T^{{\rm H}}(\mathbf{\hat{n}})$ \citep{Haslam1982} and scale it to different frequencies assuming a constant spectral index $\beta$
\begin{equation}\label{eq:T_H}
    T^{{\rm H}}_{{\rm sky}}(\nu,\mathbf{\hat{n}})=[T_{{\rm H}}(\mathbf{\hat{n}})-T_{{\rm cmb}}]\left( \frac{\nu}{408}\right)^{\beta}+T_{{\rm cmb}}.
\end{equation}
where $T_{{\rm cmb}}=2.725$~K and $\beta=-2.5$.
Note that other different sky models such as the GSM \citep{Zheng2017} or the GMOSS \citep{GMOSS2017} could be considered, including the effect of a spatially varying spectral index. Nevertheless, in this work we are mainly interested in varying the antenna simulations and we postpone the investigation of sky modelling uncertainties to a future work.
Moreover, due to the large beam, a spatially constant spectral index is a reasonable approximation \citep{Cumner2021}.

\begin{figure}
\includegraphics[width=\columnwidth]{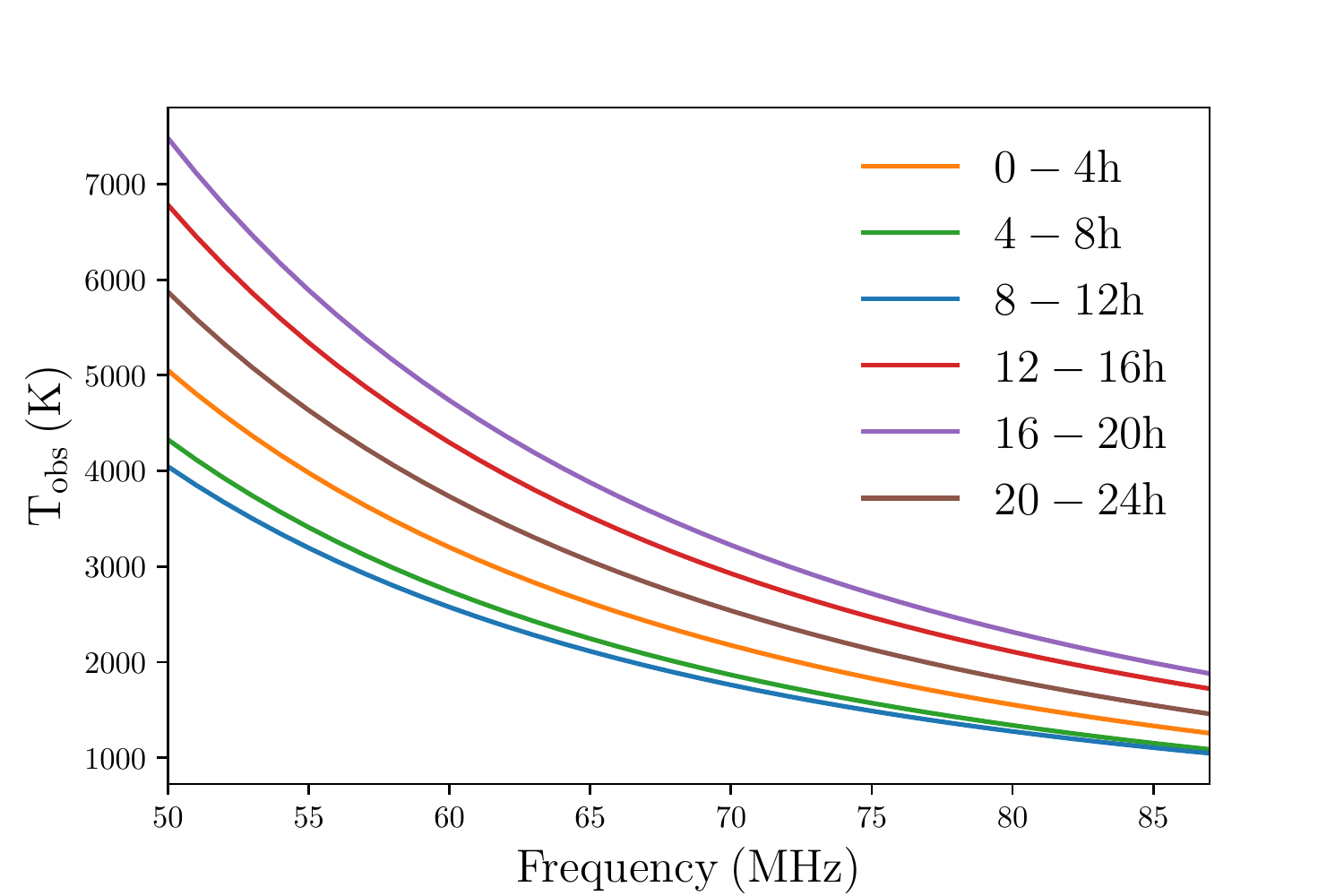}
\caption{Simulated antenna temperature as seen through the LEDA beam as a function of frequency in the 50-87 MHz range. The different colours correspond to different 4~h LST bins.}
\label{fig:tobs}
\end{figure}

We show the resulting antenna temperature averaged over 4~h LST bins in \autoref{fig:tobs}, that can be compared to Figure 13 in \citet{Mahesh2021} and shows the low foreground contamination in the LST range 8-12~h, corresponding roughly to the chosen one for our data analysis in \citet{Spinelli2021}. 

We model the 21cm global signal $T_{21}(\nu)$ with a Gaussian absorption profile \citep{Bernardi2015,Presley2015,Bernardi2016,Monsalve2017}:
\begin{equation}\label{eq:gauss}
T_{21}(\nu)=A_{21} \, e^{-\frac{(\nu-\nu_{21})^2}{2\sigma_{21}^2}},
\end{equation}
where $A_{21}$, $\nu_{21}$ and $\sigma_{21}$ are the amplitude, peak position and standard deviation of the 21cm trough, respectively.
This is an approximation and more realistic shapes for the absorption feature could be computed from numerical or semi-analytical simulations  \citep[e.g.,][]{Mirocha2014,Mirocha2015,Cohen2016,Cohen2017,Mirocha2017, Reis2021}, however, analytic expressions are useful for fast evaluation of likelihood functions.
We adopt in this work a simplistic assumption for the signal strength using $A_{21}=-180.0$~mK, $\nu_{21}=70$~MHz and $\sigma_{21}=2$~MHz. 
We do expect our results to be dependent on the parameters chosen for the signal. With these values, the absorption profile is narrow (i.e. easier to disentangle from the smooth foregrounds) and well inside our observed band. In this study we are, however, mostly interested in relative behaviours for the beam modelling allowing us to fix the input cosmological model. We discuss this approximation further in \appref{app:diffHI}.

In order to limit the parameter space to explore, we construct mock data as faithful as possible to the actual LEDA data \citep[see][]{Spinelli2021} and we construct the mock spectra $\bar{T}_{\rm ant}(\nu)$ by averaging \autoref{eq:meas} in the LST range 8.5-12~h.

To compute the noise $T_N$, we assume that it is given by the radiometer equation and it is uncorrelated in frequency and time. We assume that for each frequency channel it follows a Gaussian distribution with standard deviation: 
\begin{equation}\label{eq:noise}
\sigma^N(\nu)=\frac{\bar{T}_{\rm ant}(\nu)}{\sqrt{\Delta t \, \Delta \nu}},
\end{equation}
where we consider a $\Delta \nu = 1$~MHz channel width and a $\Delta t = 100$~hours of total integration time, in agreement to real LEDA data specifics.

\subsection{Chromaticity correction}\label{sec:chrom}

Assuming a well-behaved beam with a small degree of chromaticity, the resulting spectra of \autoref{eq:meas} are expected to be smooth in frequency. Although not visible by eye in \autoref{fig:tobs}, realistic beam shapes induce a non-smooth frequency structure in the measured sky temperature that, as we will discuss more extensively later on, complicates the signal reconstruction. As discussed in other works \citep[e.g.,][]{Mozdzen2019,Monsalve2020,Anstey2020}
a possible approach to alleviate the effect of the chromatic beam on the measured spectra is to correct the original spectra using the following factor:
\begin{equation}\label{eq:beam_chromaticity}
B_{{\rm c}}(\nu,{\rm LST})=\frac{\int_{\Omega} T_{{\rm sky}}(\nu_0,{\rm LST},\mathbf{\hat{n}'}) B(\nu,\mathbf{\hat{n}}') d\mathbf{\hat{n}}'}{\int_{\Omega} T_{{\rm sky}}(\nu_0,{\rm LST},\mathbf{\hat{n}}') B(\nu_0,\mathbf{\hat{n}}') d\mathbf{\hat{n}}'} \frac{\int_{\Omega}  B(\nu_0,\mathbf{\hat{n}}') d\mathbf{\hat{n}}'}{\int_{\Omega} B(\nu,\mathbf{\hat{n}}') d\mathbf{\hat{n}}'}  .
\end{equation}
Note that $T_{{\rm sky}}$ is a function of LST since the sky drifts with time over the antenna.
We choose $\nu_0=75$~MHz as a reference frequency  \citep{Mozdzen2019} since it is approximately central in our range. In this formulation it should also be pointed out that the beam pattern $B(\nu,\mathbf{\hat{n}}')$ used is gain, so the second term on the right hand side of the equation
is ${\eta_{\rm rad}(\nu)}/{\eta_{\rm rad}(\nu_0)}$.

\begin{figure}
\includegraphics[width=\columnwidth]{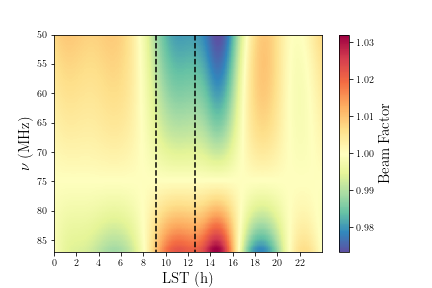}
\caption{The chromaticity correction (\autoref{eq:beam_chromaticity}) computed using the sky model of  \autoref{eq:T_H} and the baseline beam model from FEKO (one-layer, dry soil) considering the $3\times3$~m ground plane. The dashed vertical lines highlight the LST range preferred for LEDA data analysis.}
\label{fig:beamchromdry_example}
\end{figure}

We compute the beam chromaticity correction using the various beam models presented in \autoref{sec:beam}. Note that we consider LST bins of 10 minutes and the full 24~h range for completeness when computing the correction. Since our goal is to investigate the effect of the beam, we use the same sky model of equation~\ref{eq:T_H} to disentangle the two effects and to avoid introducing complications due to a different sky model than the one assumed to compute the simulated spectra. An example of the correction is reported in \autoref{fig:beamchromdry_example} where we show our baseline case of one-layer, dry soil conditions and $3\times3$~m ground plane (see \autoref{sec:beam}).

\begin{figure}
\includegraphics[width=\columnwidth]{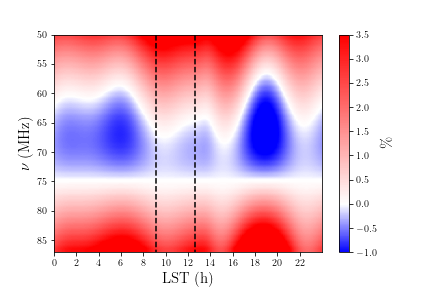}
\caption{Percentage difference between the chromaticity correction factor $B_c$  (\autoref{eq:beam_chromaticity}) computed for the new baseline beam model and the analytic beam used in previous analysis. The dashed vertical lines highlight the LST range preferred for LEDA data analysis.}
\label{fig:chrom_diff_old_new}
\end{figure}

It is interesting to evaluate, as shown in \autoref{fig:gaindb_zenith}, the difference  between the analytic beam model previously used and the new FEKO baseline simulation, computing the chromaticity correction. This is reported in \autoref{fig:chrom_diff_old_new}.
Differences reach a few percent especially around LST$\sim 18$
when the Galactic center is transiting.

\begin{figure}
\includegraphics[width=\columnwidth]{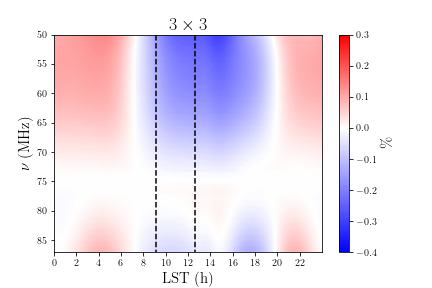}\\
\includegraphics[width=\columnwidth]{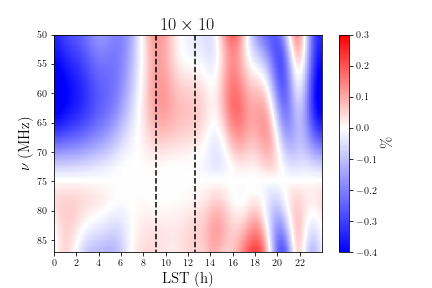}\\
\includegraphics[width=\columnwidth]{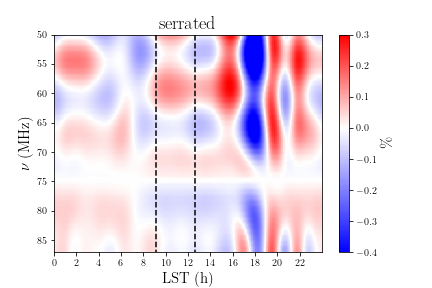}
\caption{The difference in percentage between the chromaticity correction factor $B_c$ of \autoref{eq:beam_chromaticity} computed for dry and wet soil conditions. We consider here the one-layer FEKO model and we compute the difference for the various ground planes: $3\times3$~m (top panel), $10\times10$~m (middle panel) and serrated (bottom panel). The dashed vertical lines highlight the LST range preferred for LEDA data analysis.}
\label{fig:chromdiffwet}
\end{figure}

We can repeat the same exercise for the various FEKO beam models computed for this analysis. We show in \autoref{fig:chromdiffwet} the difference between the dry and wet soil conditions for the three different ground planes. Differences are a fractional of percent but their structure present different patterns for different ground plane shapes  (see \autoref{fig:dgaindB_convergence} and \autoref{fig:gaindb_zenith}). The same structure of periodic peaks/troughs across frequency is seen in each constant LST line, as was observed for the dry/wet layer simulation, as well as the sub-layer convergence study. There is additionally a variation along LSTs in constant frequency lines that has to do with the beam coupling to a fainter/brighter sky. 

In \appref{app:chrom_diff}, we report for completeness the resulting chromaticity difference arising from small variation in soil properties and for the multi-layer parametrisation discussed in \secref{sec:FEKO}.

\begin{figure}
\includegraphics[width=\columnwidth]{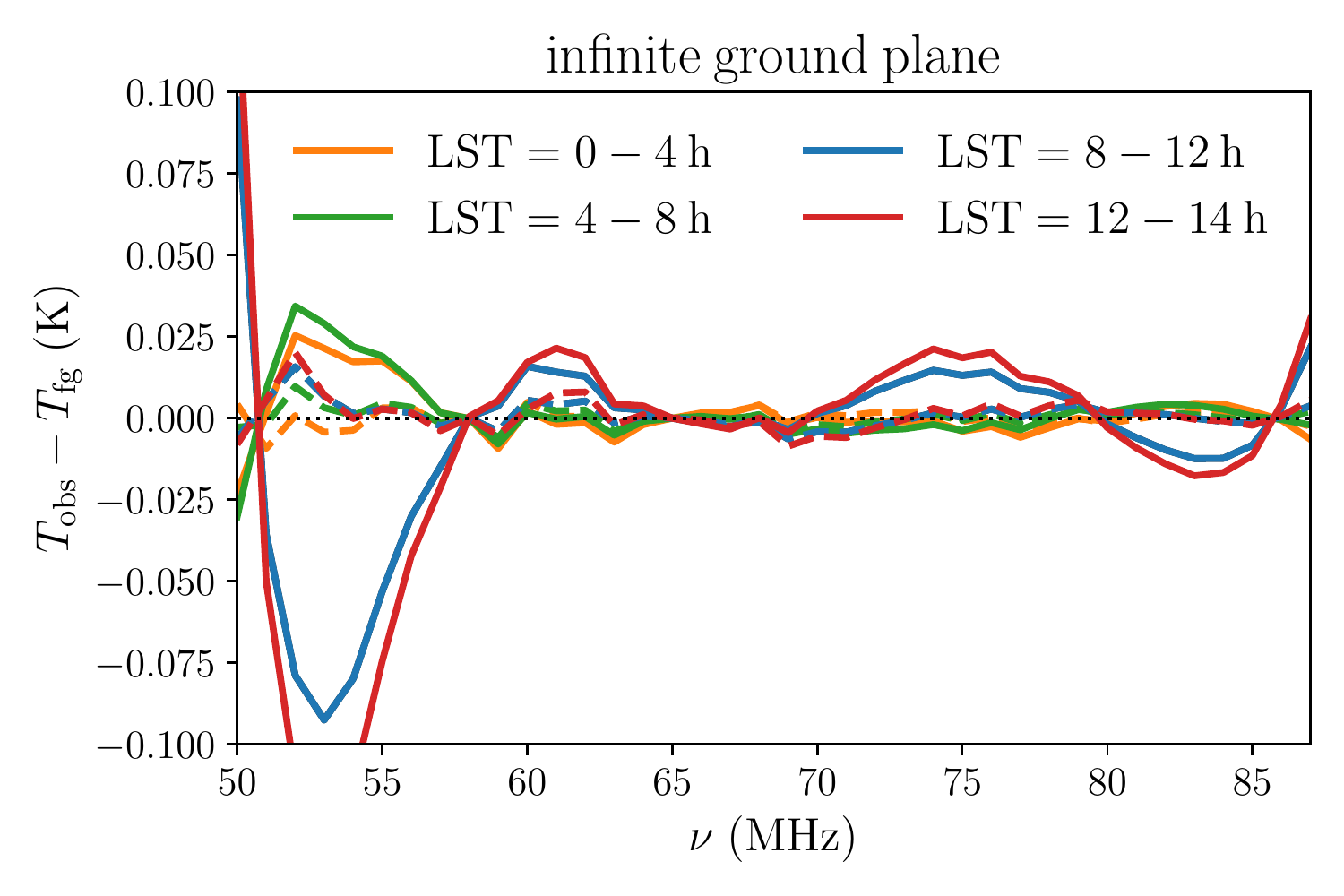}\\
\includegraphics[width=\columnwidth]{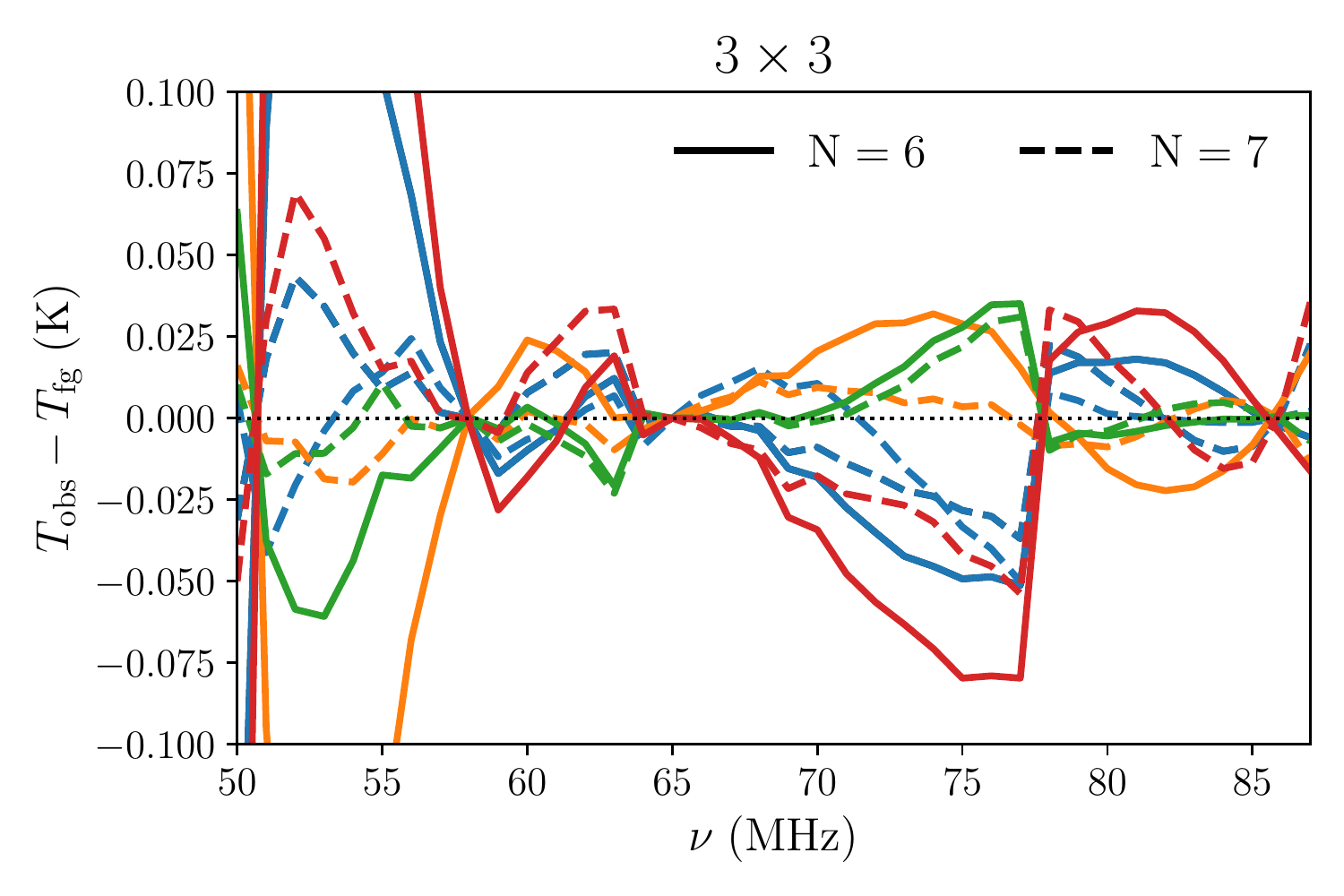}
\caption{Residual difference (in K) as a function of frequency between the simulated spectra of equation \secref{eq:meas} and the best fit model of \autoref{eq:fg}, presented for different orders of the log-polynomial ($N=6$ in solid lines and $N=7$ with dashed lines) and for different LST bins. Different ground planes are considered: an ideal infinite ground plane (top panel) and the $3\times 3$ ground plane (bottom panel). Note that the residual are computed without any chromaticity correction.}
\label{fig:inf_lst}
\end{figure}

\begin{figure*}
\includegraphics[width=0.65\columnwidth]{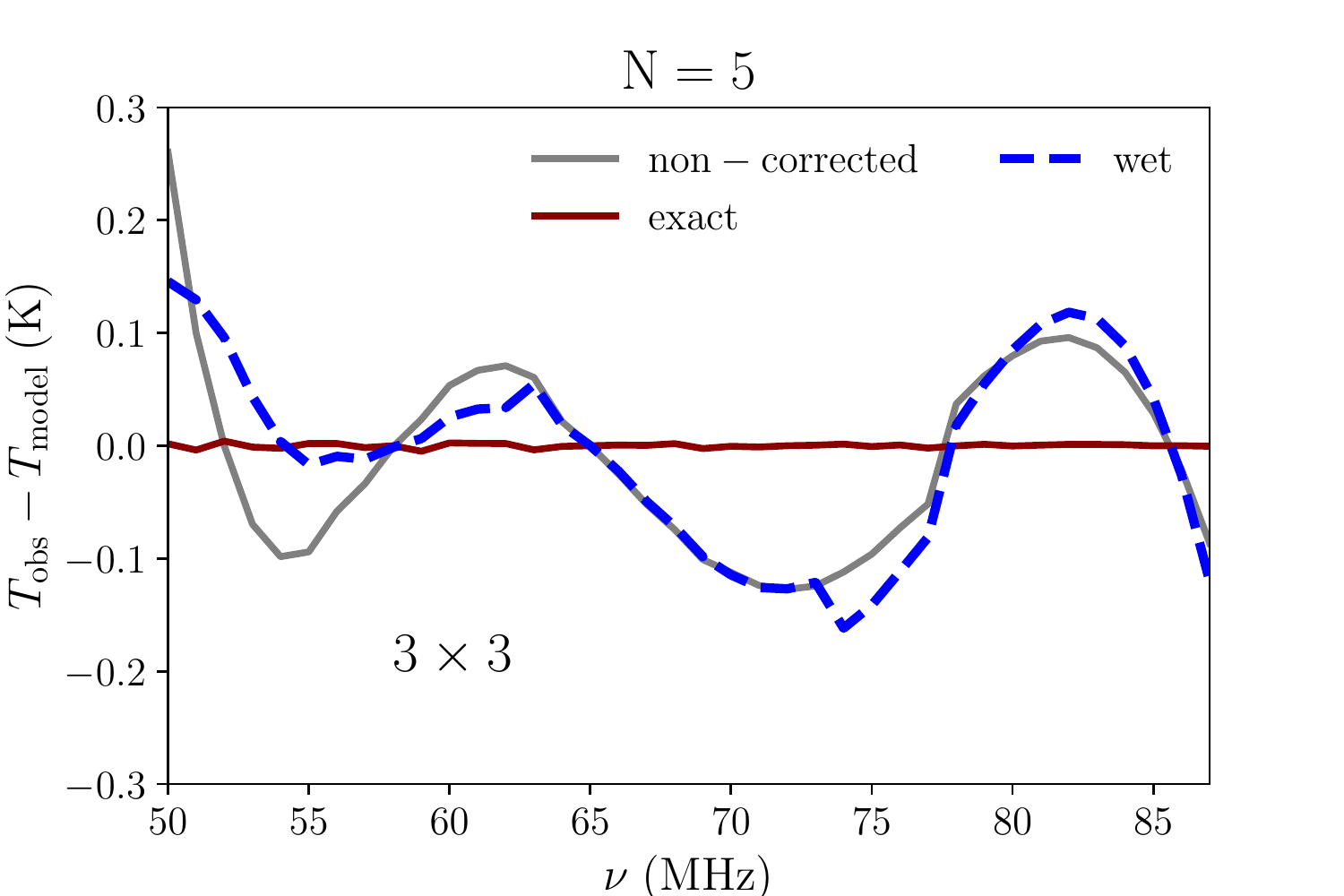}
\includegraphics[width=0.65\columnwidth]{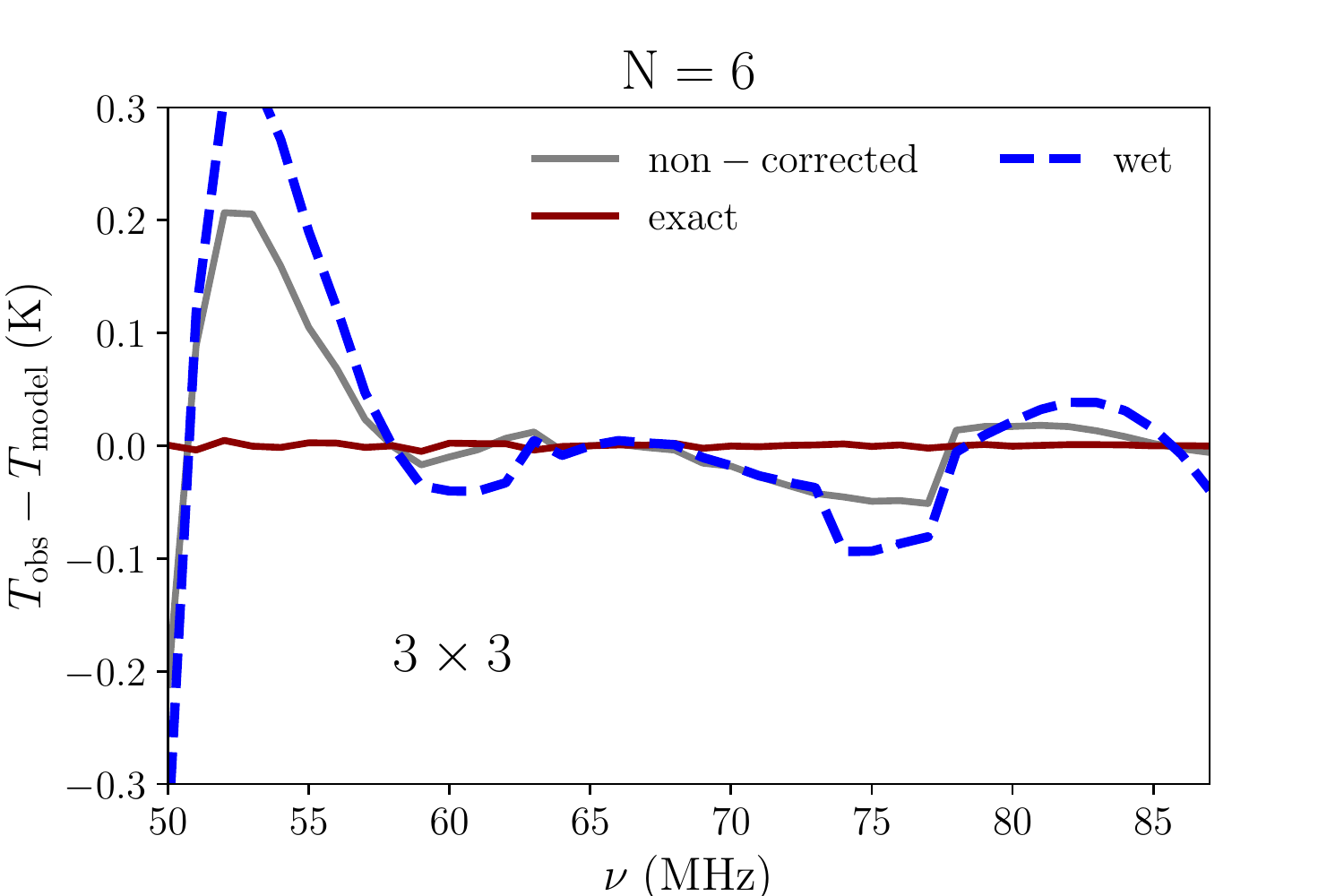}
\includegraphics[width=0.65\columnwidth]{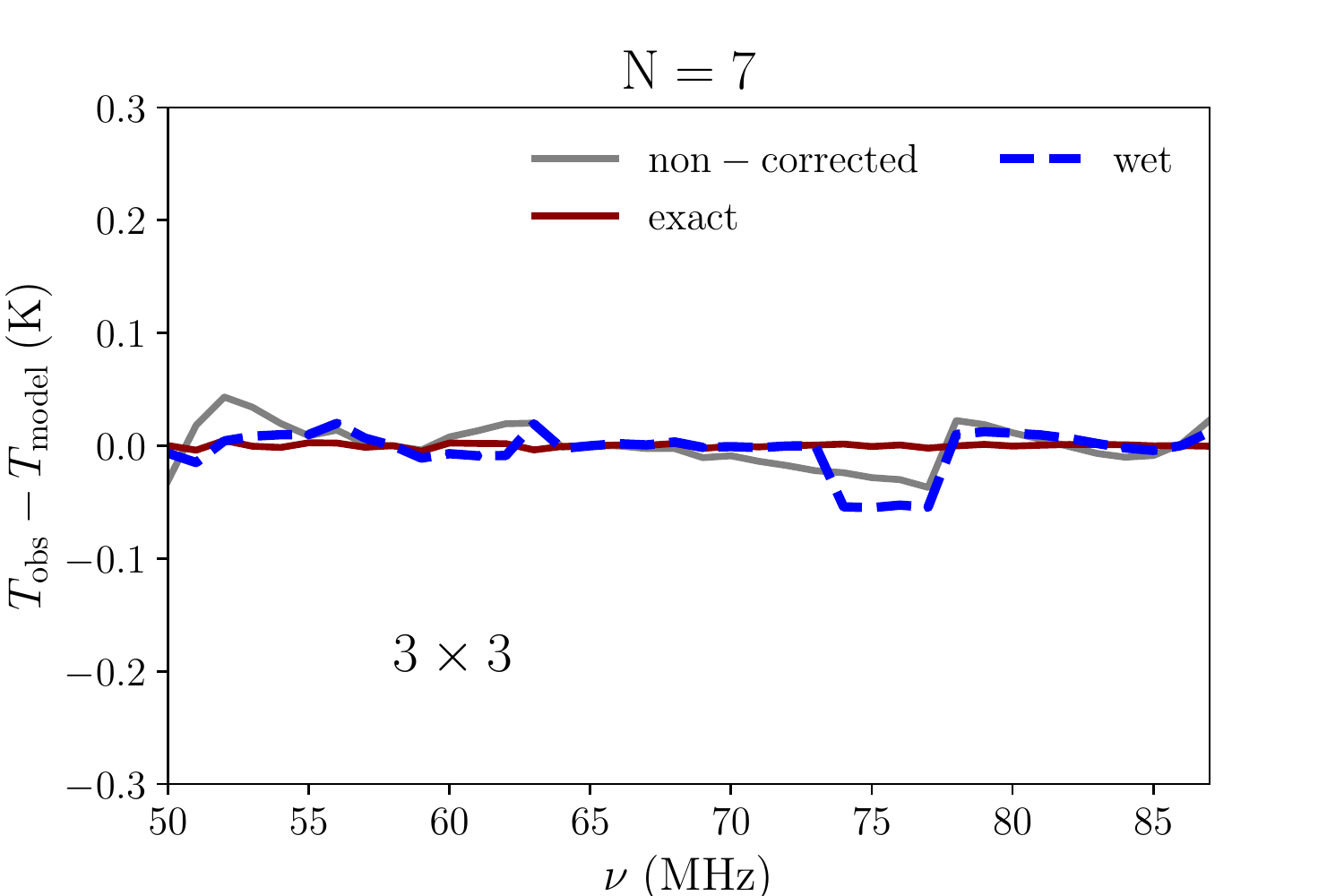}\\
\includegraphics[width=0.65\columnwidth]{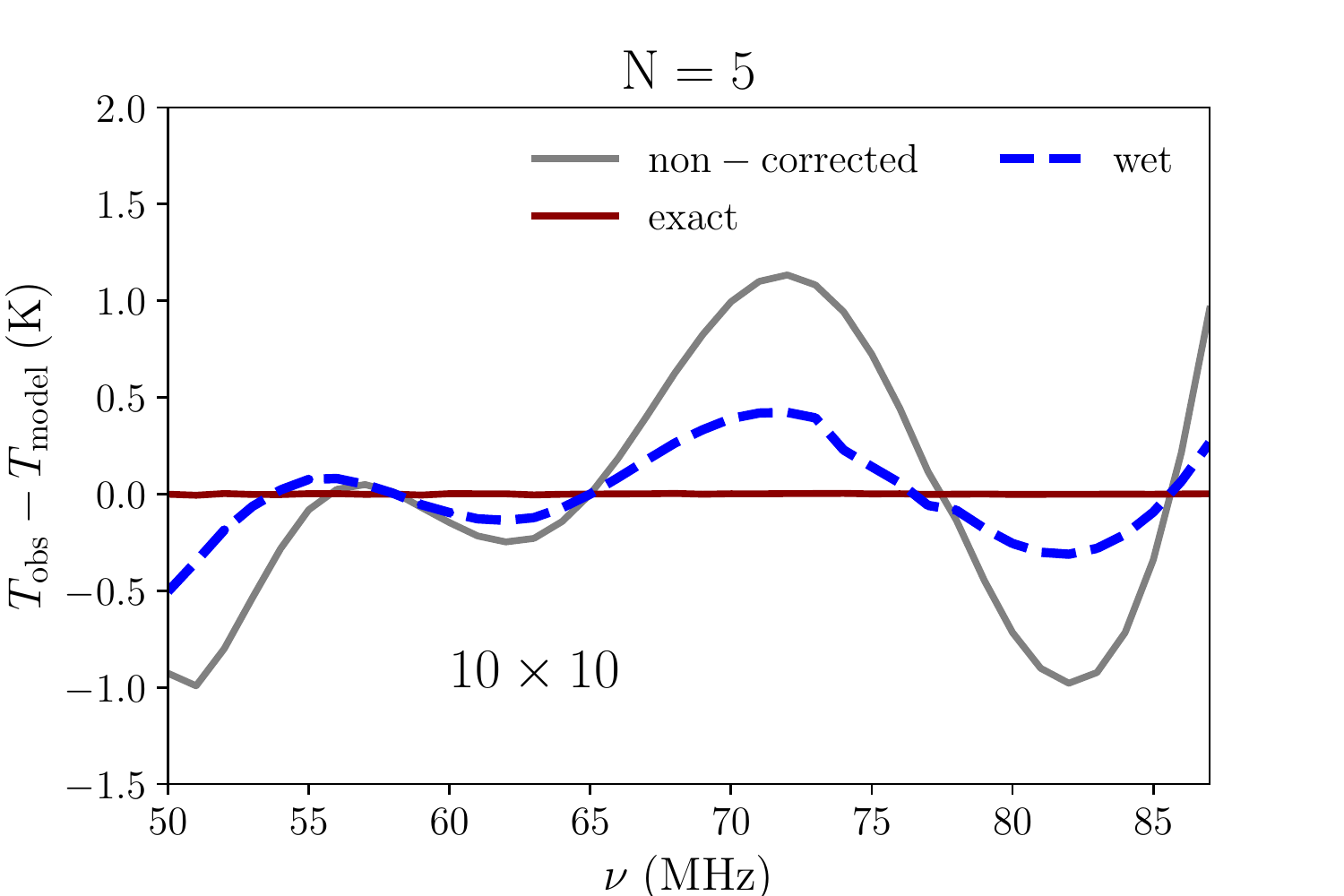}
\includegraphics[width=0.65\columnwidth]{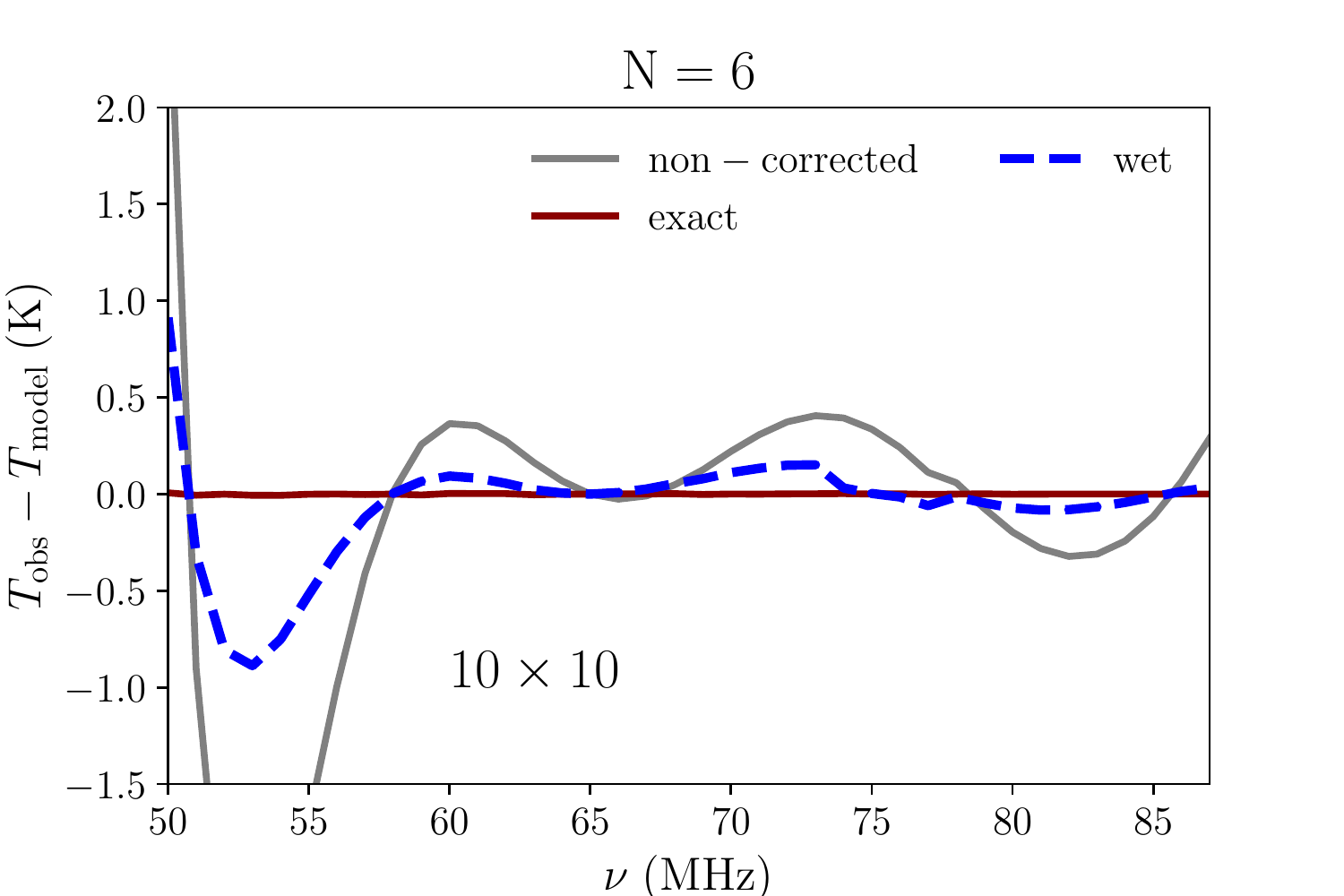}
\includegraphics[width=0.65\columnwidth]{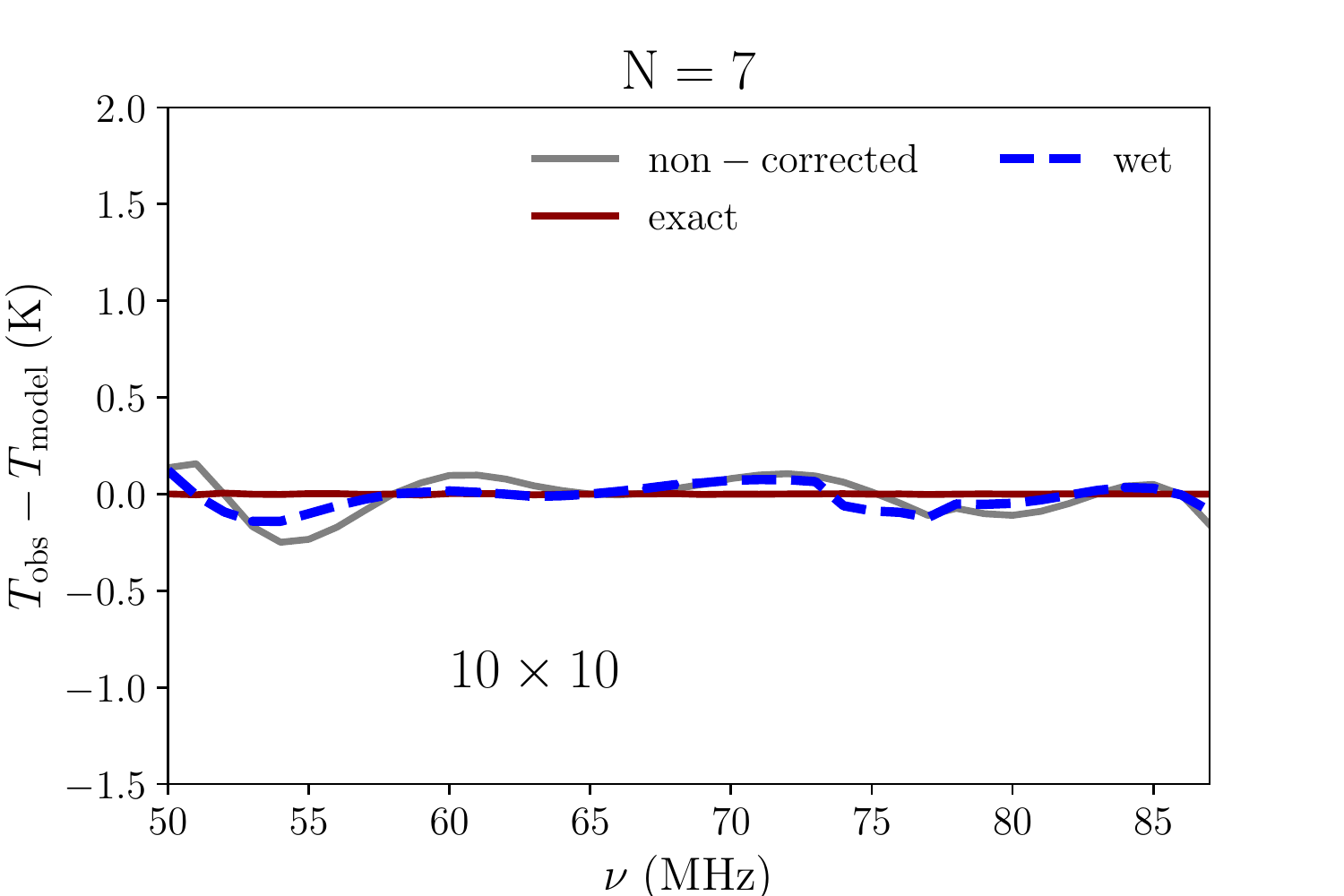}\\
\includegraphics[width=0.65\columnwidth]{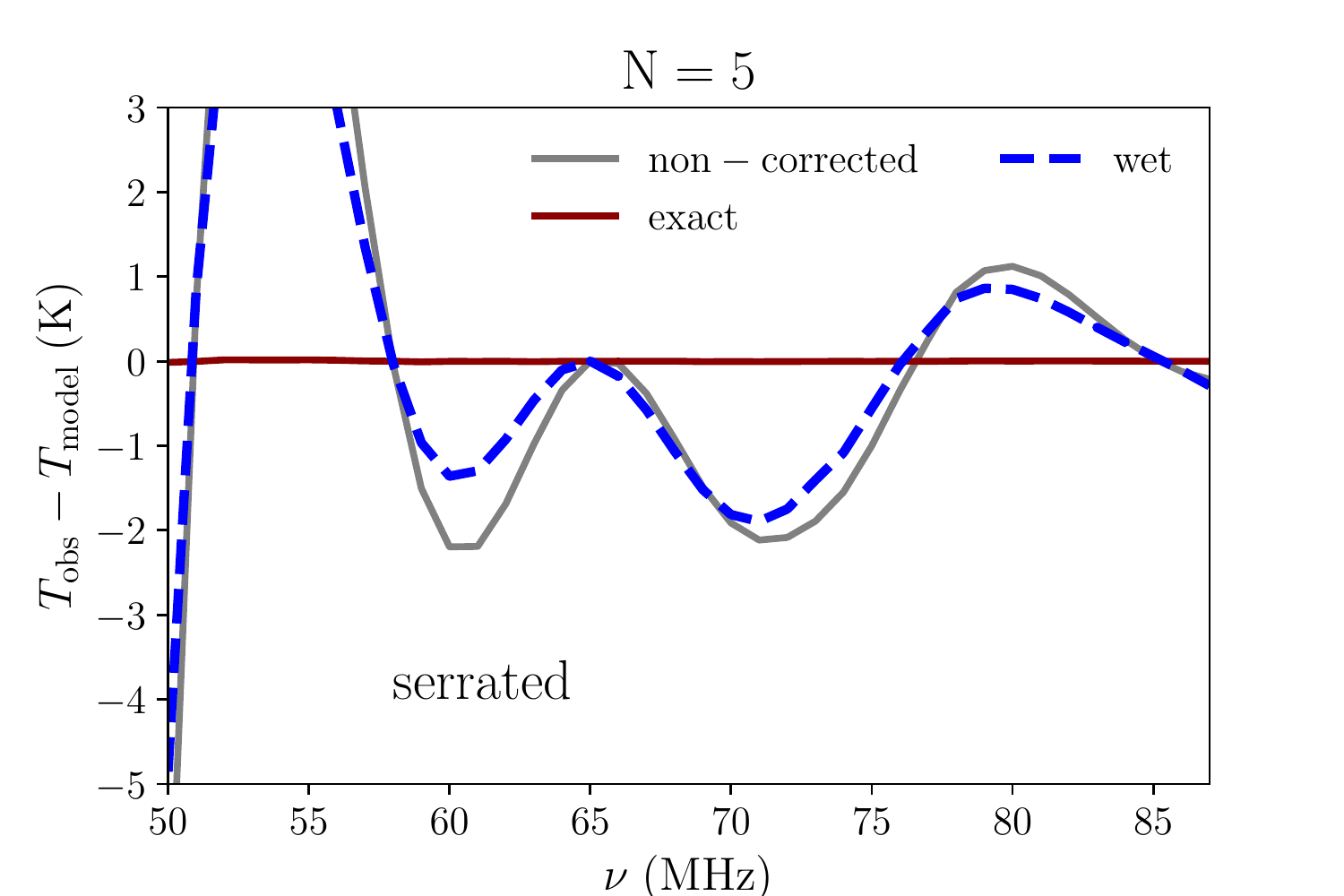}
\includegraphics[width=0.65\columnwidth]{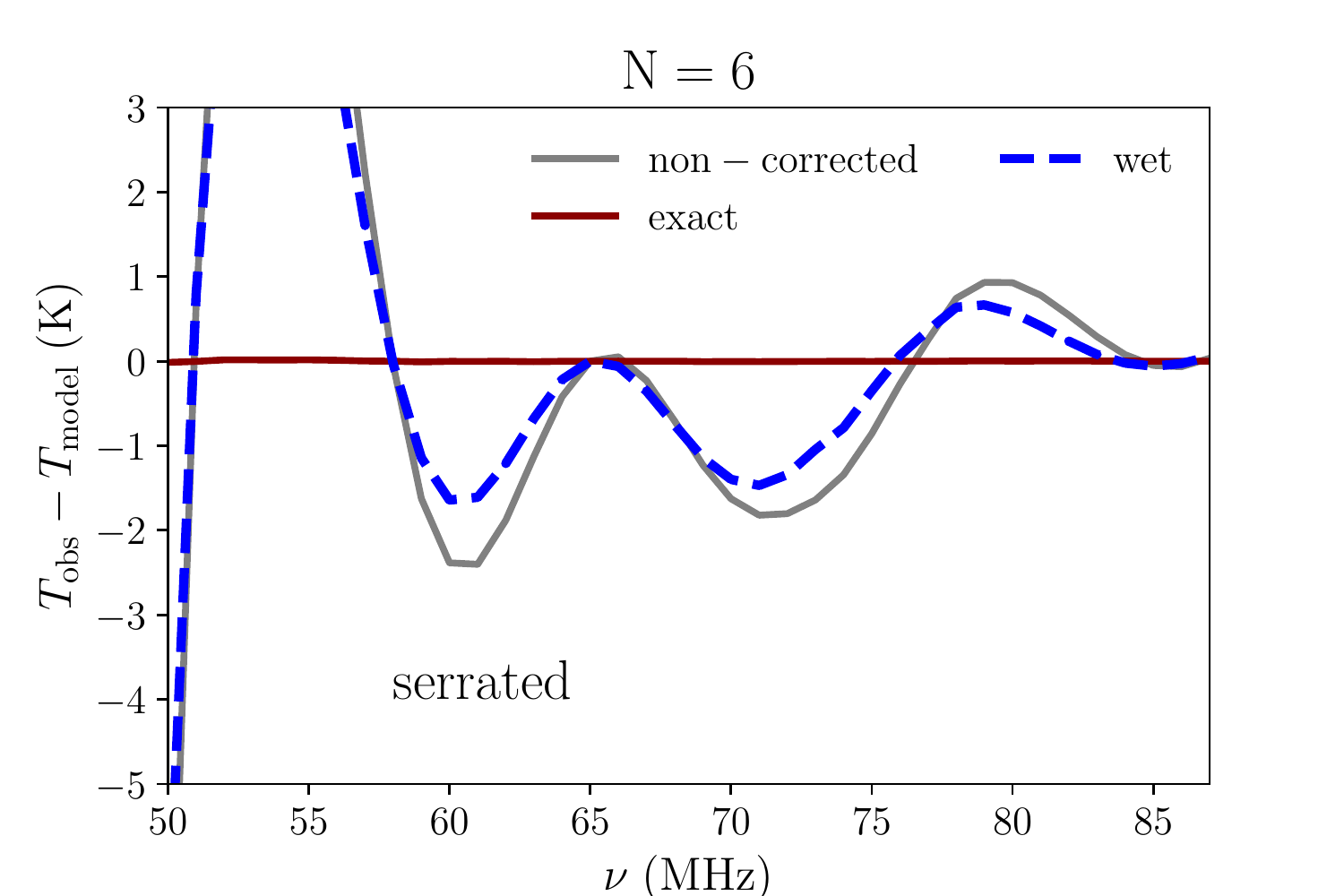}
\includegraphics[width=0.65\columnwidth]{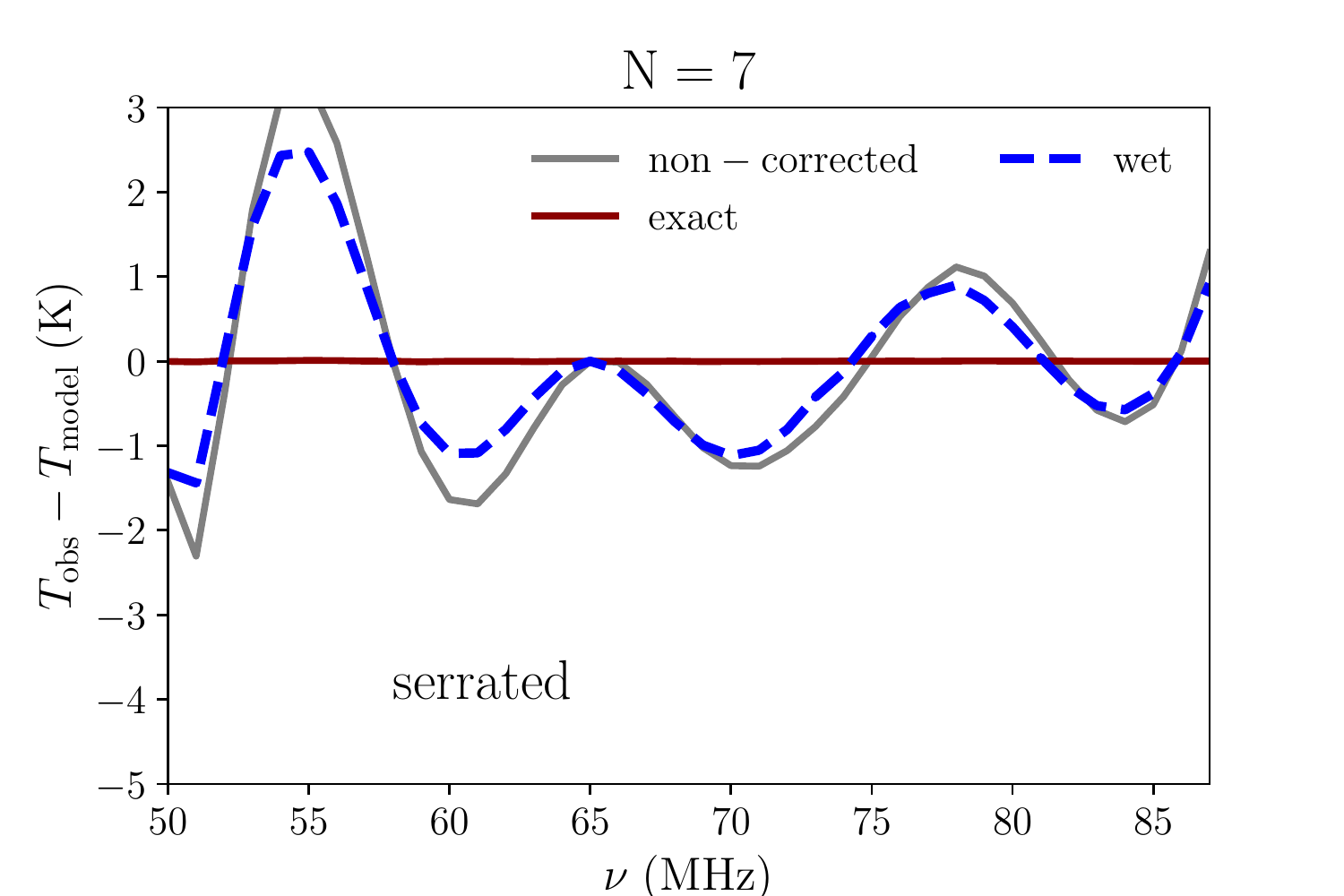}
\caption{Residual difference (in K) as a function of frequency between the best fit model of \autoref{eq:fg} and the simulated spectra obtained as described in \secref{sec:Tobs}, presented for different polynomial orders ($N=5,6,7$, left, central and right panels, respectively) and for different ground planes ($3\times 3$ top panels, $10\times 10$ central panels and serrated bottom panels). The simulated spectra are obtained considering dry condition and a one-layer description for the soil. The residuals are then computed without any chromaticity correction (grey solid lines), with an exact correction using the same beam model as for the spectra in input (dark red solid line) and for a chromaticity correction computed with wet soil condition (blue dashed lines). Note that the vertical scale changes for the different ground plane examined. Rms value for these residuals can be found in \autoref{tab:rms_soil}.}
\label{fig:res_soil}
\end{figure*}

\begin{figure*}
\includegraphics[width=0.65\columnwidth]{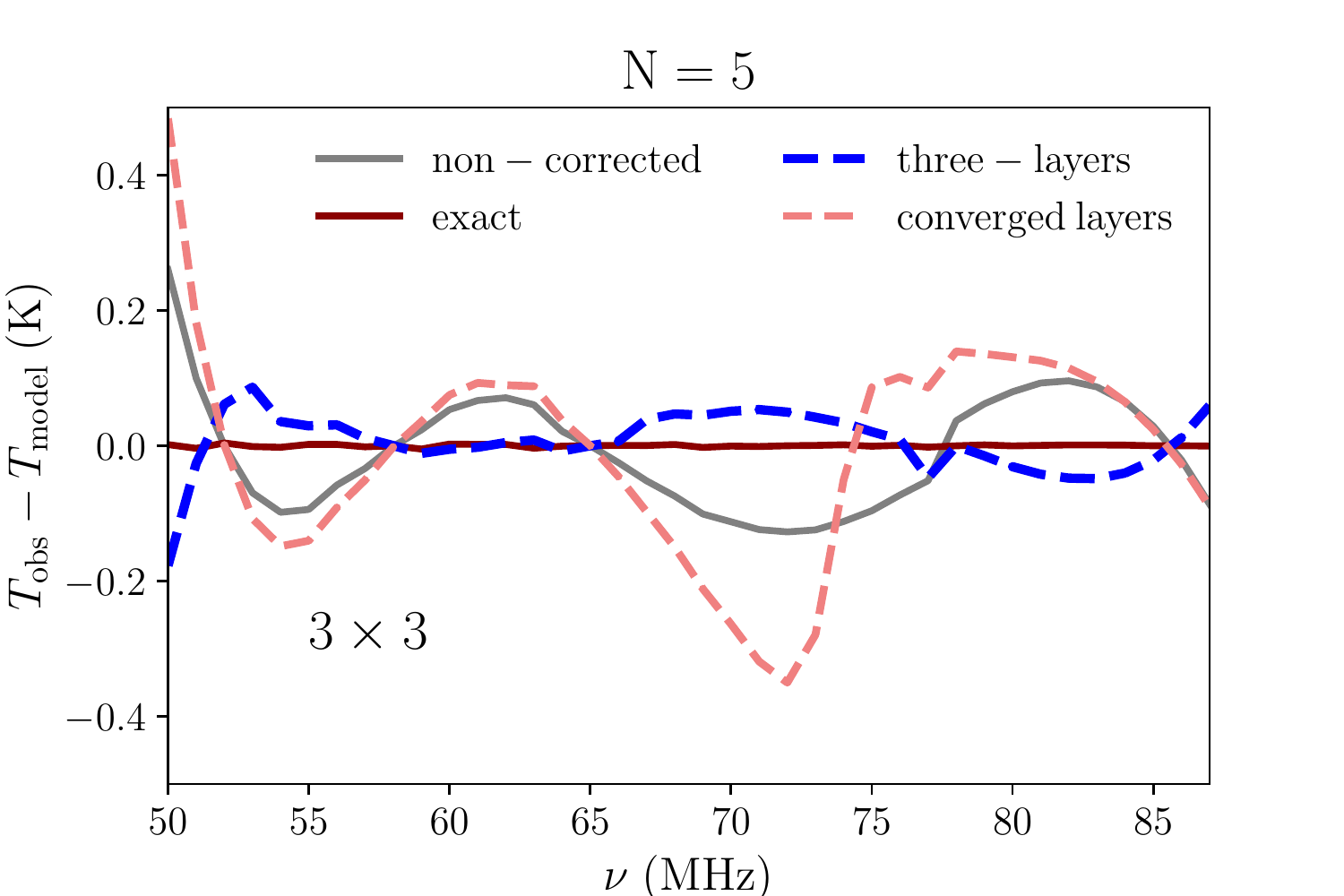}
\includegraphics[width=0.65\columnwidth]{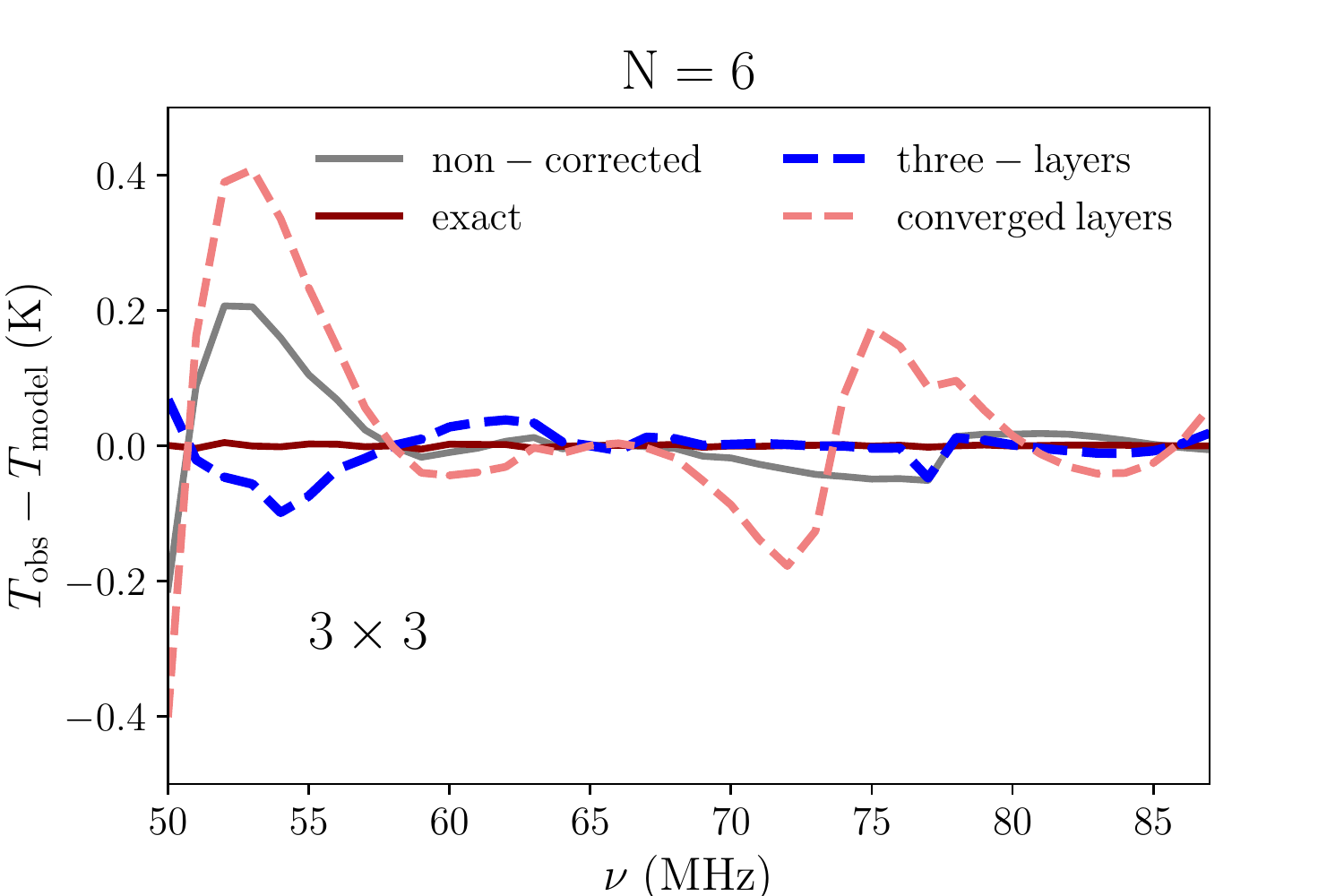}
\includegraphics[width=0.65\columnwidth]{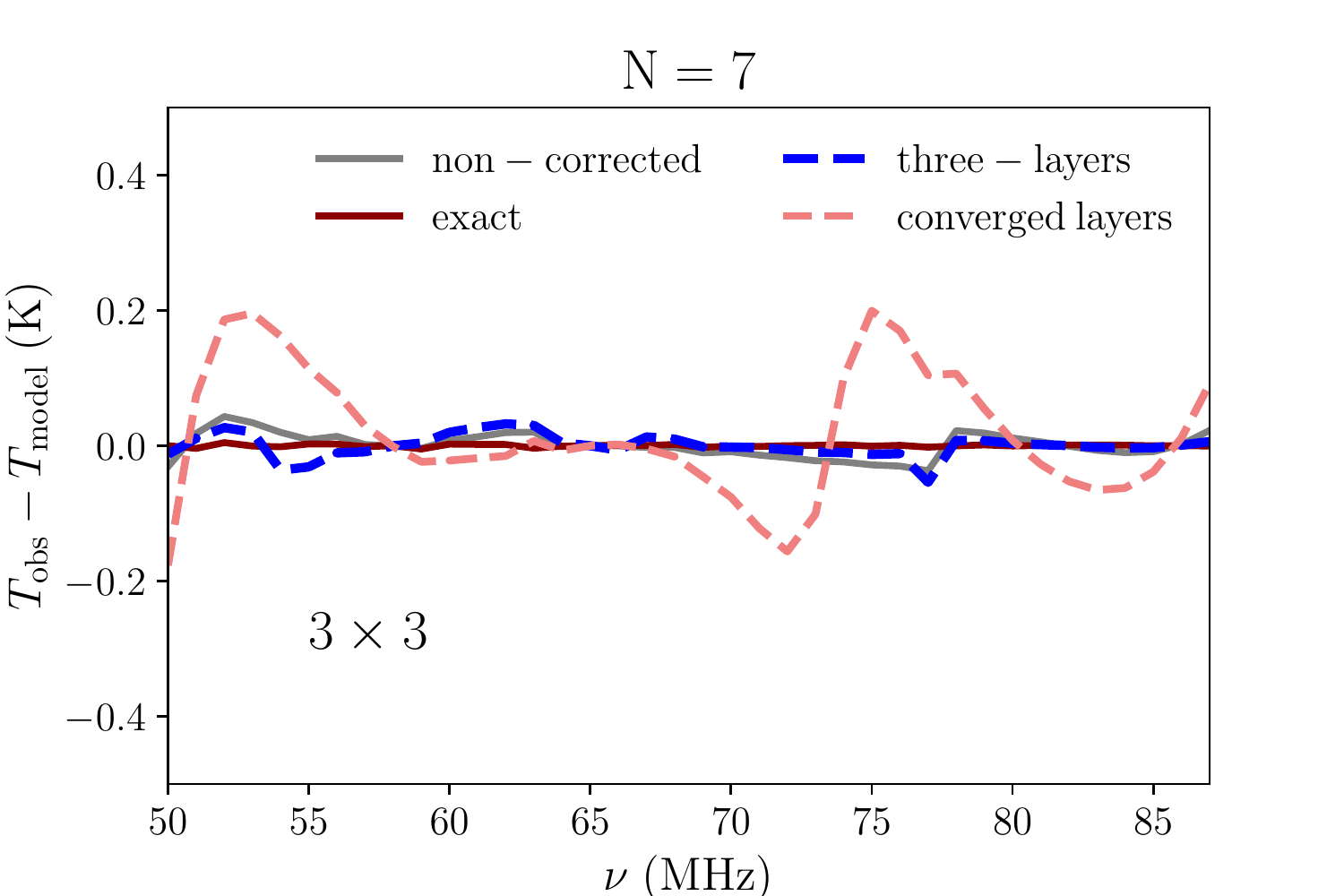}\\
\includegraphics[width=0.65\columnwidth]{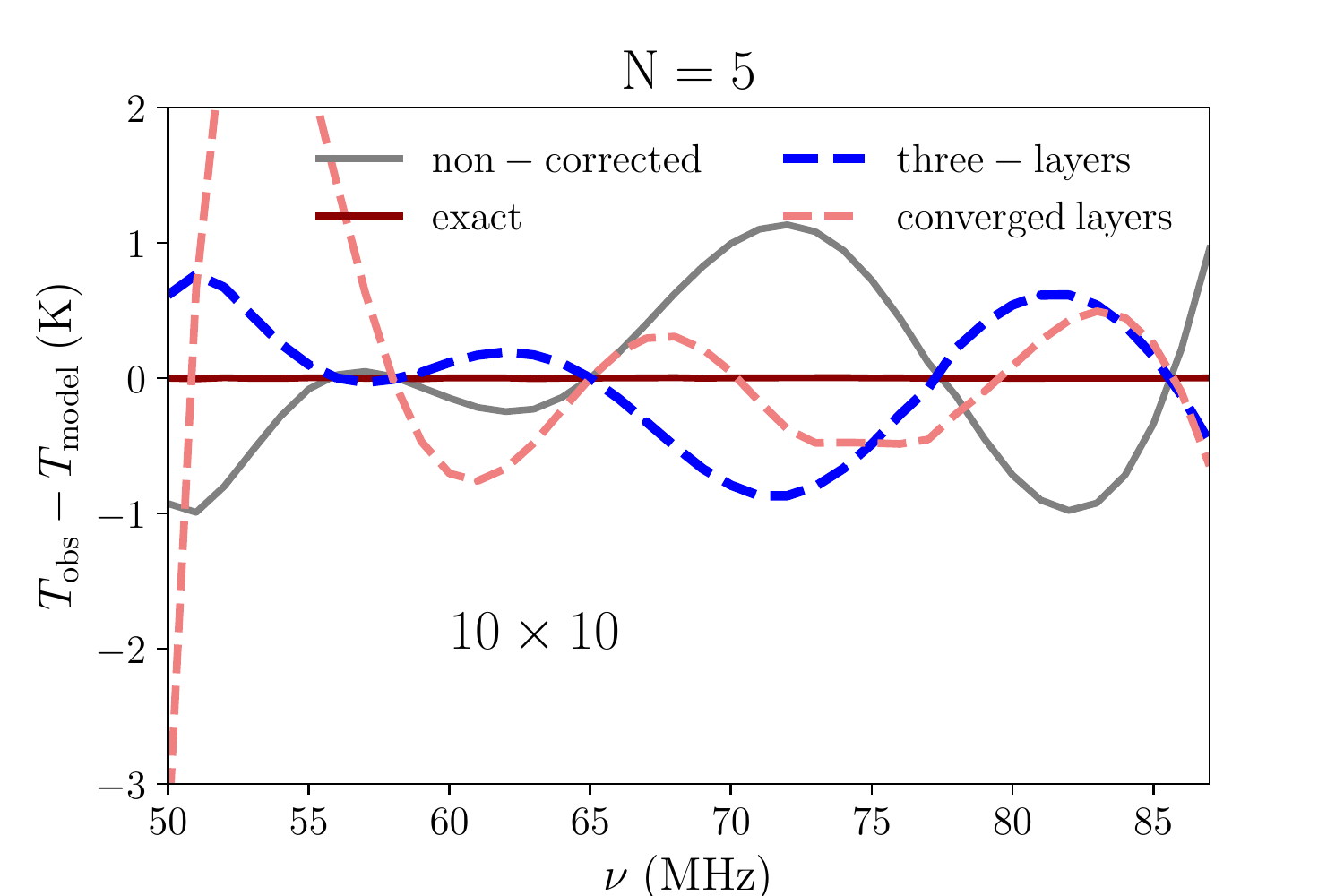}
\includegraphics[width=0.65\columnwidth]{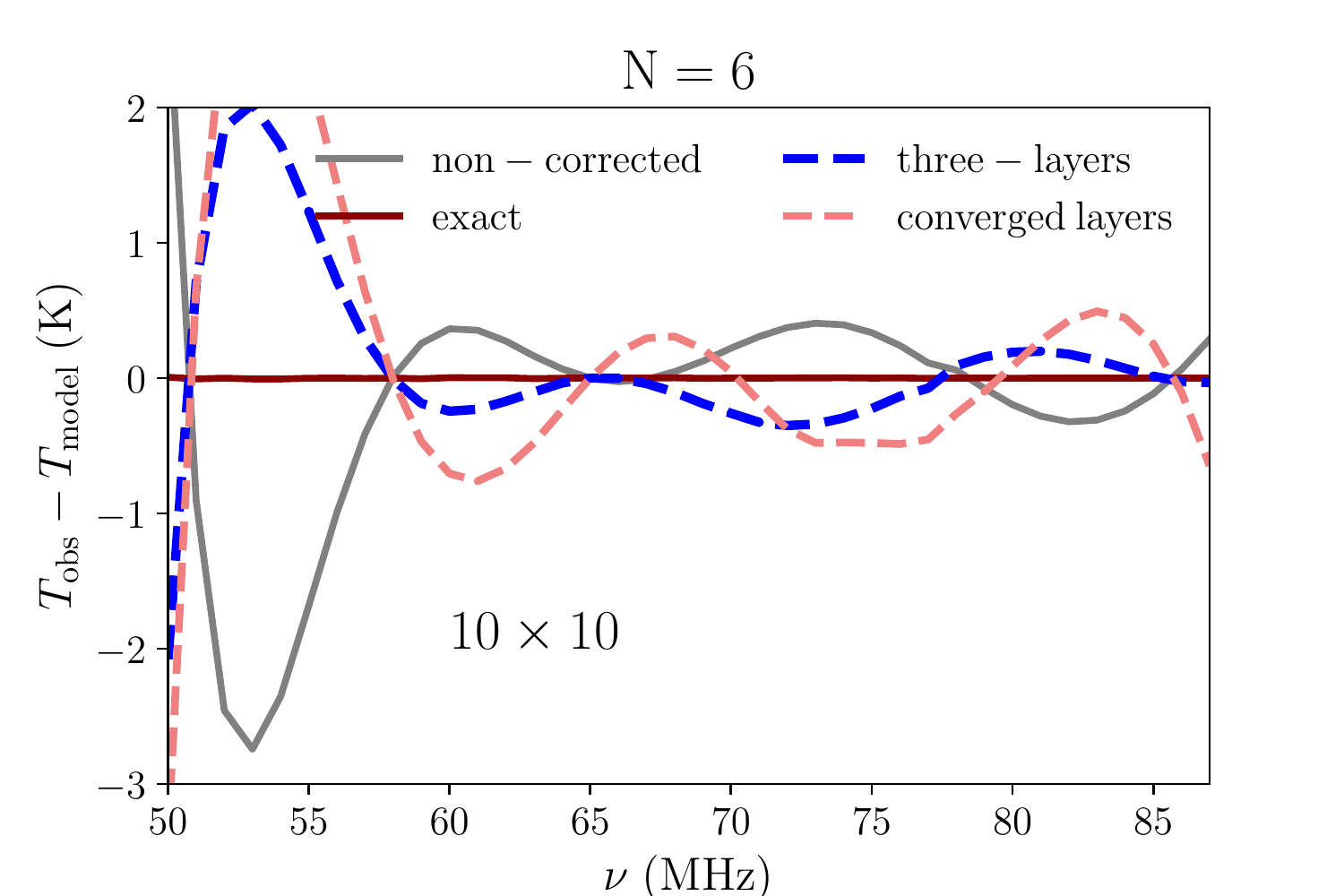}
\includegraphics[width=0.65\columnwidth]{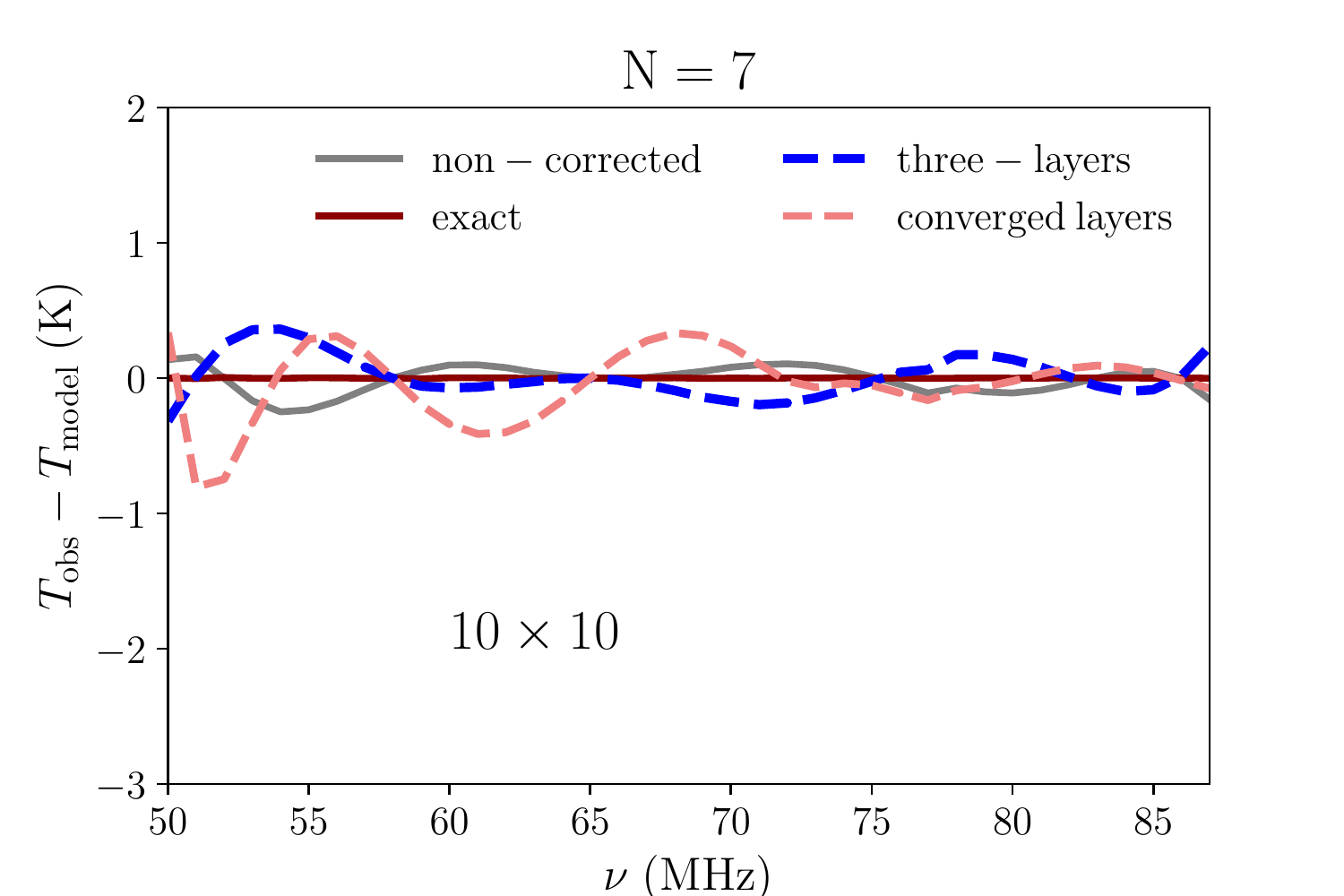}\\
\includegraphics[width=0.65\columnwidth]{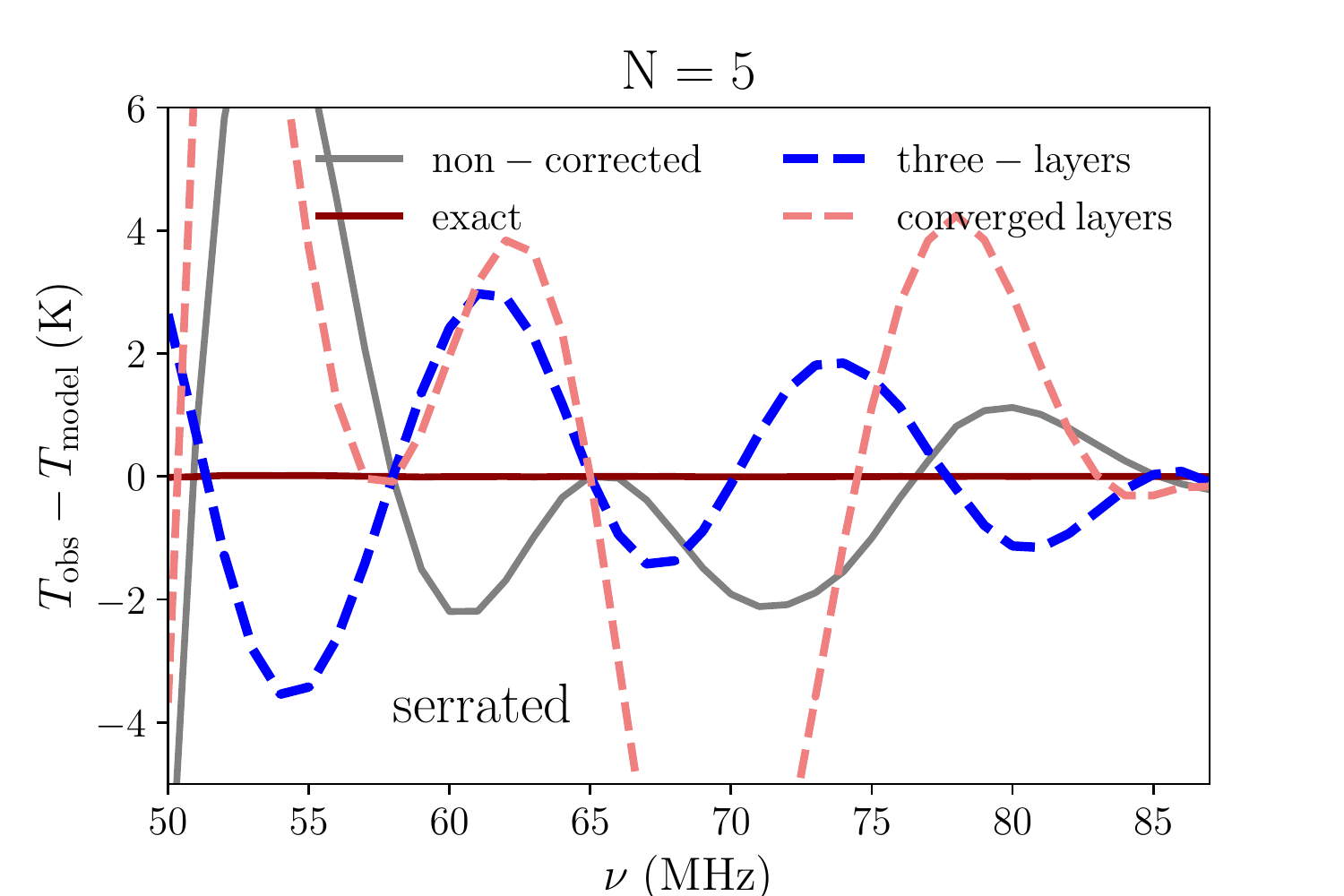}
\includegraphics[width=0.65\columnwidth]{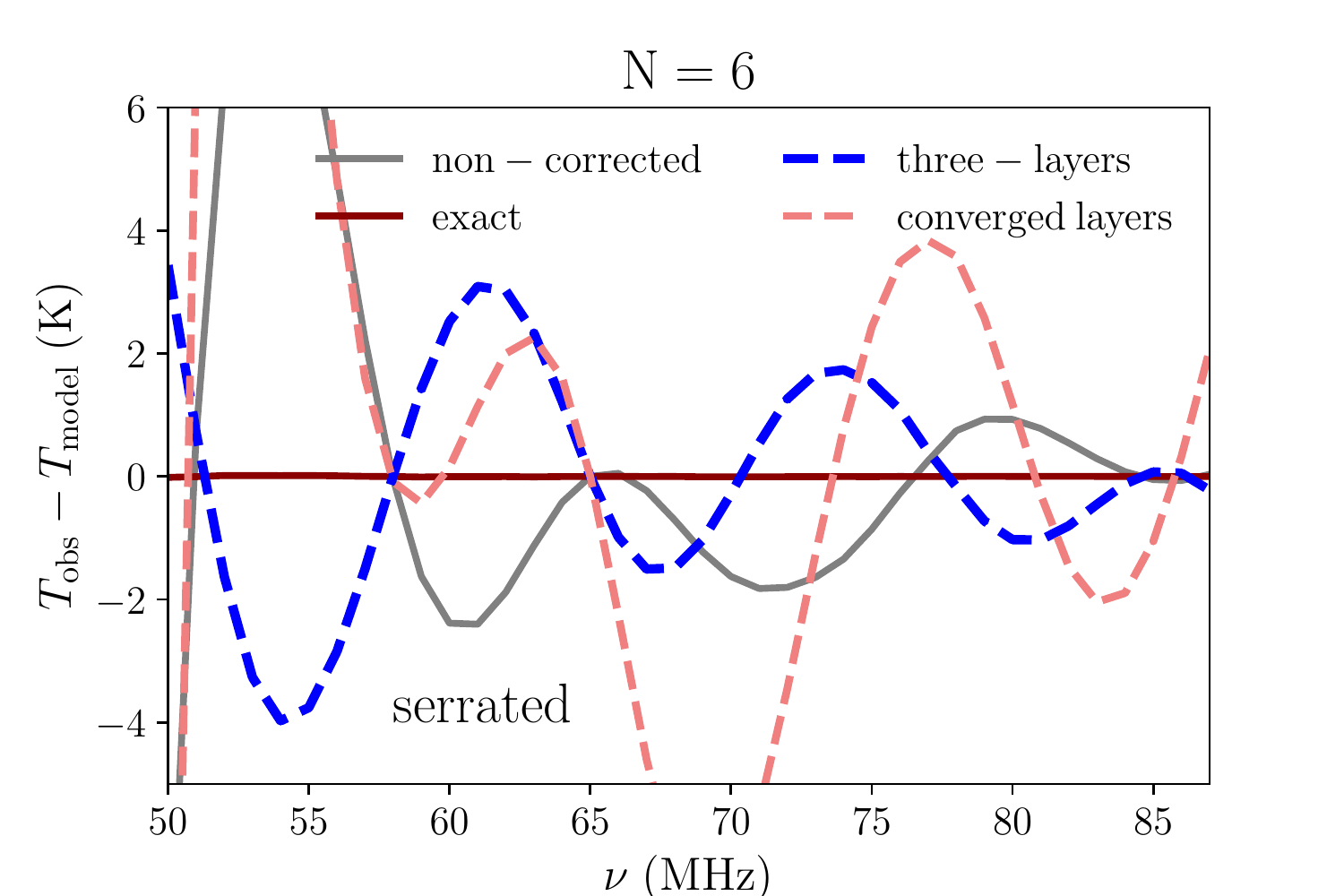}
\includegraphics[width=0.65\columnwidth]{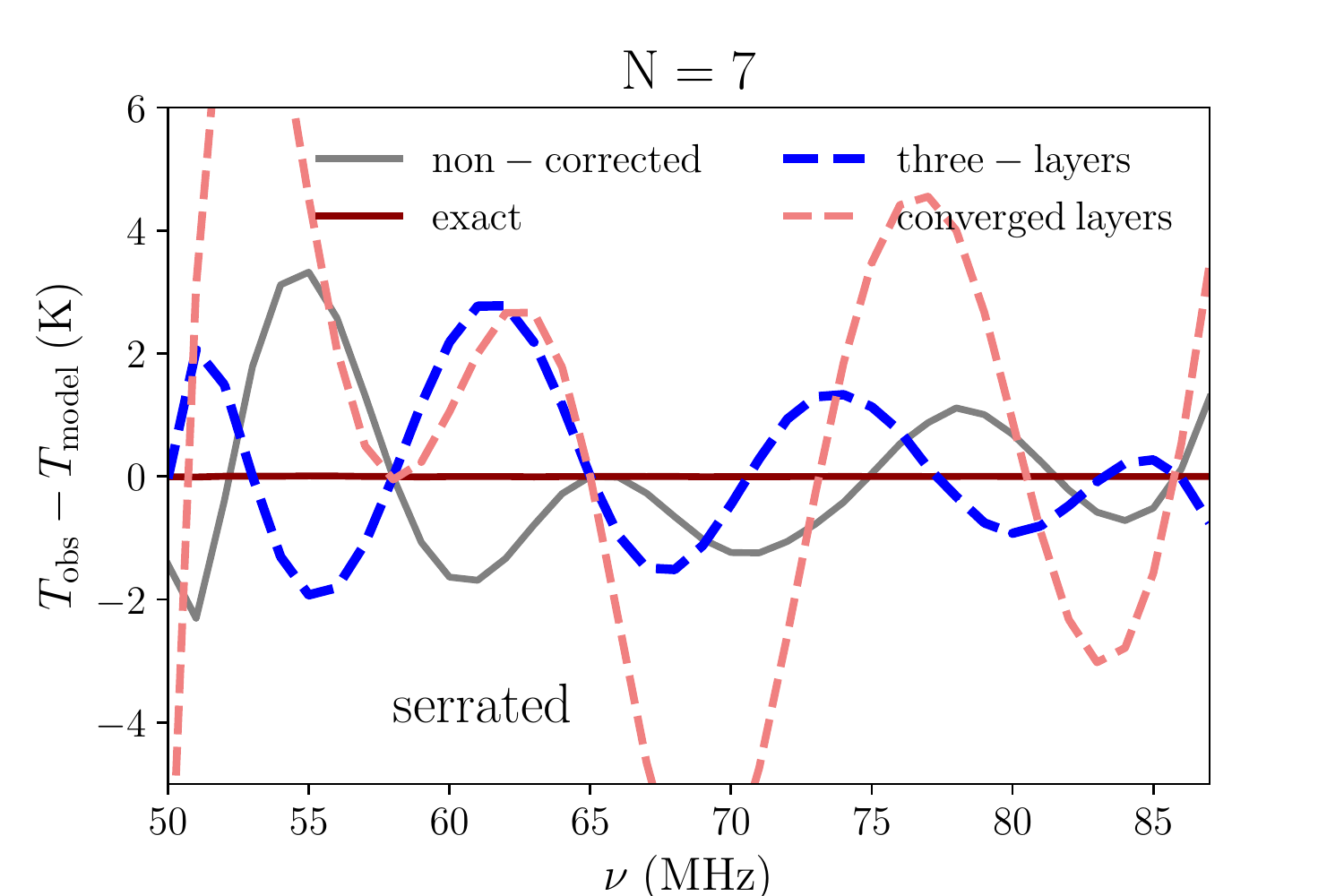}
\caption{Residual difference (in K) as a function of frequency between the best fit model of \autoref{eq:fg} and the simulated spectra obtained as described in \secref{sec:Tobs}, presented for different polynomial orders ($N=5,6,7$, left, central and right panels, respectively) and for different ground planes ($3\times 3$ top panels, $10\times 10$ central panels and serrated bottom panels). The simulated spectra are obtained considering dry condition and a one-layer description for the soil. The residuals are then computed without any chromaticity correction (grey solid lines), with an exact correction using the same beam model as for the input spectra (dark red solid line), for a chromaticity correction computed instead with the three-layer model (blue dashed lines) or the converged layer model (pink dashed line). Note that the vertical scale changes for the different ground planes examined. Rms values for these residuals can be found in \autoref{tab:rms_layers}.}
\label{fig:res_layers}
\end{figure*}

\subsection{Mock data modelling}\label{sec:model}

To construct a model for our mock data we assume that the Galactic foreground spectrum can be described as a $N$-term log-polynomial \citep[e.g.,][]{Bowman2010,Pritchard2010,Harker2012,Bernardi2015,Presley2015,Bernardi2016}:
\begin{equation}\label{eq:fg}
\log_{10}{T_{\rm fg}(\nu)} = \sum_{n=1}^{N} p_{n-1} \left[ \log_{10}{ \left(\frac{\nu}{\nu_0}\right)} \right]^{(n-1)}
\end{equation}
with $\nu_0=60$~MHz. Note that $p_1$ corresponds to the spectral index $\beta$ of \autoref{eq:T_H} since for $N=2$ we can rewrite \autoref{eq:fg} as $T_{\rm fg}(\nu)=p_0 (\nu/\nu_0)^{p_1}$.

\medskip
We are interested in assessing the impact of the beam chromaticity on this assumption, examining its effect on the frequency smoothness of our simulated spectra computed as in \autoref{eq:meas} and assuming no absorption feature, i.e. $T_{\rm HI}=0$. 
We compute the deviation of the mock measured sky temperature with respect to the smooth foreground model of \autoref{eq:fg}. Note that the best fit values for the foreground parameters are obtained with a non-linear least squares solver. 

\medskip
We first analyse the ideal case of an infinite ground plane for completeness. This antenna pattern is an ideal case which has only been used as reference for subtraction in \autoref{fig:groundplanedB_convergence}, and produces the smoothest spectral response for gain. We show in the top panel of \autoref{fig:inf_lst} that the log-polynomial model with $N=6$ is already capable of describing the spectra very well, in particular for LST$<8$~h. A higher log-polynomial order is required instead for the LST range of the 2018/2019 LEDA observing campaign (i.e. $9-12$~h). When the beam modelling includes the finite ground plane (the $3 \times 3$ ground plane is shown as an example in the bottom panel of \autoref{fig:inf_lst}) there are residual structures that are not captured by the model of \autoref{eq:fg}. These residuals again depend on the LST range considered, as the beam couples with the sky structures. 

\medskip
We investigate this further as a function of the ground plane type in \autoref{fig:res_soil}, where we use our baseline case, i.e. the dry soil one-layer model.  If we do not correct for the effect of beam chromaticity, the induced structures in the spectra prevent the smooth model from accurately describing the foregrounds, and the residuals are highly oscillating in frequency. 
The rms values for the results can be found in \autoref{tab:rms_soil} for different choices for the order of the log-polynomial. 
The effect is more prominent for the serrated ground plane (where we found residual rms values around 1K) and still important for the $10 \times 10$ ground plane. An exact chromaticity correction solves the problem for all types of ground planes describing the resulting simulated spectra with residuals of only a few mK.

If we attempt a correction assuming wet properties for the soil parameters instead of dry conditions, we obtain better residuals but still with strong features as a function of frequency.

We repeat  this same analysis considering now the impact of the multi-layer description of the soil. We show in \autoref{fig:res_layers} the residual structures for the baseline model (one-layer, dry soil condition) when correcting the effect of chromaticity assuming either the three-layer or the converged multi-layer model, for different N-term log-polynomial models (5,6 and 7) and different ground planes ($3\times 3$, $10\times10$ and serrated). The rms of the residuals are reported in  \autoref{tab:rms_layers} in \appref{app:FoM_multi}. Note that the non-corrected and the exact correction cases are the same of \autoref{fig:res_soil}.

Attempting a chromaticity correction with the three-layer model (that is similar to the baseline one) always improves the rms and the smoothness of the resulting spectra. The converged layer model instead worsens the structures in the residuals and, for the serrated case, there is very small improvement increasing the order of the log-polynomial model. These results can be compared with \autoref{fig:chromdiffgrounds}. Despite the differences in the chromaticity patterns between the three-layer and the converged layer cases not being strong, the higher contrast of the structures in the right column of  \autoref{fig:chromdiffgrounds}, foreshadows the possible struggles of the correction procedure for the converged case.

\medskip
We can in each case confirm that the presence or not of a ground plane is more important than the parametrisation of soil both for dry/wet conditions and different layer models, since the residuals are high for the more oscillating larger ground planes, and these oscillations are inherent to the antenna pattern as a function of frequency.

\medskip
These type of structures in the real data could prevent the detection of the cosmological signal or produce an erroneous detection and thus need to be investigated further. The rest of this work is dedicated to this problem.

\begin{figure}
\includegraphics[width=\columnwidth]{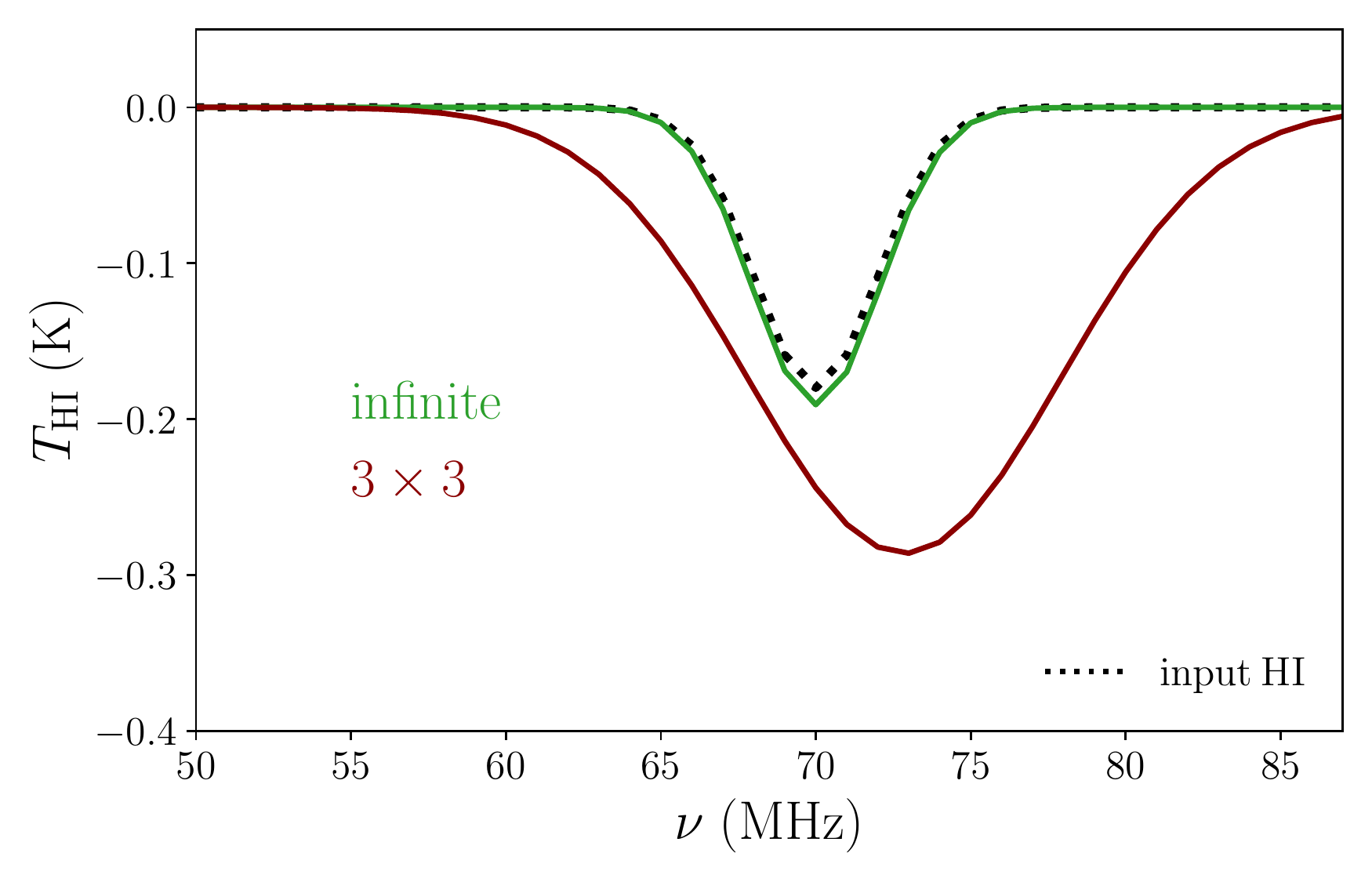}
\includegraphics[width=\columnwidth]{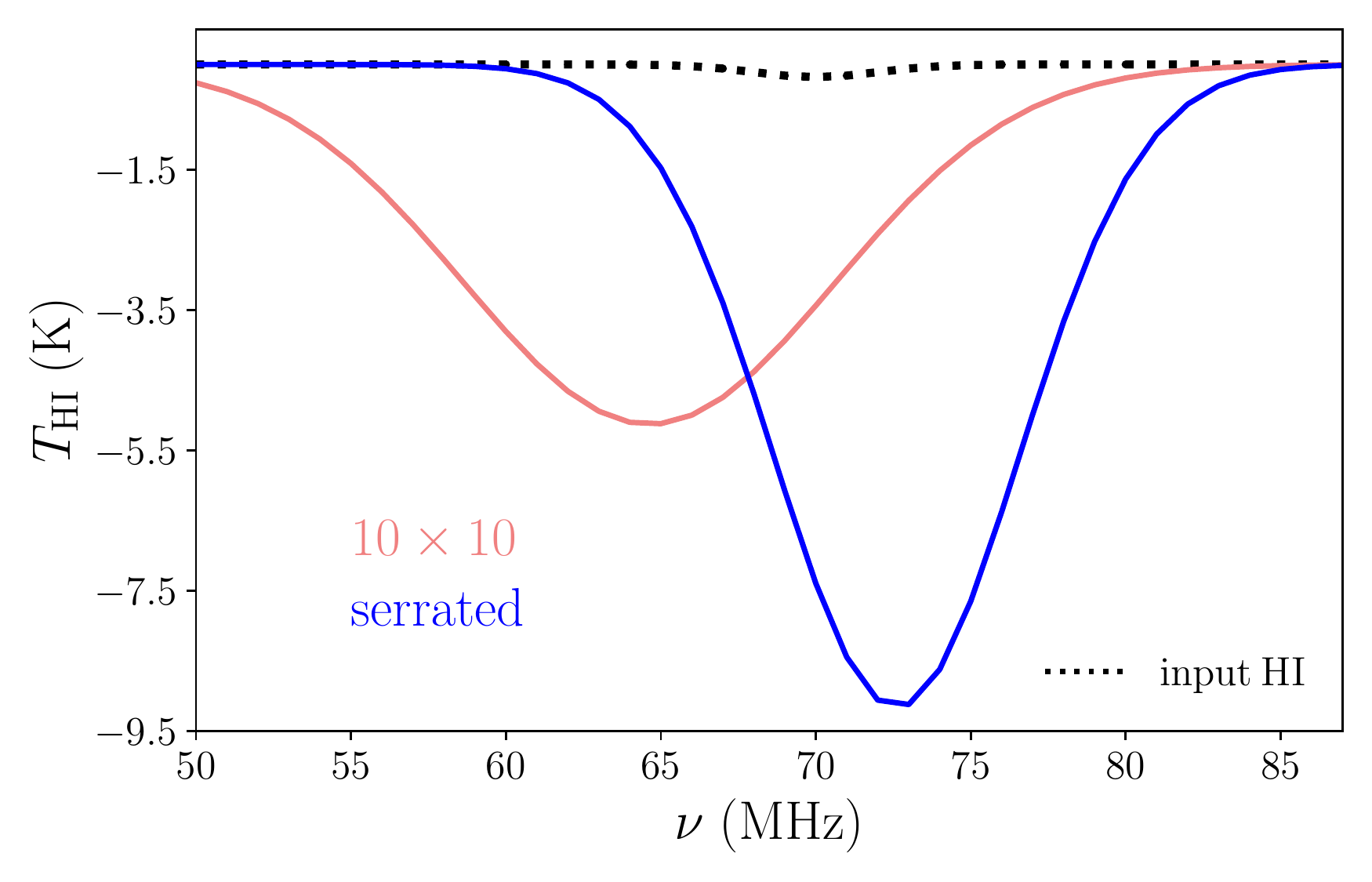}
\caption{The reconstructed $21$cm absorption profile for different ground planes considered. In the upper panel the infinite case and $3\times3$ are shown in green and dark red respectively.  The $10\times10$ and the serrated cases are instead shown in the lower panel in pink and blue, respectively. Note the different scale for the y-axis. The simulated spectra are generated for the one-layer dry soil conditions and are not corrected for the effect of chromaticity. The input model that we would like to reconstruct is shown as a black dotted line. The serrated result is shown for completeness but the amplitude of the absorption feature converged to the edge of the prior.}
\label{fig:diffground}
\end{figure}

\section{Foreground and cosmological parameter reconstruction}\label{sec:recon}

In the last session we have discussed qualitatively the impact of a realistic beam model for the LEDA antennas, which can compromise the smoothness of the measured spectra, and is thus an important assumption for discerning the 21cm signal from the foregrounds.

In this section we investigate how much these spurious structures in the simulated spectra impact the Bayesian extraction of the 21cm absorption feature \citep[e.g.][]{Bowman2018,Bernardi2016,Singh2021}.

\medskip
Our results were obtained running the \textsc{hibayes} code \citep{Bernardi2016,Zwart2016}, a fully Bayesian framework where the posterior probability distribution is explored through the \textsc{multinest} sampler \citep{Feroz2008,Feroz2009}. We assume a Gaussian Likelihood for the data and make use of the model described in \secref{sec:model}. The covariance matrix is assumed diagonal in frequency and the diagonal terms are computed with \autoref{eq:noise}. For each analysed case we run the pipeline for different order of the log-polynomial and present the results with the highest evidence.

\subsection{Finite ground plane effect}

We analyse the structure induced on the simulated spectra by the different ground planes used in LEDA observations and modelled in this work.
We show in \autoref{fig:diffground} the reconstructed $T_{\rm HI}$ obtained from the mean values of the posterior distributions, in comparison with the input HI model used for the simulated spectra. While for an infinite ground plane it is possible to reconstruct the correct input, in presence of a finite ground plane the algorithm converges towards biased solutions, preventing a correct detection of the cosmological signal.
In \autoref{fig:diffground} no correction for the effect of chromaticity is applied. 
If we divide the simulated spectra with the beam factor of \autoref{eq:beam_chromaticity}, computing it with the same beam model used for the simulated spectra, we are exactly correcting for the effect of chromaticity. The corrected spectra are much smoother as a function of frequency and they are well modelled by a low order of the log-polynomial. The input HI parameters are successfully reconstructed with residuals of the order of a few mK using only a 5-term polynomial model.

\subsection{Soil properties}

The results of the previous section suggest that, when attempting a reconstruction of the absorption feature in the LEDA data we need to correct for the effect of chromaticity. It is, however, not realistic to assume that our beam model agrees perfectly with the true beam. We investigate here the effect of a non-perfect reconstruction by varying the properties of the soil.

\begin{figure}
\includegraphics[width=\columnwidth]{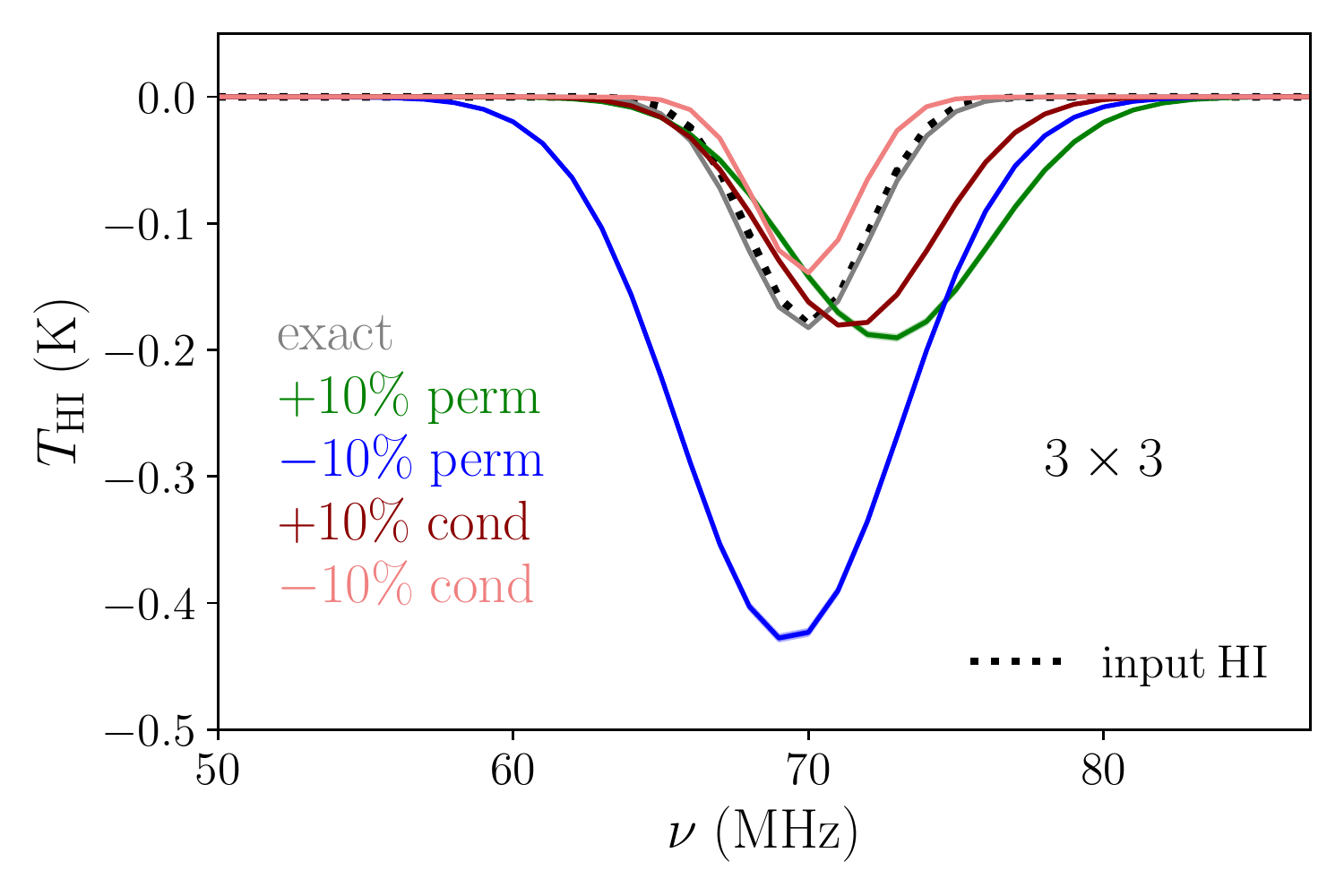}
\caption{The reconstructed $21$cm absorption profile
obtained when the simulated spectra (generated for the one-layer dry soil conditions) are corrected for the effect of chromaticity considering a $\pm 10\%$ variation for the permittivity and conductivity. The input model that we would like to reconstruct is shown as a black dotted line.}
\label{fig:diffsmall}
\end{figure}

\medskip We analyse in \autoref{fig:diffsmall}, for the $3\times3$ ground plane, the effect on the reconstructed absorption feature of small variations in the assumed value for the conductivity ($\sigma$) or the permittivity ($\epsilon_r$) of the soil. We consider a 10\% shift with respect to the dry soil condition to be compared with a factor $\sim2$ difference between the dry and wet condition (presented in \autoref{tab:soil_params}).  We report for completeness in \autoref{fig:chromdiff3x3} the variation of the beam factor in these analysed cases with respect to the baseline beam model. As can be seen also from \autoref{fig:diffsmall}, the change in conductivity biases the reconstructed absorption feature for both higher and lower values of $\sigma$. A lower value of the conductivity leads to a 20\% lower amplitude for the absorption signal while a higher conductivity bias the reconstruction of its central frequency.  
Increasing the permittivity has a similar effect while the bias gets stronger for a lower value of $\epsilon_r$, resulting in a factor $\sim2$ enhanced amplitude. These results are consistent with \autoref{fig:chromdiff3x3}. Note that a different assumption for the input model would have produced slightly different results. It is however still instructive to estimate the expected magnitude of the bias.

\begin{figure}
\includegraphics[width=\columnwidth]{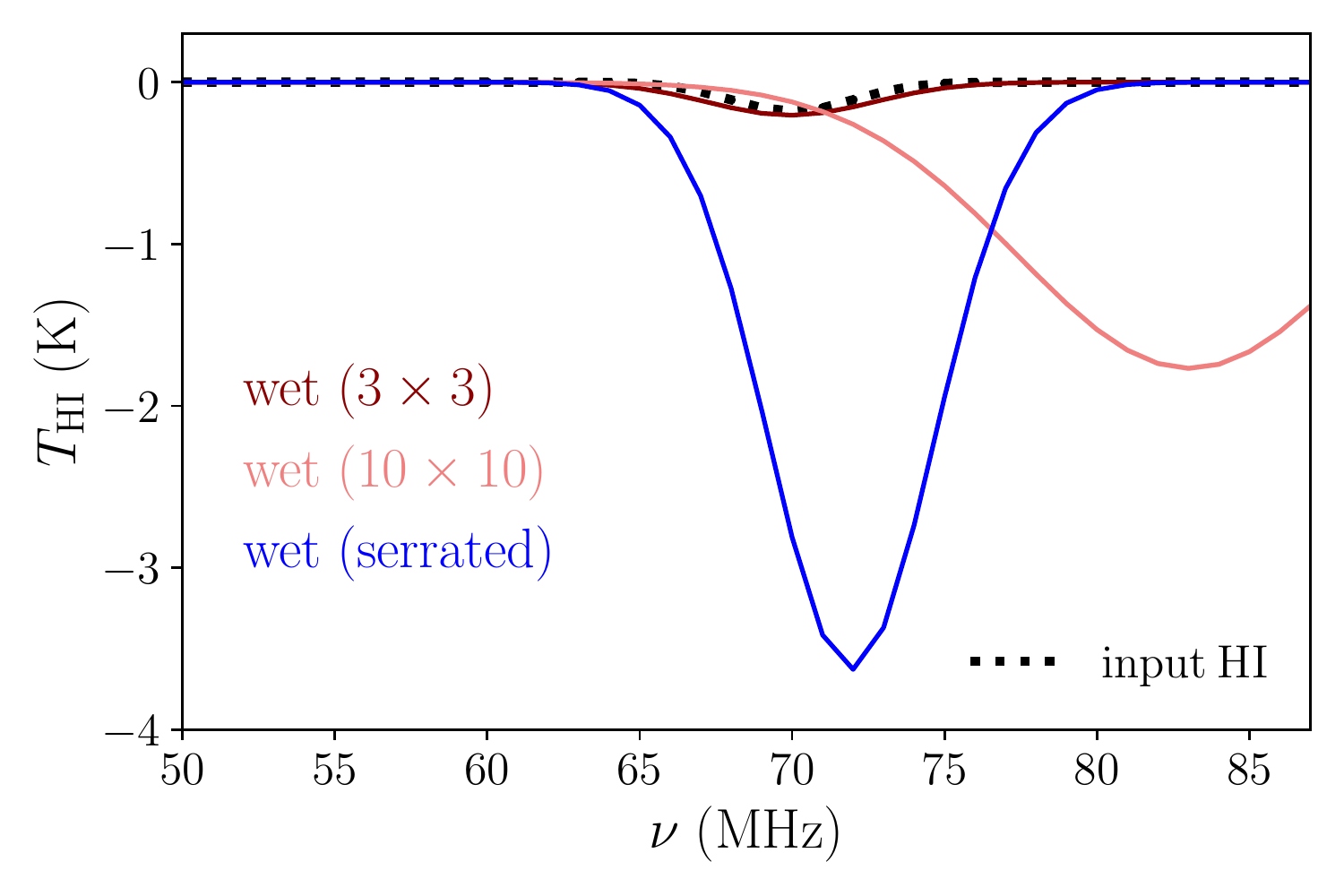}
\caption{The reconstructed $21$cm absorption profile
obtained when the simulated spectra (generated for the one-layer dry soil conditions) are corrected for the effect of chromaticity considering a wet soil moisture. We present the results for the three different ground planes considered ($3\times3$ in dark red, $10\times10$ in pink, and the serrated case in blue).}
\label{fig:diffgroundwet}
\end{figure}

\medskip
We investigate also a more drastic situation, where the soil conditions for the correcting chromaticity factor are the ``wet'' case presented  in \autoref{tab:soil_params} and \autoref{fig:chromdiffwet}. We report the reconstructed $T_{\rm HI}$ in \autoref{fig:diffgroundwet}. For the $3\times3$ ground plane, already analysed, the result is similar to the small variations in permittivity and conductivity just discussed, as could have been anticipated comparing the top panel of \autoref{fig:chromdiffwet} with \autoref{fig:chromdiff3x3}. 
The reconstruction is completely biased for the case of the larger ground planes, where the ``dry'' and ``wet'' chromaticity corrections present more structured differences (see the  middle and lower panel of \autoref{fig:chromdiffwet}). 

\begin{figure*}
\includegraphics[width=0.65\columnwidth]{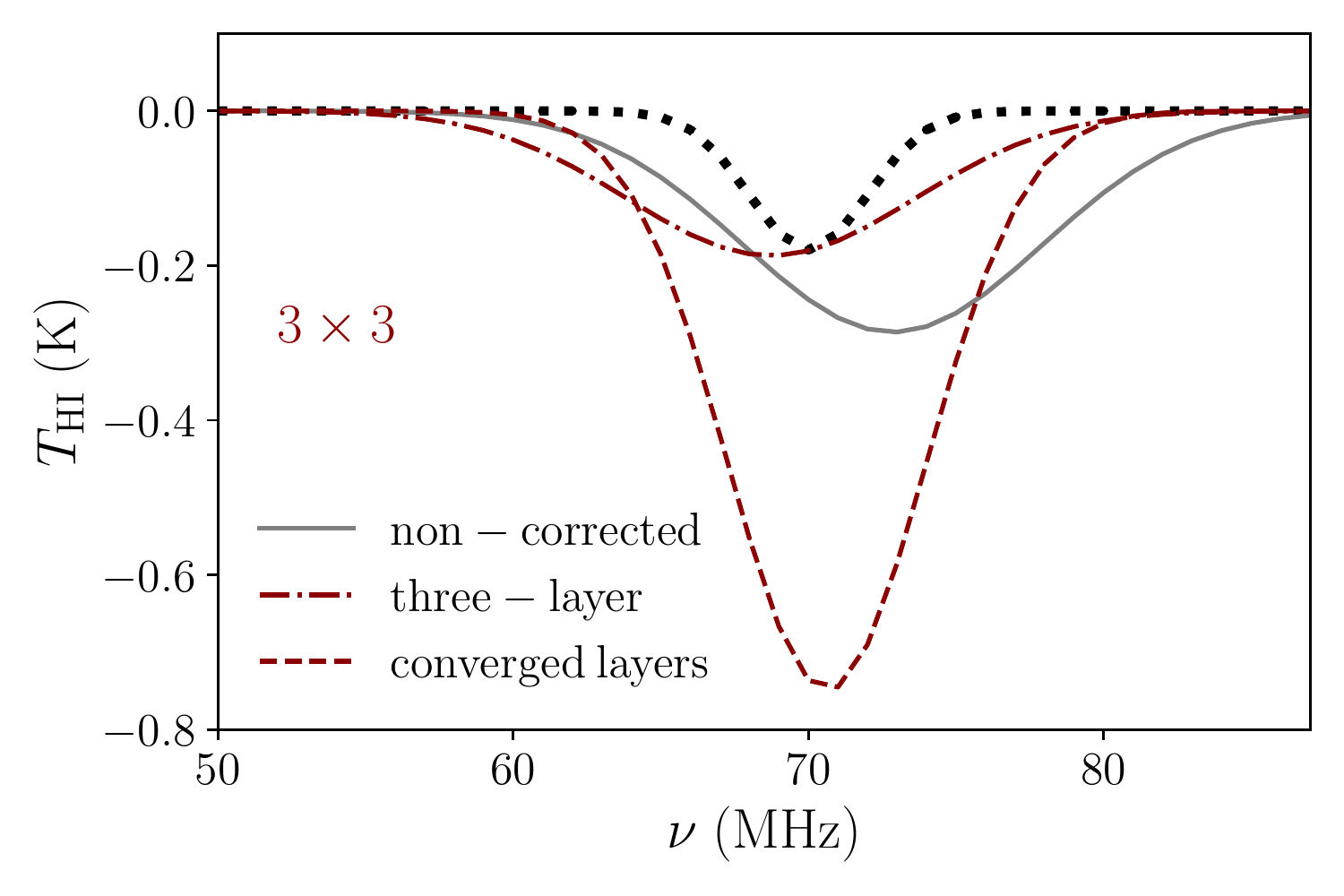}
\includegraphics[width=0.65\columnwidth]{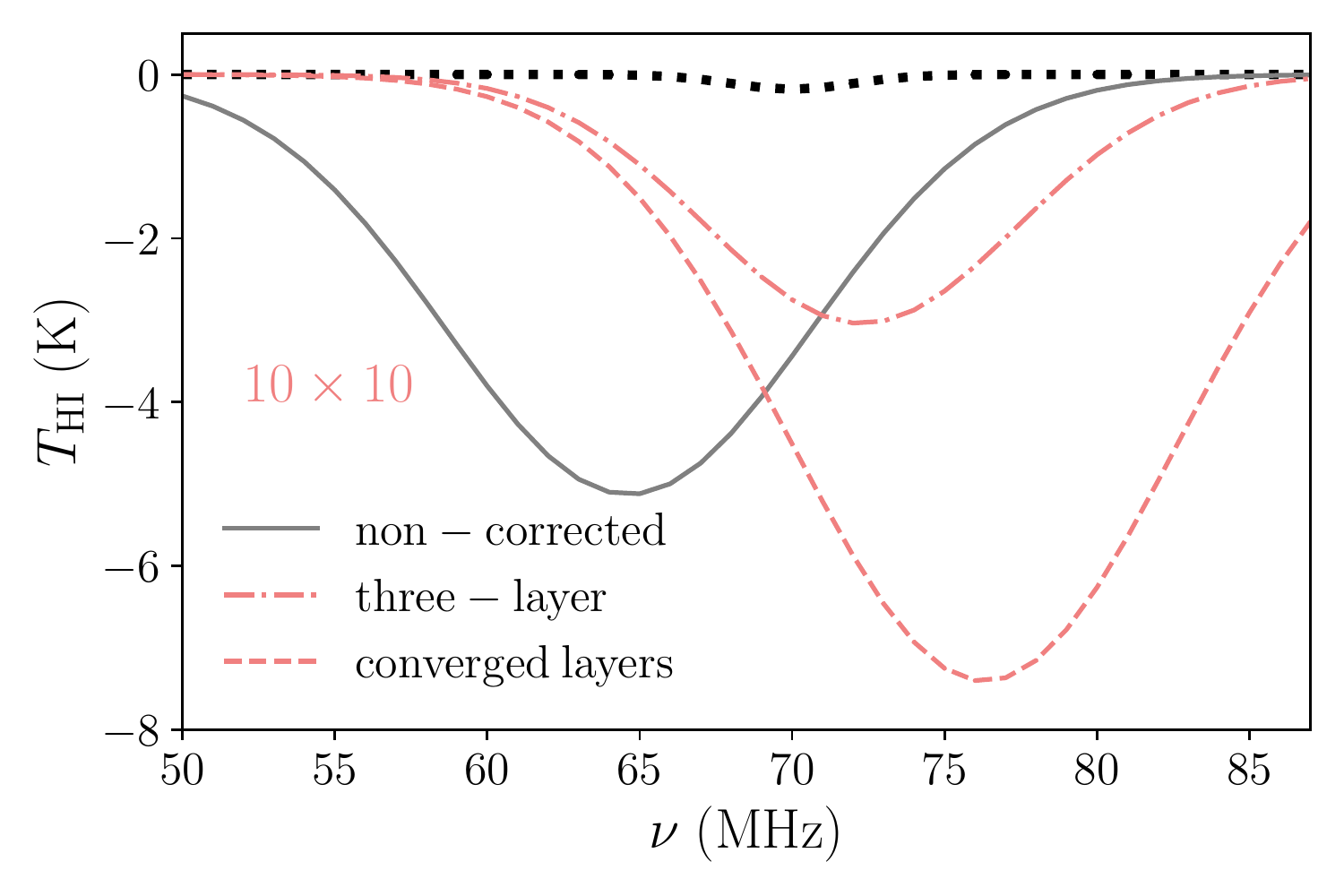}
\includegraphics[width=0.65\columnwidth]{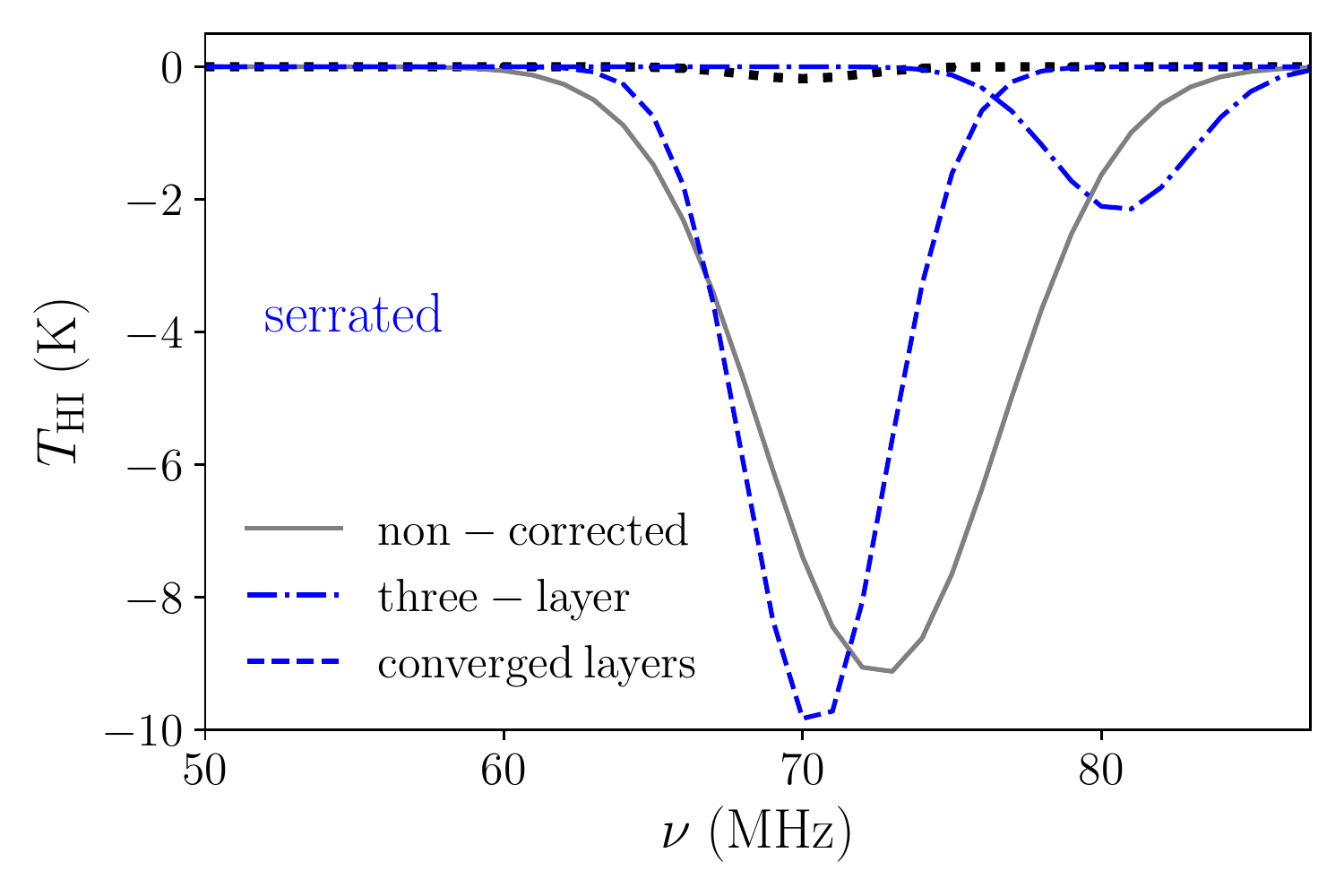}
\caption{The reconstructed absorption profile for the $3\times3$ (left panel), $10\times10$ (central panel) and serrated ground plane (right panel) for the one-layer dry soil baseline case spectra when we are correcting with a chromaticity factor computed using the three- or the converged layer model. Note the different vertical axis scale for the three panels. The dotted line is the input profile. For completeness, also the non-corrected case presented in \autoref{fig:diffground} is shown in grey in each panel.}
\label{fig:difflayers}
\end{figure*}

\medskip 
We analyse the impact of soil modelling further in \autoref{fig:difflayers} where we compare the three-layer and converged layer models against the baseline. As was hinted in \autoref{fig:res_layers}, for the $3\times 3$ case, the three-layer correction is better than no correction, while the converged layers description of the soil is already too much different from the one layer to offer better correction. The other ground plane cases show various biased results, proving that the correction is not solving the problem of the residual structures in the simulated spectra.

\subsection{Spectral index reconstruction}

\begin{figure*}
\includegraphics[width=16cm]{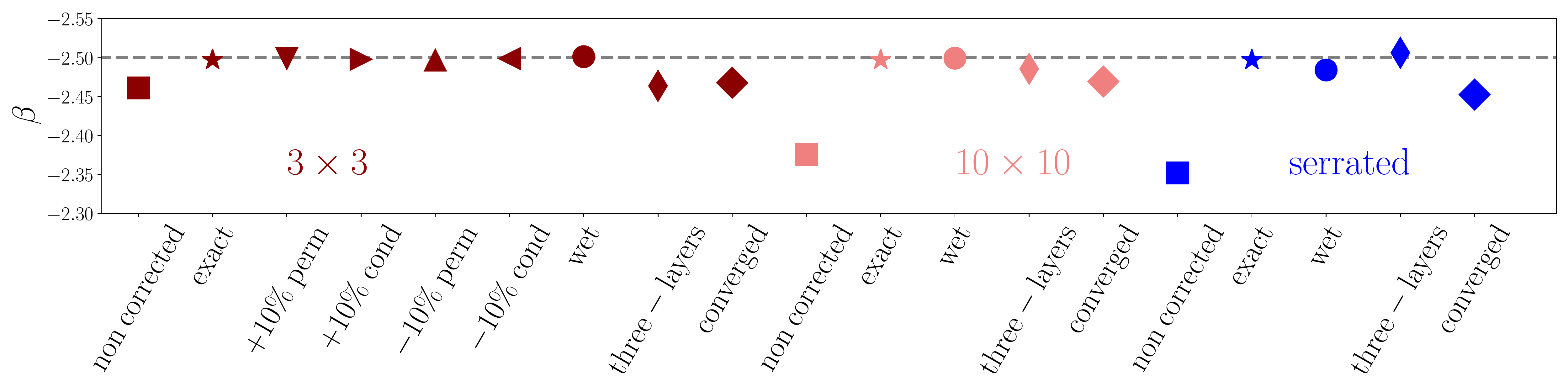}
\caption{Mean value for the posterior of the spectral index $\beta$ (corresponding to the second term of the log-polynomial model) for the different cases analysed in this work. The true input value of $\beta=-2.5$ is indicated with a dashed line to guide the eye. The results are colour-coded to distinguished the three different ground planes considered.}
\label{fig:betavar}
\end{figure*}

While studying the correct reconstruction of the HI absorption feature, we also discuss the reconstructed foreground parameters. The most informative one is the spectral index $\beta$ that can be compared with its input value $-2.5$ (see \secref{sec:Tobs}). We report the peak of the posterior distribution of the spectral index for each case in \autoref{fig:betavar}. We find that, when we do not correct for the effect of chromaticity, we recover up to a 10\% flatter $\beta$. The effect is stronger for larger ground planes. We instead always recover the right spectral index for the exact correction, or when the beam factor is computed varying the soil moisture properties. Finally, there is a tendency for a slightly flatter $\beta$ (a few \%) when we use different multi-layer modelling.
These flatter values of the spectral index are associated with strongly biased values of the amplitude of the absorption feature as can be seen in \autoref{fig:diffground} and \autoref{fig:difflayers}.

\section{Discussion and conclusions}
\label{sec:conclusions}

In this work, we presented the characterisation of the LEDA antenna beam,
with emphasis on the role of the ground plane and of the soil modelling both in terms of discretizing the semi-infinite volume it occupies, and varying its electromagnetic parameters. We used FEKO for our simulations and constructed a multi-layer model of the terrain relying on in-situ measurements of the soil complex permittivity. We used a more standard one-layer model under dry/wet soil conditions as our baseline and discussed the variations in the absolute gain pattern. when exploring more sophisticated models. The characterisation of the antenna beam is of primary importance in assuring an accurate enough control of the systematic effects in 21cm global signal analysis and has been studied in detail for other experiments \citep[e.g. ][]{Mahesh2021,Raghunathan2021}.

We explored the impact of beam modelling uncertainties due to soil moisture on the antenna chromaticity, focusing on three different ground planes used in the actual LEDA observations \citep{Spinelli2021}: a 3~m $\times$ 3~m, a 10~m $\times$ 10~m and a 10~m $\times$ 10~m with 10~m long triangular serrations. The addition of the ground plane induces a frequency ripple in the beam pattern that compromises its smoothness. The ripple amplitude is smaller for larger ground planes while its oscillation frequency is larger. The amplitude of the ripple depends also, to a lesser extent, on the number of layers that describe the permittivity as a function of depth. However, this effect saturates for a sufficiently large number of layers.

We then compared the gain at zenith for values of the complex permittivity corresponding to dry or wet terrain at the LEDA site. Surprisingly, the impact of different soil conditions is not suppressed by the presence of a larger ground plane and appears as a slight shift in the oscillatory pattern. The shift is more evident for the serrated ground plane case and affects the full shape of the beam. This makes the correction of the beam chromaticity more difficult and sensitive to electromagnetic property assumptions or measurement errors.

In future works, antenna beam simulations could be improved by
collecting more data to model the moisture conditions, leading to a better description of the complex permittivity as a function of depth \citep{Campbell1990,Bobrov2015}. Moreover, a more complex frequency dependence with respect to \autoref{eqn:complex_epsilon} could be adopted \citep[for example a Cole-Cole model as in][]{Sternberg2001}.
Note also that the finite accuracy of the numerical solver limits the refinement of the multi-layer model and different EM software solutions should be compared for benchmarking our model.

We followed with computing the beam chromaticity
factor and simulated observed beam-averaged sky spectra as
a function of frequency and LST, assuming the sky can be described as diffuse synchrotron emission scaled with a constant power law $\beta=-2.5$ across our frequency range. The differences in complex permittivity typical of dry or wet soil conditions translate to a few \textperthousand\ variations in the beam chromaticty factor with similar amplitude across ground planes, but increasingly more structured in frequency and LST for larger ground planes, as expected.

We modelled the simulated spectra with an $N$-term log-polynomial exploring N values from 5 to 7, and studied the behaviour of the residuals as a function of $N$, LST and beam model, finding negligible frequency structure only for the ideal case of an infinite ground plane.

We added a Gaussian absorption feature to the various simulated spectra to mimic the high redshift 21cm signal. The final model for the mock measured spectra consisted of the $N^{\rm th}$ term log-polynomial model plus three parameters to describe the Gaussian absorption feature (the central frequency $\nu_{21}$, the amplitude $A_{21}$ and the width $\sigma_{21}$).
We studied how beam model uncertainties propagate in the analysis, and bias the Bayesian model parameter reconstruction.

When the exact beam chromaticity correction is applied to the simulated spectra, the model parameters are reconstructed without any bias and residuals are around the mK level. 
However, we found much larger residuals when the chromaticity is not accounted for, or the beam model for the correction is obtained for different soil conditions or for the different multi-layer implementations. These residuals can reach up to a few hundreds (thousands) of mK for the $10{\rm m}\times 10{\rm m}$ (serrated) ground plane and bias the absorption signal reconstruction. Our results disfavour the use of a large ground plane coupled with the LEDA antenna 
as it seems that the interaction of the soil and the antenna itself lead to significant oscillating factors on the gain spectral response for all realistic ground plane sizes. 

The $3{\rm m}\times 3{\rm m}$ case behaves better and beam model uncertainties results in smaller parameter bias. Note however that a 10\% changes in permittivity or conductivity can enhance/reduce the parameters $A_{21}$ and $\sigma_{21}$ up to a factor of two or shift by few \% the recovered central frequency $\nu_{21}$.

Finally, we checked the effect of the various beam models on the reconstructed spectral index $\beta$ of the simulated spectra and find results in agreement with the input value. We observed 
a maximum 6\% flattening only when the correction for beam chromaticity is not applied.

Apart from some ideal cases, the smooth foreground log-polynomial model is found not to be an accurate description of the frequency structures induced in the observed spectra by realistic LEDA gain patterns, preventing the Bayesian exploration of the parameter space to converge to the expected result. In the future we will investigate how this effect can be mitigated by increasing the number of model parameters used to describe the foreground spatial distribution \citep[e.g.,][]{Anstey2020}.


\section*{acknowledgements}
LEDA research herein was supported in part by NSF grant AST/1616709. MS acknowledges support from the AstroSignals Synergia grant CRSII5\_193826 from the Swiss National Science Foundation. 
This research made use of \texttt{Numpy} \citep{Numpy2020}, \texttt{Astropy} \citep{Astropy} and \texttt{Scipy} \citep{Scipy2020}.
Some of the results in this paper have been derived using the \texttt{healpy} \citep{Zonca2019} and \textit{HEALPix} \citep{2005ApJ...622..759G} package.

\section*{Data availability}

The simulated antenna gain patterns used in this work can be found tabulated in \texttt{.mat} format in \href{https://drive.google.com/file/d/1xkxdh7N6B9w8f12ZG_UNuTzIBAKjrP5X/view?usp=sharing}{this drive folder}.

\bibliographystyle{mnras}
\bibliography{biblio}


\appendix

\section{Multi-layer implementation: choice of number of layers}\label{app:multi}

The multi-layer modelling and its fine-layer implementation, guided by the convergence of the beam gain,
have been discussed in \secref{sec:FEKO} and highlighted in \autoref{alg:sublayers} and \autoref{fig:example_layers}. Some more details on the choice of number of layers are offered here for completeness. Let us denote $\epsilon_{r,i}$ the value of the measurement of relative permittivity at a depth of $|z_i|,\: i=1,2,3$, while $z_0=0$. For the $i+1$-th sub-layer we have $|z_i|<|z|<|z_{i+1}|$ and the value of $\epsilon_r$ at a given $z$ can be obtained through a linear interpolation of the available values: 
\begin{equation}
    \epsilon_r(z)=\epsilon_{r,i}+\Delta\epsilon_r\frac{|z-z_i|}{|z_{i+1}-z_i|}.
\end{equation}
By construction, $ \epsilon_r(z)<\epsilon_{r,i+1}=\epsilon_{r,i}+\Delta\epsilon_r$ such that always
\begin{equation}
\lambda_p(z)=\frac{2\pi c_0}{\epsilon_r(z)\nu}>\frac{2\pi c_0}{\epsilon_{r,i+1}\nu}=\lambda_{p,i+1},
\end{equation}
ensuring that the $\lambda_{p}/10$ rules at the depth
$i+1$-th is stronger than the one for a sub-layer at z. 
This is important, since we can use the upper integer value $ \frac{|z_{i+1}-z_i|}{\lambda_{p,i+1}/10}$ as the number of initial sub-layers beyond the three initial $z_i$ and iteratively double this number until convergence is reached.

\section{Dependence on the assumed input absorption profile}\label{app:diffHI}

\begin{figure}
\includegraphics[width=\columnwidth]{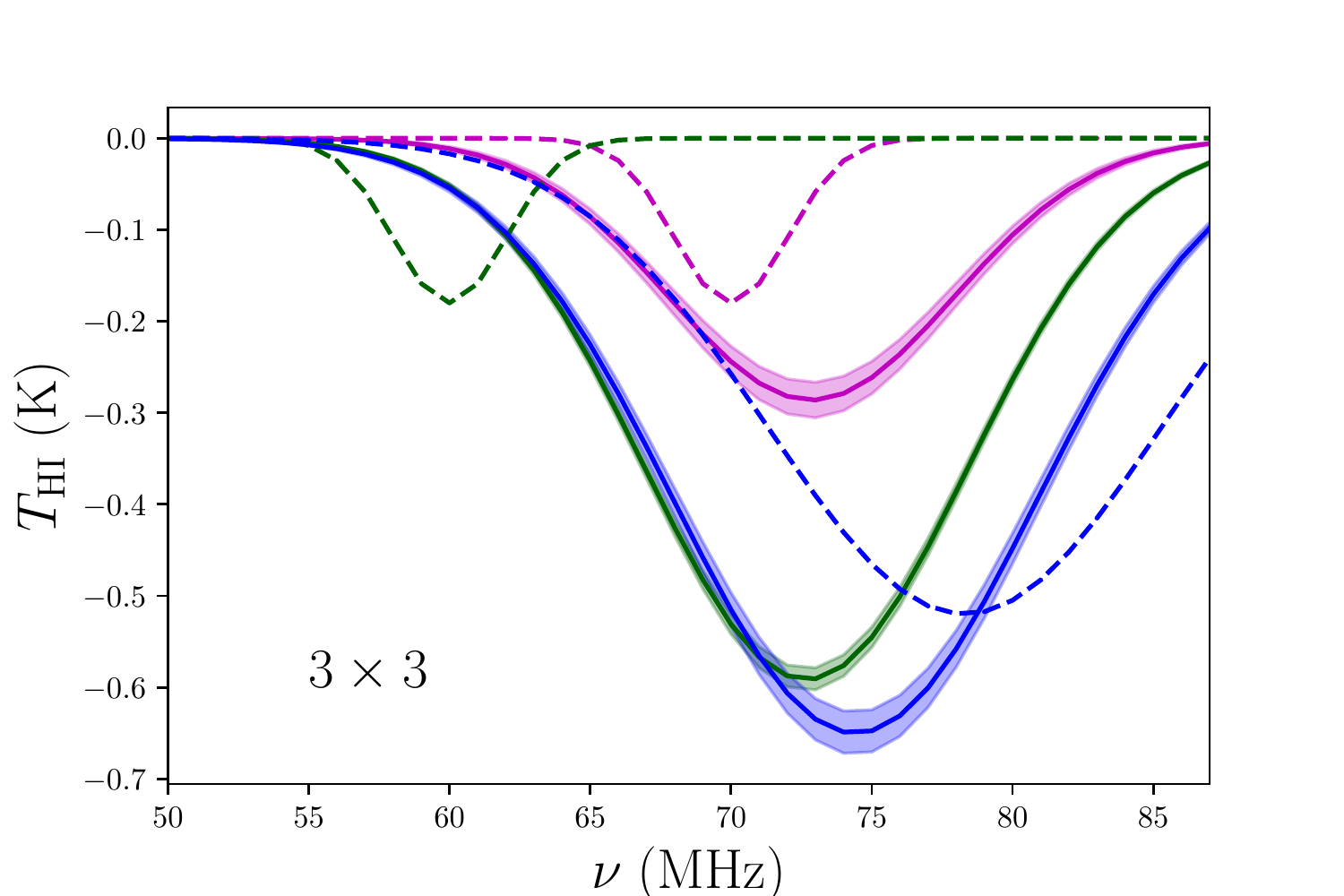}
\caption{The reconstructed $21$cm absorption profile (solid lines) considering three different input models (dashed lines). The model in blue acts for an EDGES-like type of absorption feature. The simulated spectra are generated considering a $3\times3$ ground plane and one-layer dry soil conditions and are not corrected for the effect of chromaticity. Shaded area correspond to the propagated 1-$\sigma$ uncertainties in the parameter posteriors.}
\label{fig:diffHI}
\end{figure}

As discussed in \autoref{sec:sim}, in this work we focus on the effect of the beam modelling on the analysis using simulated data. We assume for the cosmological signal a simple Gaussian absorption feature with fixed values for the parameters $A_{21}$, $\nu_{21}$  and $\sigma_{21}$ (see \autoref{eq:gauss}).
To qualitatively address the consistency of our conclusions with respect to the input 21cm signal choice, we check in this session the effect varying the parameters of \autoref{eq:gauss}.
We report in \autoref{fig:diffHI}, together with the case analysed in the rest of the paper (in magenta in the figure), the reconstructed signal considering i) a similar signal with lower $\nu_{21}$ (in green) or ii) an EDGES-type profile (in blue).
We focus on our baseline beam model for the $3 \times 3$ ground plane.
As expected, the results are not identical for the three cases although they all led to conclude that the reconstructed absorption feature is larger, deeper and not necessarily correctly located within the analysed band.
We note however that we expect a stronger signal to be less biased. This is why, to be conservative, we have assumed for the analysis in this work an absorption feature shallower than the EDGES results.

\section{Chromaticity correction plot}\label{app:chrom_diff}

We report in this appendix some useful further results to complement the discussions of \secref{sec:chrom}. We recall that in this work we have explored two main ingredients for the beam simulation: the characteristic of the soil moisture and a multi-layer approach to soil modelling. 

\begin{figure*}
\includegraphics[width=\columnwidth]{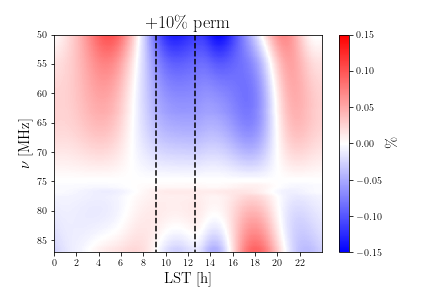}
\includegraphics[width=\columnwidth]{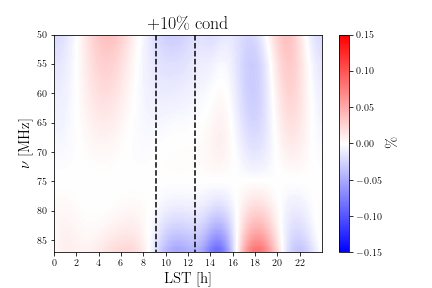}\\
\includegraphics[width=\columnwidth]{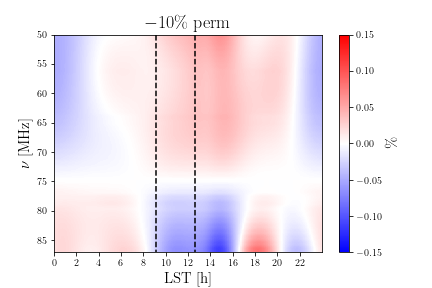}
\includegraphics[width=\columnwidth]{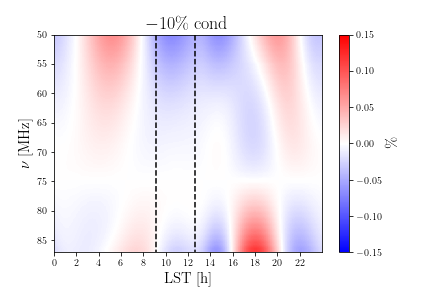}
\caption{The difference in percentage between the chromaticity factor $B_c$ (see \autoref{eq:beam_chromaticity}) computed for the one-layer dry condition FEKO model case and the ones where the or the conductivity are varied by $+10\%$ (top row) or by $-10\%$ (bottom row). We consider here the $3\times3$~m ground plane.}
\label{fig:chromdiff3x3}
\end{figure*}

We report the percentage difference in the chromaticity pattern between the baseline dry soil condition and $\pm10\%$ variation in the input values of permittivity $\epsilon_r$ and conductivity $\sigma$ for the soil in \autoref{fig:chromdiff3x3}. 
While the explored variations of the conductivity have a similar impact on the chromaticity correction factors presented in the figure, lowering the permittivity shows an almost inverted pattern for the chromaticity percentage change. Note, however, that the variations are always below 2\textperthousand. 

\begin{figure*}
\includegraphics[width=\columnwidth]{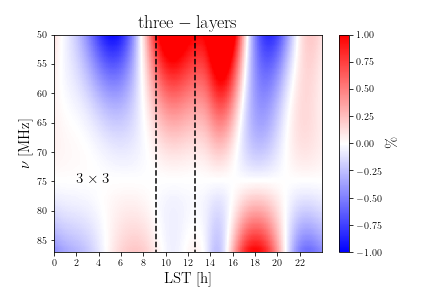}
\includegraphics[width=\columnwidth]{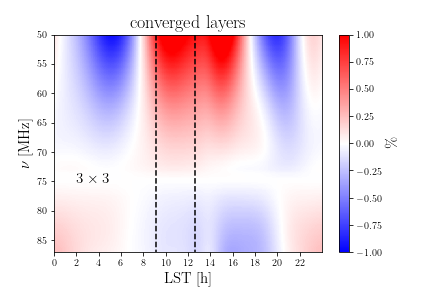}\\
\includegraphics[width=\columnwidth]{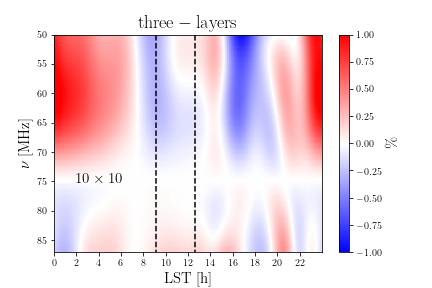}
\includegraphics[width=\columnwidth]{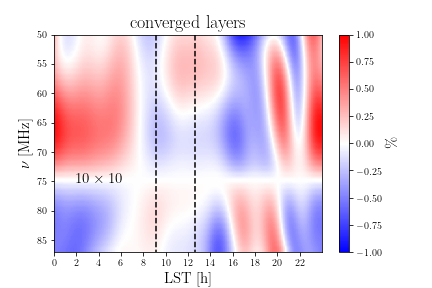}\\
\includegraphics[width=\columnwidth]{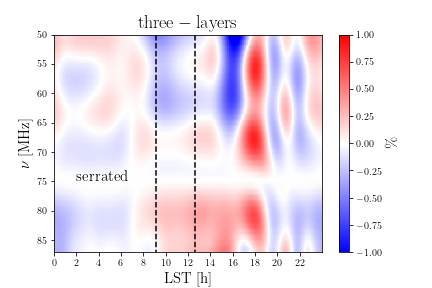}
\includegraphics[width=\columnwidth]{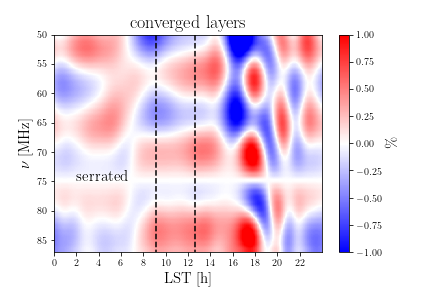}
\caption{The difference in percentage between the chromaticity factor $B_c$ (see \autoref{eq:beam_chromaticity}) computed for the one-layer dry condition FEKO model and the three-layer (left) or the converged layers case (right) for the three different type of ground planes: $3\times3$ (top), $10\times10$ (middle) and serrated (bottom).}
\label{fig:chromdiffgrounds}
\end{figure*}

In \autoref{fig:chromdiffgrounds} we report instead the difference in the beam chromaticity between the baseline dry one-layer case and the three or converged multi-layer models described in \secref{sec:FEKO} (see also \autoref{tab:beam_type}). We explore these differences as a function of the different ground planes used in LEDA analysis.

\section{Residuals with respect to the smooth foreground model}\label{app:FoM_multi}

As discussed in \secref{sec:model}, assuming no absorption feature, we can assess the impact of a non perfect beam chromaticity correction on the mock sky spectra
looking at the residual structure after subtraction of the smooth foreground model of \autoref{eq:fg}.

In this appendix, we complement the visual content of \autoref{fig:res_soil} and \autoref{fig:res_layers} reporting  the root mean square (rms) of the residuals in \autoref{tab:rms_soil} and \autoref{tab:rms_layers}. We recall that we construct the spectra with the baseline model (one-layer, dry soil condition) and then we correct for the effect of chromaticity with a slightly different beam, changing the soil moisture (\autoref{tab:rms_soil}) or the soil layering (\autoref{tab:rms_layers}), respectively. We analyse different N-term log-polynomial models (5,6 and 7) and different ground planes ($3\times 3$, $10\times10$ and serrated). 
Note that the non-corrected and the exact correction cases are the same of in both Tables.  Note that best fit values for the foreground parameters are obtained with a non-linear least squares solver.

\begin{table}
\caption{The rms (in K) of the residual difference, reported in \autoref{fig:res_soil}, between the best fit model of \autoref{eq:fg} and the simulated spectra obtained as described in \secref{sec:Tobs}, presented for different polynomial orders ($N=4,5,6$) and for different ground planes. The simulated spectra are obtained considering dry condition and a one-layer description for the soil. The residuals are then computed without any chromaticity correction, with an exact correction using the same beam model as for the spectra in input and for a chromaticity correction computed with wet soil condition instead.}\label{tab:rms_soil}
\begin{center}
\begin{tabular}{l|l|l|l|l|} 
ground plane & chrom. correction & \multicolumn{3}{c}{residual rms (K)}\\
 & & N=5 & N=6 & N=7 \\
\hline
\hline
$3\times3$ & non corrected &  0.086 &  0.071 & 0.018\\
& exact & 0.002 &  0.002 & 0.002 \\
& wet soil & 0.084 & 0.116 & 0.019\\
\hline
$10\times10$ & non corrected & 0.639  & 0.942 & 0.100\\
& exact  &  0.002 & 0.002 & 0.002\\
& wet soil & 0.226 & 0.301 & 0.064\\
\hline
serrated & non corrected   & 2.985 & 3.260 & 1.279\\
& exact &  0.006 & 0.006 & 0.003\\
& wet soil & 2.085 & 2.491 & 0.969 \\
\end{tabular}
\end{center}
\end{table}

\begin{table}
\caption{The rms (in K) of the residual difference, reported in \autoref{fig:res_layers}, between the best fit model of \autoref{eq:fg} and the simulated spectra obtained as described in \secref{sec:Tobs}, presented for different polynomial orders ($N=4,5,6$) and for different ground planes. The simulated spectra are obtained considering dry condition and a one-layer description for the soil. The residuals are then computed without any chromaticity correction, with an exact correction using the same beam model as for the input spectra, and for chromaticity corrections computed using the three-layer or converged layer model instead.}\label{tab:rms_layers}
\begin{center}
\begin{tabular}{l|l|l|l|l|} 
ground plane & chrom. correction & \multicolumn{3}{c}{residual rms (K)}\\
 & & N=5 & N=6 & N=7 \\
\hline
\hline
$3\times3$ & non corrected &  0.086 &  0.071 & 0.018\\
& exact & 0.002 &  0.002 & 0.002 \\
& three-layers & 0.046 & 0.031 & 0.017\\
& converged layers & 0.158 & 0.151 & 0.095\\
\hline
$10\times10$ & non corrected & 0.639  & 0.942 & 0.100\\
& exact  &  0.002 & 0.002 & 0.002\\
& three-layers &  0.463 & 0.696 & 0.154\\
& converged layers  & 1.151 & 1.152 & 0.268\\
\hline
serrated & non corrected   & 2.985 & 3.260 & 1.279\\
& exact &  0.006 & 0.006 & 0.003\\
& three-layers  & 1.661 & 1.788 & 1.259\\
& converged layers  & 4.802 & 6.400 & 4.076
\end{tabular}
\end{center}
\end{table}

\bsp	
\label{lastpage}
\end{document}